\pdfoutput=1

\documentclass[12pt,a4paper,hyperref]{jhepLike}
\let\ifpdf\relax
\usepackage{ifpdf,color}
\let\normalcolor\relax
\usepackage{mathtools,cite}
\newcommand{\eq}[1]{\vspace{-0.5pt}\begin{equation}#1\vspace{-0.5pt}\end{equation}}
\newcommand{\fwbox}[2]{\text{\makebox[#1][c]{$\hspace{-150pt}\displaystyle#2\hspace{-150pt}$}}}
\newcommand{\fwboxL}[2]{\text{\makebox[#1][l]{$#2$}}}
\newcommand{\fwboxR}[2]{\text{\makebox[#1][r]{$#2$}}}
\newcommand{\fig}[3]{\raisebox{#1}{\ \includegraphics[scale=#2]{#3}}}
\newcommand{\mi}{\raisebox{0.75pt}{\scalebox{0.75}{$\,-\,$}}}
\newcommand{\pl}{\raisebox{0.75pt}{\scalebox{0.75}{$\,+\,$}}}
\renewcommand{\phi}{\varphi}
\newcommand{\ab}[1]{\langle\hspace{-0.5pt}#1\hspace{-0.5pt}\rangle}
\newcommand{\bigger}[1]{\raisebox{-2.25pt}{\scalebox{1.75}{$#1$}}}
\newcommand{\x}[2]{x_{#1\hspace{0.5pt}#2}^2}
\renewcommand{\hat}{\widehat}
\definecolor{dim}{rgb}{0.75,0.75,0.75}
\definecolor{hblue}{rgb}{0.18,0.19,0.572}
\definecolor{hred}{rgb}{0.745,0.118,0.176}

\title{{\LARGE \mbox{Amplitudes and Correlators to Ten Loops}}\\ {\LARGE\mbox{Using Simple, Graphical Bootstraps}}}
\author{{\normalsize \mbox{Jacob~L.~Bourjaily$^1$, Paul~Heslop$^2$, Vuong-Viet~Tran\mbox{$^{2}$}}}\\
\mbox{{\mbox{$^1$}\ Niels Bohr International Academy and Discovery Center, University of Copenhagen}}\\
\mbox{{\mbox{$^{\phantom{1}}$}\ Blegdamsvej 17, DK-2100 Copenhagen \O, Denmark}}\\
\mbox{{\mbox{$^2$}\ Centre for Particle Theory, Department of Mathematical Sciences, Durham University,}}\\
\mbox{{\mbox{$^{\phantom{2}}$}\ South Road, Durham DH1 3LE, United Kingdom}}\\\vspace{-10pt}\\
\mbox{\hspace{-13pt}{{\it E-mails:} {\tt bourjaily@nbi.ku.dk, paul.heslop@durham.ac.uk,}}}\\\vspace{-15pt}\\\mbox{\hspace{28pt}{ {\tt vuong-viet.tran@durham.ac.uk}}}\vspace{-10pt}
}
\keywords{scattering amplitudes, correlation functions, supersymmetry}
\date{\today}
\abstract{%
We introduce two new graphical-level relations among possible contributions to the four-point correlation function and scattering amplitude in planar, maximally supersymmetric Yang-Mills theory. When combined with the rung rule, these prove powerful enough to fully determine both functions through ten loops. This then also yields the full five-point amplitude to eight loops and the parity-even part to nine loops. We derive these rules, illustrate their applications, compare their relative strengths for fixing coefficients, and survey some of the features of the previously unknown nine and ten loop expressions. Explicit formulae for amplitudes and correlators through ten loops are available at: \href{http://goo.gl/JH0yEc}{http://goo.gl/JH0yEc}.
}
\preprint{DCPT-16/31}

\begin{document}
\vspace{-27.8pt}\section{Introduction}\label{sec:introduction}\vspace{-6pt}
Among all four-dimensional quantum field theories lies a unique example singled out for its remarkable symmetry and mathematical structure as well as its key role in the AdS/CFT correspondence. This is maximally supersymmetric ($\mathcal{N}\!=\!4$) Yang-Mills theory (SYM) in the planar limit \cite{ArkaniHamed:2008gz}. It has been the subject of great interest over recent years, and the source of many remarkable discoveries that may extend to much more general quantum field theories. These features include a connection to Grassmannian geometry \cite{ArkaniHamed:2009dn,ArkaniHamed:2009sx,ArkaniHamed:2009dg,ArkaniHamed:2012nw,ArkaniHamed:book}, extra simplicity for planar theories' loop integrands \cite{ArkaniHamed:2010gh,Bourjaily:2011hi,Bourjaily:2015jna}, the existence of all-loop recursion relations \cite{ArkaniHamed:2010kv}, and the existence of unanticipated symmetries \cite{Drummond:2007cf,Drummond:2008vq,Brandhuber:2008pf,Drummond:2009fd} and related dualities between observables in the theory \cite{Alday:2007hr,Drummond:2007aua,Brandhuber:2007yx,Alday:2010zy,Eden:2010zz,Mason:2010yk,CaronHuot:2010ek,Eden:2011yp,Adamo:2011dq,Eden:2011ku}. Of these, the duality between scattering amplitudes and correlation functions, will play a fundamental role throughout this work. 

Much of this progress has been fueled through concrete theoretical data: heroic efforts of computation are made to determine observables (with more states, and at higher orders of perturbation); and this data leads to the discovery of new patterns and structures that allow these efforts to be extended even further. This virtuous cycle---even when applied merely to the `simplest' quantum field theory---has taught us a great deal about the structure of field theory in general, and represents an extremely fruitful way to improve our ability to make predictions for experiments. 

In this paper, we greatly extend the reach of this theoretical data by computing a particular observable in this simple theory to {\it ten} loops---mere months after eight loops was first determined. This is made possible through the use of powerful new {\it graphical} rules described in this work. The observable in question is the four-point correlation function among scalars---the simplest operator that receives quantum corrections in planar SYM. This correlation function is closely related to the four-particle scattering amplitude, as reviewed below. But the information contained in this single function is vastly more general: it contains information about all scattering amplitudes in the theory---including those involving more external states (at lower loop-orders). As such, our determination of the four-point correlator at ten loops immediately provides information about the five-point amplitude at nine loops, the six-point amplitude at eight loops, etc.\ \cite{Ambrosio:2013pba}. 

Before we review this correspondence and describe the rules used to obtain the ten loop correlator, it is worth taking a moment to reflect on the history of our knowledge about it. Considered as an amplitude, it has been the subject of much interest for a long time. The tree-level amplitude was first expressed in supersymmetric form by Nair in \mbox{ref.\ \cite{Nair:1988bq}}. It was computed using unitarity to two loops in 1997 \cite{Bern:1997nh} (see also \cite{Anastasiou:2003kj}), to three loops in 2005 \cite{Bern:2005iz}, to five loops in 2007---first at four loops \cite{Bern:2006ew}, and five quickly thereafter \cite{Bern:2007ct}---and to six loops around 2009 \cite{Bern:2012di} (although published later). The extension to seven loops required significant new technology. This came from the discovery of the soft-collinear bootstrap in 2011 \cite{Bourjaily:2011hi}. Although not known at the time, the soft-collinear bootstrap method (as described in \mbox{ref.\ \cite{Bourjaily:2011hi}}), would have failed beyond seven loops; but luckily, the missing ingredient would be supplied by the duality between amplitudes and correlation functions discovered in \cite{Eden:2010zz,Alday:2010zy} and elaborated in \cite{Eden:2010ce,Eden:2011yp,Eden:2011ku,Adamo:2011dq}. The determination of the four-point correlator in planar SYM followed a somewhat less linear trajectory. One and two loops were obtained soon after (and motivated by) the AdS/CFT correspondence between 1998 and 2000 \cite{GonzalezRey:1998tk,Eden:1998hh,Eden:1999kh,Eden:2000mv,Bianchi:2000hn}. But despite a great deal of effort by a number of groups, the three loop result had to wait over 10 years until 2011---at which time the four, five, and six loop results were found in quick succession \cite{Eden:2011we,Eden:2012tu,Ambrosio:2013pba,Drummond:2013nda}; seven loops was reached in 2013 \cite{Ambrosio:2013pba}. 

The breakthrough for the correlator, enabling this rapid development, was the discovery of a hidden symmetry \cite{Eden:2011we,Eden:2012tu}. On the amplitude side, the extension of the above methods to eight loops also required the exploitation of this symmetry via the duality between amplitudes and correlators. This hidden symmetry (reviewed below) greatly simplifies the work required to extend the soft-collinear bootstrap, making it possible to determine the eight loop functions in 2015 \cite{Bourjaily:2015bpz}. 

While the eight loop amplitude and correlator were determined (the `hard way',) using just the soft-collinear bootstrap and hidden symmetry, we had already started exploring alternative methods to find these functions which seemed quite promising. These were mentioned in the conclusions of \mbox{ref.\ \cite{Bourjaily:2015bpz}}---the details of which we describe in this note. This new approach, based not on algebraic relations but graphical ones, has allowed for a watershed of new theoretical data similar to that of 2007: within a few short months, we were able to fully determine both the nine and ten loop correlation functions. The reason for this great advance---the (computational) advantages of graphical rules---will be discussed at the end of this introduction. 

Our work is organized as follows. In \mbox{section \ref{sec:review_of_duality}} we review the representation of amplitudes and correlation functions, and the duality between them. This will include a summary of the notation and conventions used throughout this paper, and also a description of the way that the terms involved are represented both algebraically and graphically. We elaborate on how the plane embedding of the terms that contribute to the correlator (viewed as graphs) allow for the direct extraction of amplitudes at corresponding (and lower) loop-orders---including amplitudes involving more than four external states---in \mbox{section \ref{subsec:amplitude_extraction}}. The three graphical rules sufficient to fix all possible contributions (at least through ten loops) are described in \mbox{section \ref{sec:graphical_bootstraps}}. We will refer to these as the triangle, square, and pentagon rules. 

The triangle and the square rules relate terms at different loop orders, while the pentagon rule relates terms at a given loop-order. While the square rule is merely the graphical manifestation of the so-called `rung' rule \cite{Bern:1997nh,Eden:2012tu} (generalized by the hidden symmetry of the correlator), the triangle and pentagon rules are new. We provide illustrations of each and proofs of their validity in \mbox{section \ref{sec:graphical_bootstraps}}. These rules have varying levels of strength. While the square rule is well-known to be insufficient to determine the amplitude or correlator at all orders (and the same is true for the pentagon rule), we expect that the combination of the square and triangle rules {do} prove sufficient---but only after their consequences at higher loop-orders are taken also into account. (For example, the pentagon rule was not required for us to determine the nine loop correlator---but the constraints that follow from the square and triangle rules at ten loops were necessary.) In \mbox{section \ref{sec:results}} we describe the varying strengths of each of these rules, and summarize the expressions found for the correlation function and amplitude through ten loops in \mbox{section \ref{subsec:statistical_tour}}. The explicit expressions for the ten loop correlator and amplitude have been made available at  \href{http://goo.gl/JH0yEc}{http://goo.gl/JH0yEc}. Details on how this data can be obtained and the functionality provided (as part of a bare-bones {\sc Mathematica} package) are described in \mbox{Appendix \ref{appendix:mathematica_and_explicit_results}}.

Before we begin, however, it seems appropriate to first describe what accounts for the advance---from eight to ten loops---in such a short interval of time. This turns out to be entirely a consequence of the computational power of working with graphical objects over algebraic expressions. The superiority of a graphical framework may not be manifest to all readers, and so it is worth describing why this is the case---and why a direct extension of the soft-collinear bootstrap beyond eight loops (implemented algebraically) does not seem within the reach of existing resources.

\paragraph{Why {Graphical} Rules?}~\\
\indent It is worth taking a moment to describe the incredible advantages of {\it graphical} methods over analytic or algebraic ones. The integrands of planar amplitudes or correlators can only meaningfully be defined if the labels of the internal loop momenta are fully symmetrized. Only then do they become well-defined, rational functions. But this means that, considered as algebraic functions, even {\it evaluation} of an integrand requires summing over all the permuted relabelings of the loop momenta (not to mention any cyclic or dihedral symmetrization of the legs that is also required). Thus, any analysis that makes use of evaluation will be rendered computationally intractable beyond some loop-order by the simple factorial growth in the time required by symmetrized evaluation. 

This is the case for the soft-collinear bootstrap as implemented in \mbox{ref.\ \cite{Bourjaily:2015bpz}}. At eight loops, the system of equations required to find the coefficients is a relatively straight-forward problem in linear algebra; and solving this system of equations is well within the limits of a typical laptop computer. However, {\it setting up} this linear algebra problem requires the evaluation of many terms---each at a sufficient number of points in loop-momentum space. And even with considerable ingenuity (and access to dozens of CPUs), these evaluations required more than two weeks to complete. Extending this method to nine loops would cost an additional factor of 9 from the combinatorics, and also a factor of 15 from the growth in the number of unknowns. This seems well beyond the reach of present-day computational resources. 

However, when the terms involved in the representation of an amplitude or correlator are considered more abstractly as {\it graphs}, the symmetrization required by evaluation becomes irrelevant: relabeling the vertices of a graph clearly leaves the {\it graph} unchanged. And it turns out that graphs can be compared with remarkable efficiency. Indeed, {\sc Mathematica} has built-in (and impressive) functionality for checking if two graphs are isomorphic (providing all isomorphisms that may exist). This means that relations among terms, when expressed as identities among graphs, can be implemented well beyond the limits faced for any method requiring evaluation.

We do not yet know of how the soft-collinear bootstrap can be translated as a graphical rule. And this prevents its extension beyond eight loops---at least at any time in the near future. However, the graphical rules we describe here prove sufficient to uniquely fix the amplitude and correlator through at least ten loops, and reproduce the eight loop answer in minutes rather than weeks. The extension of these ideas---perhaps amended by a broader set of analogous rules---to higher loops seems plausible using existing computational resources. Details of what challenges we expect in going to higher orders will be described in the conclusions.

\newpage
\vspace{-6pt}\section{Review of Amplitude/Correlator Duality}\label{sec:review_of_duality}\vspace{-6pt}
Let us briefly review the functional forms of the four-particle amplitude and correlator in planar maximally supersymmetric ($\mathcal{N}\!=\!4$) Yang-Mills theory (SYM), the duality that exists between these observables, and how each can be represented analytically as well graphically at each loop-order. This will serve as a casual review for readers already familiar with the subject; but for those less familiar, we will take care to be explicit about the (many, often implicit) conventions. 

The most fundamental objects of interest in any conformal field theory are gauge-invariant operators and their correlation functions. Perhaps the simplest operator in planar SYM is $\mathcal{O}(x)\!\equiv\!\mathrm{Tr}(\phi(x)^2)$, where $\phi$ is one of the six scalars of the theory and the trace is taken over gauge group indices (in the adjoint representation). This is a very special operator: it is related by (dual) superconformal symmetry to both the stress-energy tensor and the on-shell Lagrangian, is dual to supergravity states on AdS$_5$, protected from renormalization, and annihilated by half of the supercharges of the theory. Moreover, its two- and three-point correlation functions are protected from perturbative corrections. 

The four-point correlator involving $\mathcal{O}(x)$,
\eq{\mathcal{G}_4(x_1,x_2,x_3,x_4)\equiv\langle\mathcal{O}(x_1)\overline{\mathcal{O}}(x_2)\mathcal{O}(x_3)\overline{\mathcal{O}}(x_4)\rangle,\label{definition_of_correlator}}
is therefore the first non-trivial observable of interest in the theory. This correlator, computed perturbatively in loop-order and divided by the tree-level correlator is related to the four-particle amplitude (also divided by the tree) in a simple way \cite{Alday:2010zy,Eden:2010zz}:
\eq{\lim_{\substack{\text{4-point}\\\text{light-like}}}\left(\frac{\mathcal{G}_4^{}(x_1,x_2,x_3,x_4)}{\mathcal{G}_4^{(0)}\!(x_1,x_2,x_3,x_4)}\right)=\mathcal{A}^{}_4(x_1,x_2,x_3,x_4)^2,\label{correlator_amplitude_relation}}
where the amplitude is represented in dual-momentum coordinates, \mbox{$p_a\!\equiv\!x_{a+1}\mi x_a$}, and the light-like limit corresponds to taking the four (otherwise generic) points $x_a\!\in\!\mathbb{R}^{3,1}$ to be light-like separated: defining $x_{ab}\!\equiv\!x_b\mi x_a$, this corresponds to the limit where $\x{1}{2}\!=\!\x{2}{3}\!=\!\x{3}{4}\!=\!\x{1}{4}\!=\!0$. Importantly, while the correlator is generally finite upon integration, the limit taken in (\ref{correlator_amplitude_relation}) is divergent; however, the correspondence exists at the level of the loop {\it integrand}---both of which can be uniquely defined in any (planar) quantum field theory upon symmetrization in (dual) loop-momentum space. 

As a loop integrand, both sides of the identity (\ref{correlator_amplitude_relation}) are rational functions in $(4\pl\ell)$ points in $x$-space---to be integrated over the $\ell$ additional points, which we will (suggestively) denote as $x_{4+1},\ldots,x_{4+\ell}$. While the external points $x_1,\ldots,x_4$ would seem to stand on rather different footing relative to the loop momenta, it was noticed in \mbox{ref.\ \cite{Eden:2011we}} that this distinction disappears completely if one considers instead the function (appropriate for the component of the supercorrelator in (\ref{definition_of_correlator})),
\eq{\mathcal{F}^{(\ell)}(x_1,\ldots,x_4,x_5,\ldots,x_{4+\ell})\equiv\frac{1}{2}\left(\frac{G_4^{(\ell)}(x_1,x_2,x_3,x_4)}{G_4^{(0)}\!(x_1,x_2,x_3,x_4)}\right)/\xi^{(4)},\label{definition_of_f}}
where $\xi^{(4)}$ is defined to be $\x{1}{2}\x{2}{3}\x{3}{4}\x{1}{4}(\x{1}{3}\x{2}{4})^2$. As the attentive reader may infer, we will later have use to generalize this---yielding $\xi^{(4)}$ as a particular instance of,
\eq{\xi^{(n)}\equiv\prod_{a=1}^n\x{a}{a+1}\x{a}{a+2},\label{definition_of_general_xi}}
where cyclic ordering on $n$ points $x_a$ is understood (as well as the symmetry \mbox{$\x{a}{b}\!=\!\x{b}{a}$}). With this slight modification, it was discovered in \mbox{ref.\ \cite{Eden:2011we}} that the function $\mathcal{F}^{(\ell)}$ is fully {\it permutation invariant in its arguments}. This hidden symmetry is quite remarkable, and is responsible for a dramatic simplification in the representation of both the amplitude and the correlator. Because of the close connection between $\mathcal{F}^{(\ell)}$ and the correlation function defined via (\ref{definition_of_f}), we will frequently refer to $\mathcal{F}^{(\ell)}$ as `the $\ell$ loop correlation function' throughout the rest of this work; we hope this slight abuse of language will not lead to any confusion to the reader.

\vspace{-6pt}\subsection{$f$-Graphs: Their Analytic and Graphical Representations}\label{subsec:fgraphs_and_conventions}\vspace{-0pt}

Considering the full symmetry of $\mathcal{F}^{(\ell)}$ among its $(4\pl\ell)$ arguments, we are led to think of the possible contributions more as graphs than algebraic expressions. Conformality requires that any such contribution must be weight $\!\mi4$ in each of its arguments; locality ensures that only factors of the form $\x{a}{b}$ can appear in the denominator; analyticity requires that there are at most single poles in these factors (for the amplitude---for the correlator, analysis of OPE limits); and finally, planarity informs us that these factors must form a plane graph. The denominator of any possible contribution, therefore, can be encoded as a plane graph with edges $a\!\leftrightarrow\!b$ for each factor $\x{a}{b}$. (Because $\x{a}{b}\!\!=\!\x{b}{a}$, these graphs are naturally {\it undirected}.) 

We are therefore interested in plane graphs involving $(4\pl\ell)$ points, with valency at least 4 in each vertex. Excess conformal weight from vertices with higher valency can be absorbed by factors in the numerator. Conveniently, it is not hard to enumerate all such plane graphs---one can use the program {\tt CaGe} \cite{CaGe}, for example. Decorating each of these plane graphs with all inequivalent numerators capable of rending the net conformal weight of every vertex to be $\!\mi4$ results in the space of so-called `$f$-graphs'. The enumeration of the possible $f$-graph contributions that result from this exercise (through eleven loop-order) is given in \mbox{Table \ref{f_graph_statistics_table}}. Also in the Table, we have listed the number of (graph-inequivalent) planar, (dual-)conformally invariant (`DCI') integrands that exist. (The way in which these contributions to the four-particle amplitude are obtainable from each $f$-graph is described below.)

\newpage
\begin{table}[t]$\hspace{1.5pt}\begin{array}{|@{$\,$}c@{$\,$}|@{$\,$}r@{$\,$}|@{$\,$}r@{$\,$}|@{$\,$}r@{$\,$}|@{$\,$}r@{$\,$}|}\multicolumn{1}{@{$\,$}c@{$\,$}}{\begin{array}{@{}l@{}}\\[-4pt]\text{$\ell\,$}\end{array}}&\multicolumn{1}{@{$\,$}c@{$\,$}}{\!\begin{array}{@{}c@{}}\text{number of}\\[-4pt]\text{plane graphs}\end{array}}\,&\multicolumn{1}{@{$\,$}c@{$\,$}}{\begin{array}{@{}c@{}}\text{number of graphs}\\[-4pt]\text{admitting decoration}\end{array}}\,&\multicolumn{1}{@{$\,$}c@{$\,$}}{\begin{array}{@{}c@{}}\text{number of decorated}\\[-4pt]\text{plane graphs ($f$-graphs)}\end{array}}\,&\multicolumn{1}{@{$\,$}c@{$\,$}}{\begin{array}{@{}c@{}}\text{number of planar}\\[-4pt]\text{DCI integrands}\end{array}}\,\\[-0pt]\hline1&0&0&0&1\\\hline2&1&1&1&1\\[-0pt]\hline3&1&1&1&2\\\hline4&4&3&3&8\\\hline5&14&7&7&34\\\hline6&69&31&36&284\\\hline7&446&164&220&3,\!239\\\hline8&3,\!763&1,\!432&2,\!709&52,\!033\\\hline9&34,\!662&13,\!972&43,\!017&1,\!025,\!970\\\hline10&342,\!832&153,\!252&900,\!145&24,\!081,\!425\\\hline11&3,\!483,\!075&1,\!727,\!655&22,\!097,\!035&651,\!278,\!237\\\hline\end{array}$\vspace{-6pt}\caption{Statistics of plane graphs, $f$-graphs, and DCI integrands through $\ell\!=\!11$ loops.\label{f_graph_statistics_table}}\vspace{-10pt}\end{table}
\noindent(To be clear, \mbox{Table \ref{f_graph_statistics_table}} counts the number of {\it plane} graphs---that is, graphs with a fixed plane embedding. The distinction here is only relevant for graphs that are not 3-vertex connected---which are the only planar graphs that admit multiple plane embeddings. We have found that no such graphs contribute to the amplitude or correlator through ten loops---and we strongly expect their absence can be proven. However, because the graphical rules we describe are sensitive to the plane embedding, we have been careful about this distinction in our analysis---without presumptions on their irrelevance.)

When representing an $f$-graph graphically, we use solid lines to represent every factor in the denominator, and dashed lines (with multiplicity) to indicate the factors that appear in the numerator. For example, the possible $f$-graphs through four loops are as follows:
\eq{\begin{array}{rc@{$\;\;\;\;\;$}rc@{$\;\;\;\;\;$}rc}\\[-40pt]f^{(1)}_1\equiv&\fwbox{75pt}{\fig{-54.75pt}{1}{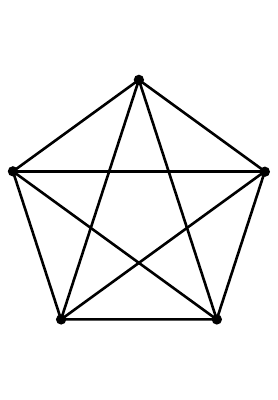}}&f^{(2)}_1\equiv&\fwbox{75pt}{\fig{-54.75pt}{1}{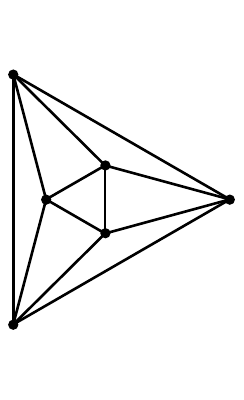}}&f^{(3)}_1\equiv&\fwbox{75pt}{\fig{-54.75pt}{1}{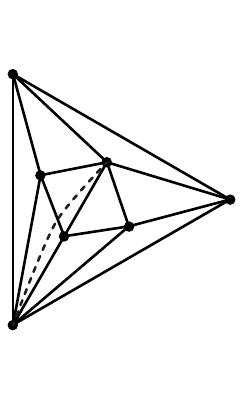}}\\[-26pt]f^{(4)}_1\equiv&\fwbox{75pt}{\fig{-54.75pt}{1}{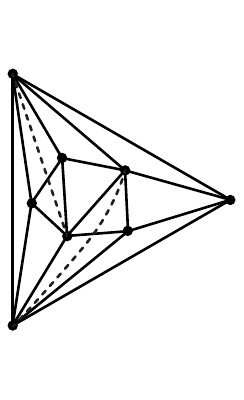}}&f^{(4)}_2\equiv&\fwbox{75pt}{\fig{-54.75pt}{1}{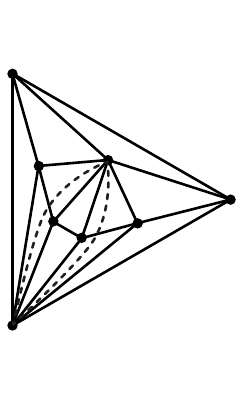}}&f^{(4)}_3\equiv&\fwbox{75pt}{\fig{-54.75pt}{1}{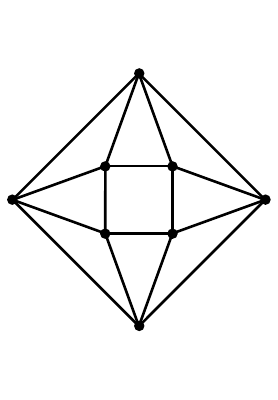}}\\[-35pt]\end{array}\vspace{20pt}\label{one_through_four_loop_f_graphs}}
In terms of these, the loop-level correlators $\mathcal{F}^{(\ell)}$ would be expanded according to:
\eq{\mathcal{F}^{(1)}=f^{(1)}_1,\quad \mathcal{F}^{(2)}=f^{(2)}_1,\quad \mathcal{F}^{(3)}=f^{(3)}_1,\quad \mathcal{F}^{(4)}=f^{(4)}_1+f^{(4)}_2-f^{(4)}_3.\label{correlators_through_four_loops}}
(Notice that $f^{(1)}_1$ in (\ref{one_through_four_loop_f_graphs}) is not planar; this is the only exception to the rule; however, it does lead to planar contributions to $\mathcal{G}^{(1)}_4$ and $\mathcal{A}_4^{(1)}$ after multiplication by $\xi^{(4)}$.)

In general, we can always express the $\ell$ loop correlator $\mathcal{F}^{(\ell)}$ in terms of the \mbox{$f$-graphs} $f^{(\ell)}_i$ according to,
\eq{\mathcal{F}^{(\ell)}\equiv\sum_{i}c^{\ell}_i\,f^{(\ell)}_i\,,\label{general_correlator_expansion}\vspace{-5pt}}
where the coefficients $c_i^{\ell}$ (indexed by the complete set of $f$-graphs at $\ell$ loops) are rational numbers---to be determined using principles such as those described below. At eleven loops, for example, there will be $22,\!097,\!035$ coefficients $c_i^{11}$ that must be determined (see \mbox{Table \ref{f_graph_statistics_table}}).

Analytically, these graphs correspond to the product of factors $\x{a}{b}$ in the denominator for each solid line in the figure, and factors $\x{a}{b}$ in the numerator for each dashed line in the figure. This requires, of course, a choice of the labels for the vertices of the graph. For example, 
\vspace{-5pt}\eq{\hspace{-85.5pt}\fig{-54.75pt}{1}{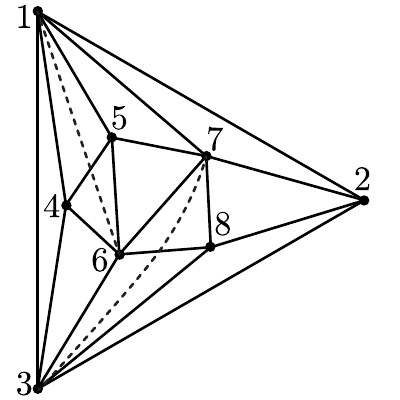}\hspace{-8.5pt}\equiv\!\!\frac{\x{1}{6}\x{3}{7}}{\x{1}{2}\x{1}{3}\x{1}{4}\x{1}{5}\x{1}{7}\x{2}{3}\x{2}{7}\x{2}{8}\x{3}{4}\x{3}{6}\x{3}{8}\x{4}{5}\x{4}{6}\x{5}{6}\x{5}{7}\x{6}{7}\x{6}{8}\x{7}{8}}\!.\hspace{-50pt}\vspace{-2pt}}
But any other choice of labels would have corresponded to the same graph, and so we must sum over all the (distinct) relabelings of the function. Of the $8!$ such relabelings, many leave the corresponding function unchanged---resulting (for this example) in 8 copies of each function. Thus, had we chosen to na\"{i}vely sum over all permutations of labels, we would over-count each graph, requiring division by a compensatory `symmetry factor' of 8 in the analytic expression contributing to the amplitude or correlation function. (This symmetry factor is easily computed as the size of the automorphism group of the graph.) However, we prefer not to include such symmetry factors in our expressions, which is why we write the coefficient of this graph in (\ref{correlators_through_four_loops}) as `$\!\pl\!$1' rather than `$\!\pl\!$1/8'. 

And so, to be perhaps overly explicit, we should be clear that this will always be our convention. Contributions to the amplitude or correlator, when converted from graphs to analytic expressions, should be symmetrized and summed; but we will always (implicitly) consider the summation to include only the {\it distinct} terms that result from symmetrization. Hence, no (compensatory) symmetry factors will appear in our coefficients. Had we instead used the convention where $f$-graphs' analytic expressions should be generated by summing over {\it all} terms generated by $\mathfrak{S}_{4+\ell}$, the coefficients of the four loop correlator, for example, would have been $\{\!\pl1/8,\!\pl1/24,\!\mi1/16\}$ instead of $\{\!\pl1,\!\pl1,\!\mi1\}$ as written in (\ref{correlators_through_four_loops}).

\newpage
\vspace{-6pt}\subsection{Four-Particle Amplitude Extraction via Light-Like Limits Along Faces}\label{subsec:amplitude_extraction}\vspace{-0pt}
When the correlation function $\mathcal{F}^{(\ell)}$ is expanded in terms of plane graphs, it is very simple to extract the $\ell$ loop scattering amplitude through the relation (\ref{correlator_amplitude_relation}). To be clear, upon expanding the square of the amplitude in powers of the coupling (and dividing by the tree amplitude), we find that:
\eq{\lim_{\substack{\text{4-point}\\\text{light-like}}}\!\!\Big(\xi^{(4)}\mathcal{F}^{(\ell)}\Big)=\frac{1}{2}\big((\mathcal{A}_4)^2\big)^{(\ell)}=\left(\mathcal{A}_{4}^{(\ell)}+\mathcal{A}_4^{(\ell-1)}\mathcal{A}_4^{(1)}+\mathcal{A}_{4}^{(\ell-2)}\mathcal{A}_4^{(2)}+\ldots\right).\label{f_to_4pt_amp_map_with_series_expansion}}
Before we describe how each term in this expansion can be extracted from the contributions to $\mathcal{F}^{(\ell)}$, let us first discuss which terms survive the light-like limit. Recall from equation (\ref{definition_of_general_xi}) that $\xi^{(4)}$ is proportional to $\x{1}{2}\x{2}{3}\x{3}{4}\x{1}{4}$---each factor of which vanishes in the light-like limit. Because $\xi^{(4)}$ identifies four specific points $x_a$, while $\mathcal{F}^{(\ell)}$ is a permutation-invariant sum of terms, it is clear that these four points can be arbitrarily chosen among the $(4\pl\ell)$ vertices of any $f$-graph; and thus the light-like limit will be non-vanishing iff the graph contains an edge connecting each of the pairs of vertices: $1\!\leftrightarrow\!2$, $2\!\leftrightarrow\!3$, $3\!\leftrightarrow\!4$, $1\!\leftrightarrow\!4$. Thus, terms that survive the light-like limit are those corresponding to a 4-cycle of the (denominator of the) graph. 

Any $n$-cycle of a plane graph divides it into an `interior' and `exterior' according to the plane embedding (viewed on a sphere). And this partition exactly corresponds to that required by the products of amplitudes appearing in (\ref{f_to_4pt_amp_map_with_series_expansion}). We can illustrate this partitioning with the following example of a ten loop $f$-graph (ignoring any factors that appear in the numerator):
\eq{\fig{-54.75pt}{1}{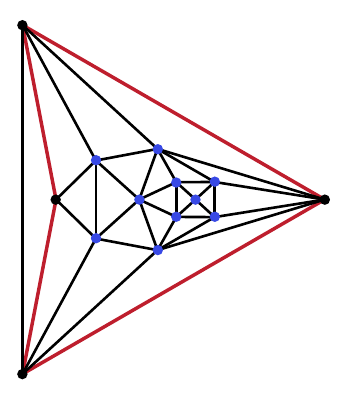}\qquad\fig{-54.75pt}{1}{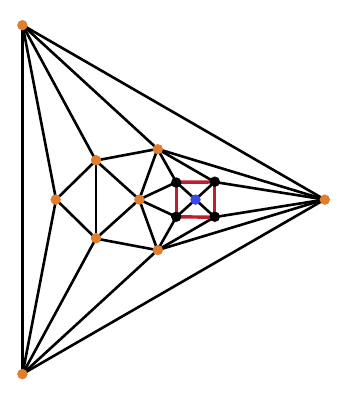}\qquad\fig{-54.75pt}{1}{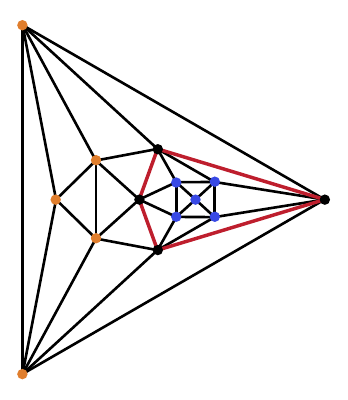}\label{example_cycles}}
These three 4-cycles would lead to contributions to $\mathcal{A}_4^{(10)}$, $\mathcal{A}_4^{(9)}\mathcal{A}_4^{(1)}$, and $\mathcal{A}_4^{(5)}\mathcal{A}_4^{(5)}$, respectively. Notice that we have colored the vertices in each of the examples above according to how they are partitioned by the cycle indicated. The fact that the $\ell$ loop correlator $\mathcal{F}^{(\ell)}$ contains within it complete information about lower loops will prove extremely useful to us in the next section. For example, the square (or `rung') rule follows immediately from the requirement that the $\mathcal{A}_4^{(\ell-1)}\mathcal{A}_4^{(1)}$ term in the expansion (\ref{f_to_4pt_amp_map_with_series_expansion}) is correctly reproduced from the representation of $\mathcal{F}^{(\ell)}$ in terms of $f$-graphs. 

The leading term in (\ref{f_to_4pt_amp_map_with_series_expansion}) is arguably the most interesting. As illustrated above, these contributions arise from any 4-cycle of an $f$-graph encompassing no internal vertices. Such cycles correspond to {\it faces} of the graph---either a single square face, or two triangular faces which share an edge. This leads to a direct projection from $f$-graphs into planar `amplitude' graphs that are manifestly dual conformally invariant (`DCI'). Interestingly, the graphs that result from taking the light-like limit along each face of the graph can appear surprisingly different. 

Consider for example the following five loop $f$-graph, which has four non-isomorphic faces, resulting in four rather different DCI integrands:
\vspace{-20pt}\eq{\fwbox{0pt}{\hspace{-225pt}\fwboxR{0pt}{\fig{-54.75pt}{1}{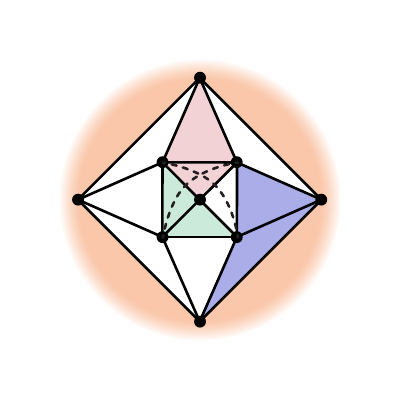}}\fwboxL{0pt}{\hspace{-15pt}\bigger{\Rightarrow}\!\left\{\!\rule{0pt}{40pt}\right.\hspace{-12.5pt}\fig{-54.75pt}{1}{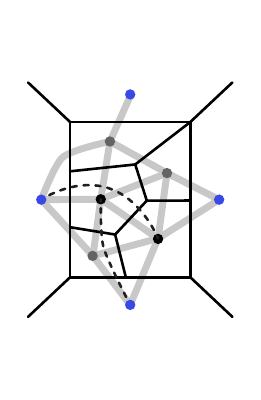}\hspace{-10pt}\fig{-54.75pt}{1}{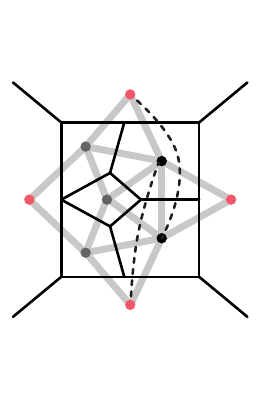}\hspace{-12.5pt}\fig{-54.75pt}{1}{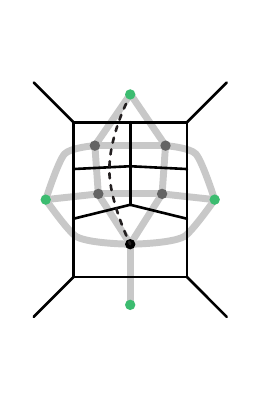}\hspace{-12.5pt}\fig{-54.75pt}{1}{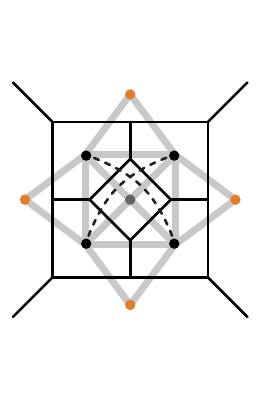}\hspace{-5pt}\left.\rule{0pt}{40pt}\right\}}}\label{five_loop_planar_projections_example}\vspace{-20pt}}
Here, we have drawn these graphs in both momentum space and dual-momentum space---with black lines indicating ordinary Feynman propagators (which may be more familiar to many readers), and grey lines indicating the dual graphs (more directly related to the $f$-graph). We have not drawn any dashed lines to indicate factors of $s\!\equiv\!\x{1}{3}$ or $t\!\equiv\!\x{2}{4}$ in numerators that would be uniquely fixed by dual conformal invariance. Notice that one of the faces---the orange one---corresponds to the `outer' four-cycle of the graph as drawn; also, the external points of each planar integrand have been colored according to the face involved. As one further illustration of this correspondence, consider the following seven loop $f$-graph, which similarly leads to four inequivalent DCI integrands (drawn in momentum space):
\vspace{2pt}\eq{\fwbox{0pt}{\hspace{-235pt}\fwboxR{0pt}{\fig{-34.75pt}{1}{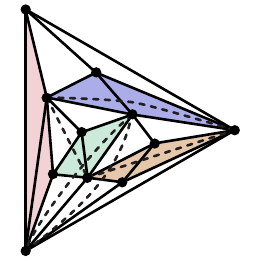}}\fwboxL{0pt}{\hspace{-0pt}\bigger{\Rightarrow}\left\{\rule{0pt}{40pt}\right.\hspace{-10pt}\fig{-34.75pt}{1}{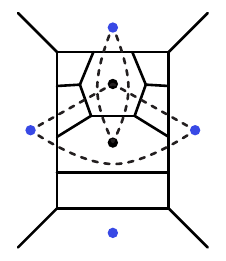}\hspace{-5pt}\fig{-34.75pt}{1}{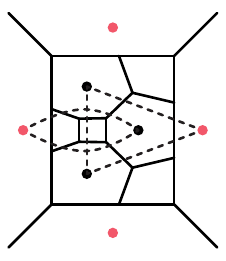}\hspace{-2.5pt}\fig{-34.75pt}{1}{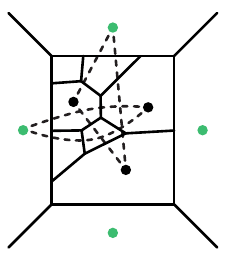}\hspace{-10.pt}\fig{-34.75pt}{1}{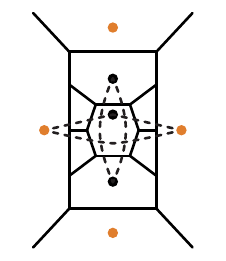}\hspace{-10pt}\left.\rule{0pt}{40pt}\right\}}}\label{seven_loop_planar_projections_example}\vspace{-0pt}}

Before moving on, it is worth a brief aside to mention that these projected contributions are to be symmetrized according to the same convention discussed above for $f$-graphs---namely, when considered as analytic expressions, only distinct terms are to be summed. This follows directly from our convention for $f$-graphs and the light-like limit, without any relative symmetry factors required between the coefficients of $f$-graphs and the coefficients of each distinct DCI integrand obtained by taking the light-like limit.

\newpage 
\vspace{-2pt}\subsection{Higher-Point Amplitude Extraction from the Correlator}\label{subsec:higher_point_amplitude_extraction}\vspace{-0pt}
Remarkably enough, although the correlation function $\mathcal{F}^{(\ell)}$ was defined to be closely related to the (actual) four-point correlation function $\mathcal{G}^{(\ell)}_4$ in planar SYM, which accounts for its relation to the four-particle scattering amplitude $\mathcal{A}_4^{(\ell)}$, it turns out that interesting combinations of {\it all} higher-point amplitudes can also be obtained from it \cite{Eden:2011yp,Eden:2011ku,Ambrosio:2013pba}. Perhaps this should not be too surprising, as $\mathcal{F}^{(\ell)}$ is a symmetrical function on $(4\pl\ell)$ points $x_a$; but it is an incredibly powerful observation: it implies that $\mathcal{F}^{(\infty)}$ contains information about {\it all} scattering amplitudes in planar SYM! 

The way in which higher-point, lower-loop amplitudes are encoded in the function $\mathcal{F}^{(\ell)}$ is a consequence of the fully supersymmetric amplitude/correlator duality \cite{Eden:2010zz,Alday:2010zy,Eden:2010ce,Eden:2011yp,Adamo:2011dq,Eden:2011ku} which was unpacked in \mbox{ref.\ \cite{Ambrosio:2013pba}}:
\vspace{-5pt}\eq{\lim_{\substack{\text{n-point}\\\text{light-like}}}\!\!\Big(\xi^{(n)}\mathcal{F}^{(\ell)}\Big)=\frac{1}{2}\sum_{k=0}^{n-4}\mathcal{A}_n^{k}\,\mathcal{A}_n^{n-4-k}/(\mathcal{A}_n^{n-4,(0)}).\label{f_to_npt_amp_map}\vspace{-5pt}}
Here, we have used the notation $\mathcal{A}_n^{k,(\ell)}$ to represent the $\ell$-loop $n$-particle N$^k$MHV amplitude divided by the $n$ particle MHV tree-amplitude. We should point out that division in (\ref{f_to_npt_amp_map}) by the N$^{n-4}$MHV ($\overline{\text{MHV}}$) tree-amplitude is required to absorb the Grassmann $\eta$ weights---resulting in a purely bosonic sum of terms from which all amplitudes can be extracted. 

It is worth mentioning that while for four particles, the $\ell$ loop amplitude can be directly extracted from $\mathcal{F}^{(\ell)}$, and for five-points one can also extract the full amplitude, for higher-point amplitudes it is not yet clear if or how one can obtain full information about amplitudes from the combination on the left-hand side of \eqref{f_to_npt_amp_map}. Elaboration of how this works in detail is beyond the scope of our present work, but because the case of $n\!=\!5$ will play an important role in motivating (and proving) the `pentagon rule' described in the next section, it is worth illustrating at least this case in some detail.

\paragraph{The Pentagonal Light-Like Limit:}~\\
\indent In addition to being the simplest example of how higher-point amplitudes can be extracted from $\mathcal{F}^{(\ell)}$ via (\ref{f_to_npt_amp_map}), the case of five particles will prove quite useful to us in our discussion of the pentagon rule described in the next section. Therefore, let us briefly summarize how this works in practice. 

In the case of five particles, the right-hand side of (\ref{f_to_npt_amp_map}) is simply the product of the MHV and $\overline{\text{MHV}}$ amplitudes---divided by the $\overline{\text{MHV}}$ tree-amplitude (with division by $\mathcal{A}_5^{0,(0)}$ left implicit, as always). Conventionally defining \mbox{$\mathcal{M}_5^{}\!\equiv\!\mathcal{A}_5^{0}/\mathcal{A}_{5}^{0,(0)}$} and \mbox{$\overline{\mathcal{M}}_5^{}\!\equiv\!\mathcal{A}_5^{1}/\mathcal{A}_{5}^{1,(0)}$}, and expanding each in powers of the coupling, the relation (\ref{f_to_npt_amp_map}) becomes more symmetrically expressed as:
\vspace{-6.5pt}\eq{\lim_{\substack{\text{5-point}\\\text{light-like}}}\!\!\Big(\xi^{(5)}\mathcal{F}^{(\ell+1)}\Big)=\sum_{m=0}^{\ell}\mathcal{M}_5^{(m)}\overline{\mathcal{M}}_5^{(\ell-m)}.\label{f_to_5pt_amp_map}\vspace{-0pt}}
Moreover, because the parity-even contributions to the loop integrands $\mathcal{M}_5^{(\ell)}$ and $\overline{\mathcal{M}}_5^{(\ell)}$ are the same, it is further convenient to define:
\eq{\mathcal{M}_{\text{even}}^{(\ell)}\equiv\frac{1}{2}\left(\mathcal{M}_5^{(\ell)}+\overline{\mathcal{M}}_5^{(\ell)}\right)\quad\text{and}\quad\mathcal{M}_{\text{odd}}^{(\ell)}\equiv\frac{1}{2}\left(\mathcal{M}_5^{(\ell)}-\overline{\mathcal{M}}_5^{(\ell)}\right).\label{5pt_even_and_odd_definitions}}

Because any integrand constructed out of factors $\x{a}{b}$ will be manifestly parity-even, it is not entirely obvious how the parity-odd contributions to loop integrands should be represented. Arguably, the most natural way to represent parity-odd contributions is in terms of a six-dimensional  formulation of dual momentum space (essentially the Klein quadric) which was first introduced in this context in \mbox{ref.\ \cite{Mason:2009qx}} following the introduction of momentum twistors in \mbox{ref.\ \cite{Hodges:2009hk}}. Each point $x_a$ is represented by a (six-component) bi-twistor $X_a$. The (dual) conformal group $SO(2,4)$ acts linearly on this six-component object and so  it is natural to define a fully antisymmetric epsilon-tensor, $\epsilon_{abcdef}\!\equiv\!\det\{X_a,\ldots,X_f\}$, in which the parity-odd part of the $\ell$ loop integrand can be represented \cite{Ambrosio:2013pba}:
\eq{\mathcal{M}_{\text{odd}}\equiv i\epsilon_{12345\ell}\,\widehat{\mathcal{M}}_{\text{odd}},\label{definition_of_epsilon_prefactors_for_odd_integrands}}
where $\widehat{\mathcal{M}}_{\text{odd}}$ is a parity-even function, directly expressible in terms of factors $\x{a}{b}$. 

Putting everything together, the expansion (\ref{f_to_5pt_amp_map}) becomes:
\eq{\hspace{-75pt}\lim_{\substack{\text{5-point}\\\text{light-like}}}\!\!\Big(\xi^{(5)}\mathcal{F}^{(\ell+1)}\Big)=\sum_{m=0}^{\ell}\left(\mathcal{M}_{\text{even}}^{(m)}\mathcal{M}_{\text{even}}^{(\ell-m)}+\epsilon_{123456}\epsilon_{12345(m+6)}\widehat{\mathcal{M}}_{\text{odd}}^{(m)}\widehat{\mathcal{M}}_{\text{odd}}^{(\ell-m)}\right).\label{f_to_5pt_amp_map2}\hspace{-40pt}\vspace{-5pt}}

The pentagon rule we derive in the next section amounts to the equality between two different ways to extract the $\ell$-loop 5-particle integrand from $\mathcal{F}^{(\ell+2)}$, by identifying, as part of the contribution, the one loop integrand. As such, it is worthwhile to at least quote these contributions. They are as follows:
\eq{\mathcal{M}_{\text{even}}^{(1)}\equiv\fig{-34.75pt}{1}{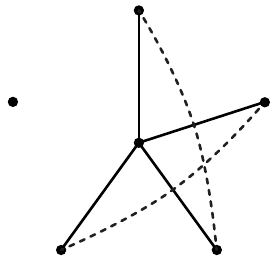}\qquad\text{and}\qquad\mathcal{M}_{\text{odd}}^{(1)}\equiv\fig{-34.75pt}{1}{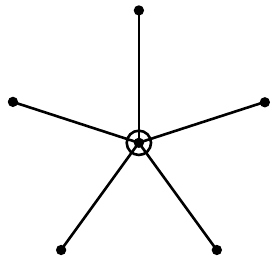}\label{five_point_one_loop_terms}}
where the circled vertex in the right-hand figure indicates the last argument of the epsilon-tensor. When converted into analytic expressions, these correspond to:
\eq{\fwbox{0pt}{\fig{-34.75pt}{1}{five_point_one_loop_even}\equiv\frac{\x{1}{3}\x{2}{4}}{\x{1}{6}\x{2}{6}\x{3}{6}\x{4}{6}}+\text{cyclic},\quad\fig{-34.75pt}{1}{five_point_one_loop_odd}\equiv\frac{i\epsilon_{123456}}{\x{1}{6}\x{2}{6}\x{3}{6}\x{4}{6}\x{5}{6}}},\label{five_point_one_loop_terms_analytic}\nonumber}
where the cyclic sum of terms involves only the 5 external vertices. 

\newpage
\vspace{-6pt}\section{(Graphical) Rules For Bootstrapping Amplitudes}\label{sec:graphical_bootstraps}\vspace{-6pt}
As described above, the correlator $\mathcal{F}^{(\ell)}$ can be expanded into a basis of $\ell$ loop \mbox{$f$-graphs} according to (\ref{general_correlator_expansion}). The challenge, then, is to determine the coefficients $c_i^{\ell}$.  We take for granted that the one loop four-particle amplitude integrand may be represented in dual momentum coordinates as:
\vspace{-5pt}\eq{\mathcal{A}_4^{(1)}\equiv\fig{-34.75pt}{1}{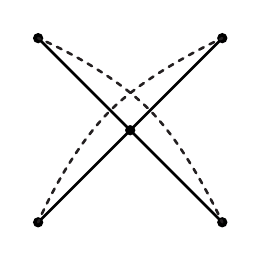}\equiv\frac{\x{1}{3}\x{2}{4}}{\x{1}{5}\x{2}{5}\x{3}{5}\x{4}{5}},\label{four_point_one_loop_integrand_in_x_space}\vspace{-5pt}}
with which we expect most readers will be familiar. This formula in fact {\it defines} the one loop $f$-graph $f^{(1)}_1$---as there does not exist any planar graph involving five points each having valency at least 4. As such, it is defined so as to ensure that equation (\ref{f_to_4pt_amp_map_with_series_expansion}) holds:
\vspace{-26pt}\eq{f^{(1)}_1\equiv\mathcal{A}_4^{(1)}/\xi^{(4)}\equiv\!\!\fwbox{90pt}{\fig{-54.75pt}{1}{one_loop_f_graph_1}}\equiv\frac{1}{\x{1}{2}\x{1}{3}\x{1}{4}\x{1}{5}\x{2}{3}\x{2}{4}\x{2}{5}\x{3}{4}\x{3}{5}\x{4}{5}}.\label{definition_of_f1}\vspace{-24pt}}
This effectively defines $\mathcal{F}^{(1)}\!\equiv\!f_1^{(1)}$, with a coefficient $c_1^{1}\!\equiv\!\pl1$. Given this seed, we will see that consistency among the products of lower-loop amplitudes in (\ref{f_to_4pt_amp_map_with_series_expansion})---as well as those involving more particles (\ref{f_to_npt_amp_map})---will be strong enough to uniquely determine the coefficients of all $f$-graphs in the expansion for $\mathcal{F}^{(\ell)}$ in terms of lower loop-orders. 

In this section we describe how this can be done in practice through three simple, graphical rules that allow us to `bootstrap' all necessary coefficients through at least ten loops. To be clear, the rules we describe are merely three among many that follow from the self-consistency of equations (\ref{f_to_4pt_amp_map_with_series_expansion}) and (\ref{f_to_npt_amp_map}); they are not obviously the strongest or most effective of such rules; but they are {\it necessary} conditions of any representation of the correlator, and we have found them to be {\it sufficient} to uniquely fix the expansion of $\mathcal{F}^{(\ell)}$ into $f$-graphs, (\ref{general_correlator_expansion}), through at least ten loops. 

Let us briefly describe each of these three rules in qualitative terms, before giving more detail (and derivations) in the following subsections. We refer to these as the `triangle rule', the `square rule', and the `pentagon rule'. Despite the natural ordering suggested by their names, it is perhaps best to start with the square rule---which is simply a generalization of what has long been called the `rung' rule \cite{Bern:1997nh}. 

\paragraph{The Square (or `Rung') Rule:}~\\
\indent The square rule is arguably the most powerful of the three rules, and provides the simplest constraints---directly fixing the coefficients of certain $f$-graphs at $\ell$ loops to be equal to the coefficients of $f$-graphs at $(\ell\mi1)$ loops. 

Roughly speaking, the square rule follows from the requirement that whenever an $f$-graph {\it has} a contribution to $\mathcal{A}_4^{(\ell-1)}\mathcal{A}_4^{(1)}$, this contribution must be correct. It simply reflects the translation of what has long been known as the `rung' rule \cite{Bern:1997nh} into the language of the correlator and $f$-graphs \cite{Eden:2012tu}; however, this translation proves much more powerful than the original, as described in more detail below. As will be seen in the \mbox{section \ref{sec:results}}, for example, the square rule fixes $\sim\!95\%$ of all $f$-graph coefficients at eleven loops---the only coefficients not fixed by the square rule are those of $f$-graphs which do not contribute any terms to $\mathcal{A}_4^{(\ell-1)}\mathcal{A}_4^{(1)}$. 

~\\[-36pt]\paragraph{The Triangle Rule:}~\\
\indent Simply put, the triangle rule states that shrinking triangular faces at $\ell$ loops is equivalent to shrinking edges at $(\ell\mi1)$ loops. By this we mean simply identifying the three vertices of any triangular face of an $f$-graph at $\ell$ loops and identifying two vertices connected by an edge of an $f$-graph at $(\ell\mi1)$ loops, respectively. The result of either operation is never an $f$-graph (as it will not have correct conformal weights, and will often involve vertices connected by more than one edge), but this does not prevent us from implementing the rule graphically. Typically, there are many fewer inequivalent graphs involving shrunken faces/edges, and so the triangle rule typically results in relations involving many $f$-graph coefficients. This makes the equations relatively harder to solve.

As described in more detail below, the triangle rule follows from the Euclidean short distance \cite{Eden:2012tu,Eden:2012fe} limit of correlation functions. We will prove this in the following subsection, and describe more fully its strength in fixing coefficients in \mbox{section \ref{sec:results}}. But it is worth mentioning here that when combined with the square rule, the triangle rule is sufficient to fix $\mathcal{F}^{(\ell)}$ completely through seven loops; and the implications of the triangle rule applied at {\it ten} loops is sufficient to fix $\mathcal{F}^{(\ell)}$ through {\it nine} loops (although the triangle and square rules alone, when imposed at nine loops, would not suffice).

~\\[-36pt]\paragraph{The Pentagon Rule:}~\\
\indent The pentagon rule is the five-particle analog of the square rule---following from the requirement that the $\mathcal{M}^{(\ell-1)}\mathcal{M}^{(1)}$ terms in the expansion (\ref{f_to_5pt_amp_map}) are correct. Unlike the square rule, however, it does not make use of knowing lower-loop five-particle amplitudes; rather, it simply requires that the odd contributions to the amplitude are consistent. We will describe in detail how the pentagon rule is derived below, and give examples of how it fixes coefficients. 

One important aspect of the pentagon rule, however, is that it relates coefficients at a {\it fixed loop-order}. Indeed, as an algebraic constraint, the pentagon rule always becomes the requirement that the sum of some subset of coefficients $c_i^\ell$ is zero (without any relative factors ever required).\\

Before we describe and derive each of these three rules in detail, it is worth mentioning that they lead to mutually overlapping and individually {\it over-constrained} relations on the coefficients of $f$-graphs. As such, the fact that any solution exists to these equations---whether from each individual rule or in combination---strongly implies the correctness of our rules (and the correctness of their implementation in our code). And of course, the results we find are consistent with all known results through eight loops, which have been found using a diversity of other methods. 

\vspace{-0pt}\subsection{The Square (or `Rung') Rule: Removing One Loop Squares}\label{subsec:square_rule}\vspace{-0pt}
Recall from \mbox{section \ref{subsec:amplitude_extraction}} that, upon taking the 4-point light-like limit, an $f$-graph contributes a term to $\mathcal{A}_4^{(\ell-1)}\mathcal{A}_4^{(1)}$ in the expansion (\ref{f_to_4pt_amp_map_with_series_expansion}) if (and only if) there exists a 4-cycle that encloses a single vertex. See, for example, the second illustration given in (\ref{example_cycles}). Because of planarity, the enclosed vertex must have valency exactly 4, and so any such cycle must form a face with the topology:
\vspace{-7pt}\eq{\fig{-34.75pt}{1}{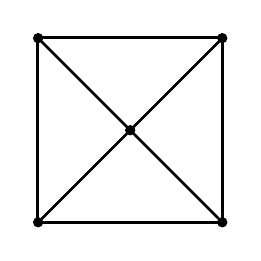}\label{square_rule_face_toplogy}\vspace{-7pt}}
Whenever an $f$-graph has such a face, it will contribute a term of the form $\mathcal{A}_4^{(\ell-1)}\mathcal{A}_4^{(1)}$ in the light-like limit. If we define the operator $\mathcal{S}(\mathcal{F})$ to be the projection onto such contributions, then the rung rule states that $\mathcal{S}(\mathcal{F}^{(\ell)})/\mathcal{A}_4^{(1)}\!\!=\!\mathcal{A}_4^{(\ell-1)}$. Graphically, division of (\ref{square_rule_face_toplogy}) by the graph for $\mathcal{A}_4^{(1)}$ in (\ref{four_point_one_loop_integrand_in_x_space}) would correspond to the graphical replacement:
\vspace{-0pt}\eq{\hspace{-120pt}\fig{-34.75pt}{1}{square_rule_face_topology}\bigger{\Rightarrow}\left(\fig{-34.75pt}{1}{square_rule_face_topology}\bigger{\times}\fig{-34.75pt}{1}{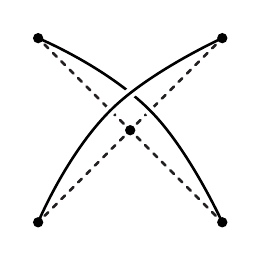}\right)\bigger{=}\fig{-34.75pt}{1}{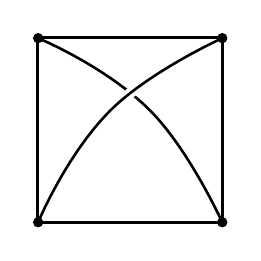}\hspace{-100pt}\label{graphical_square_rule}\vspace{-5pt}}
(Here, we have illustrated division by the graph for $\mathcal{A}_4^{(1)}$---shown in (\ref{four_point_one_loop_integrand_in_x_space})---as multiplication by its inverse.) 

Importantly, the image on the right hand side of (\ref{graphical_square_rule}) resulting from this operation is not always planar! For it to be planar, there must exist a numerator factor connecting any two of the vertices of the square face---to cancel against one or both of the `new' factors in the denominator appearing in (\ref{graphical_square_rule}). When the image is non-planar, however, the graph {\it cannot} contribute to $\mathcal{A}_4^{(\ell-1)}$,\footnote{There is an exception to this conclusion when $\ell\!=\!2$---because $f_1^{(1)}$ is not itself planar.} and thus the coefficient of such an $f$-graph must vanish. For example, consider the following six loop $f$-graph which has a face with the topology (\ref{square_rule_face_toplogy}), and so its contribution to $\mathcal{F}^{(6)}$ would be constrained by the square rule:
\vspace{-14pt}\eq{\fig{-54.75pt}{1}{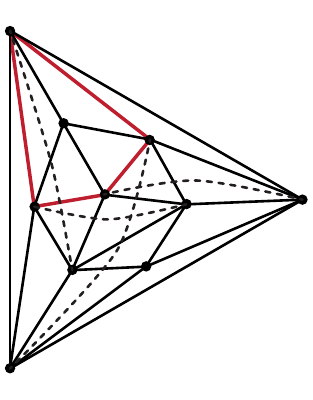}\label{six_loop_vanishing_by_square_rule_example}\vspace{-15pt}}
In this case, because there are no numerator factors (indicated by dashed lines) connecting the vertices of the highlighted 4-cycle, its image under (\ref{graphical_square_rule}) would be non-planar, and hence this term cannot appear in $\mathcal{A}_4^{(5)}$. Therefore, the coefficient of this $f$-graph must be zero. (In fact, this reasoning accounts for 8 of the 10 vanishing coefficients that first appear at six loops.) As discussed in \mbox{ref.\ \cite{Bourjaily:2015bpz}}, this immediately implies that there are no possible contributions with `$k\!=\!4$' divergences. 

More typically, however, there is at least one numerator factor in the $\ell$ loop \mbox{$f$-graph} that connects vertices of the one loop square face (\ref{square_rule_face_toplogy}) in order to cancel one or both of the new denominator factors in (\ref{graphical_square_rule}). When this is the case, the image is an $(\ell\mi1)$ loop \mbox{$f$-graph}, and the square rule states that their coefficients are identical. For example, the coefficient of the five loop \mbox{$f$-graph} shown in (\ref{five_loop_planar_projections_example}) is fixed by the square rule to have the same coefficient as $f_3^{(4)}$ shown in (\ref{one_through_four_loop_f_graphs}):
\vspace{-7pt}\eq{\fig{-34.75pt}{1}{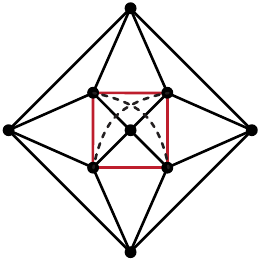}\,\,\bigger{\Rightarrow}\fig{-34.75pt}{1}{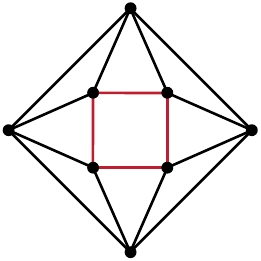}\label{five_loop_square_rule_example}\vspace{-7pt}}

In summary, the square rule fixes the coefficient of any $f$-graph that has a face with the topology (\ref{square_rule_face_toplogy}) directly in terms of lower-loop coefficients. And this turns out to constrain the vast majority of possible contributions, as summarized in \mbox{Table \ref{square_rule_strength_table}}. And it is worth emphasizing that the square rule described here is in fact substantially stronger than what has been traditionally called the `rung' rule \cite{Bern:1997nh} for two reasons: first, the square rule unifies collections of planar DCI contributions to amplitudes according to the hidden symmetry of the correlator---allowing us to fix coefficients of even the `non-rung-rule'  integrands such as those appearing in (\ref{five_loop_planar_projections_example}); secondly, the square rule allows us to infer the vanishing of certain coefficients due to the non-existence of lower-loop graphs (due to non-planarity). 

\begin{table}[b]\vspace{-20pt}$$\fwbox{0pt}{\begin{array}{|r|r|r|r|r|r|r|r|r|r|}\cline{2-10}\multicolumn{1}{r}{\ell\!=}&\multicolumn{1}{|c|}{3}&\multicolumn{1}{c|}{4}&\multicolumn{1}{c|}{5}&\multicolumn{1}{c|}{6}&\multicolumn{1}{c|}{7}&\multicolumn{1}{c|}{8}&\multicolumn{1}{c|}{9}&\multicolumn{1}{c|}{10}&\multicolumn{1}{c|}{11}\\\hline\text{number of $f$-graph coefficients:}&1&3&7&36&220&2,\!709&43,\!017&900,\!145&22,\!097,\!035\\\hline\text{number unfixed by square rule:}&0&1&1&5&22&293&2,\!900&52,\!475&1,\!017,\!869\\\hline\hline\text{percent fixed by square rule (\%):}&100&67&86&86&90&89&93&94&95\\\hline\end{array}}$$\vspace{-18pt}\caption{Statistics of correlator coefficients fixed by the square rule through $\ell\!=\!11$ loops.\label{square_rule_strength_table}}\vspace{-35pt}\end{table}

\newpage
\vspace{-6pt}\subsection{The Triangle Rule: Collapsing Triangles and Edges}\label{subsec:triangle_shrink}\vspace{-6pt}
The triangle rule relates the coefficients of $f$-graphs at $\ell$ loops to those at $(\ell\mi1)$ loops. Simply stated, collapsing triangles (to points) at $\ell$ loops is equivalent to collapsing edges of graphs at $(\ell\mi1)$ loops. More specifically, we can define an operation $\mathcal{T}$ that projects all $f$-graphs onto their triangular faces (identifying the points of each face), and another operation $\mathcal{E}$ that collapses all edges of $f$-graphs (identifying points). Algebraically, the triangle rule corresponds to,
\vspace{-5pt}\eq{\mathcal{T}(\mathcal{F}^{(\ell)})=2\,\mathcal{E}(\mathcal{F}^{(\ell-1)}).\label{algebraic_but_figurative_triangle_rule}\vspace{-6pt}}
Under either operation, the result is some non-conformal (generally) multi-graph with fewer vertices, with each image coming from possibly many $f$-graphs; thus, (\ref{algebraic_but_figurative_triangle_rule}) gives a linear relation between the $\ell$ loop coefficients of $\mathcal{F}^{(\ell)}$---those that project under $\mathcal{T}$ to the same image---and the $(\ell\mi1)$ loop coefficients of $\mathcal{F}^{(\ell-1)}$. (It often happens that an image of $\mathcal{F}^{(\ell)}$ under $\mathcal{T}$ is not found among the images of $\mathcal{F}^{(\ell-1)}$ under $\mathcal{E}$; in this case, the right-hand side of (\ref{algebraic_but_figurative_triangle_rule}) will be zero.) 

One small subtlety that is worth mentioning is that we must be careful about symmetry factors---as the automorphism group of the pre-image may not align with the image. To be clear, $\mathcal{T}$ acts on {\it each} triangular face of a graph (not necessarily inequivalent), and $\mathcal{E}$ acts on {\it each} edge of a graph (again, not necessarily inequivalent); each term in the image is then summed with a factor equal to the ratio of symmetry factor of the image to that of the pre-image. In both cases, this amounts to including a symmetry factor that compensates for the difference between the symmetries of an ordinary $f$-graph and the symmetries of $f$-graphs with a {\it decorated} triangle or edge. 

Let us illustrate this with an example from the seven loop correlation function. The image of $\mathcal{F}^{(7)}$ under $\mathcal{T}$ includes $433$ graph-inequivalent images---each resulting in one identity among the coefficients $c_i^{7}$ and $c_i^6$. One of these inequivalent images results in the identity: 
\vspace{-7.5pt}\begin{align}\fwbox{0pt}{\mathcal{T}\hspace{-4pt}\left(\rule{0pt}{40pt}\right.\hspace{-5pt}c^{7}_1\!\!\fwbox{70pt}{\fig{-37.25pt}{1}{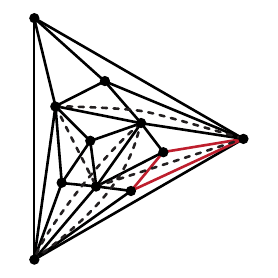}}\!+\!c^7_2\!\!\fwbox{70pt}{\fig{-37.25pt}{1}{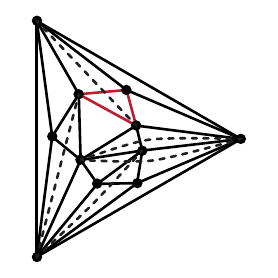}}\!+\!c^7_3\fwbox{70pt}{\fig{-37.25pt}{1}{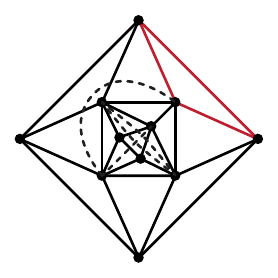}}\,\,+\!\ldots\hspace{-5pt}\left.\rule{0pt}{40pt}\right)\hspace{-4pt}=2\,\mathcal{E}\hspace{-4pt}\left(\rule{0pt}{40pt}\right.\hspace{-6pt}c^6_1\!\!\fwbox{70pt}{\fig{-37.25pt}{1}{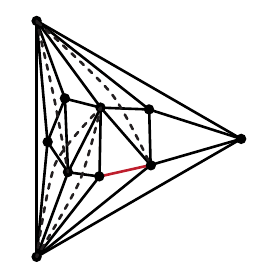}}\!+\!\ldots\hspace{-5pt}\left.\rule{0pt}{40pt}\right)}\nonumber\\[-32pt]\fwbox{0pt}{\hspace{-15pt}\bigger{\Rightarrow}\left(c_1^7+2\,c_2^7+c_3^7\right)\fwbox{70pt}{\fig{-37.25pt}{1}{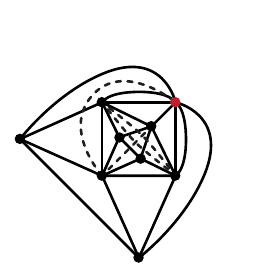}}\hspace{-10pt}=2\,c_1^6\fwbox{70pt}{\fig{-37.25pt}{1}{seven_loop_triangle_image}}\hspace{-7.5pt}\bigger{\Rightarrow}\,c_1^7+2\,c_2^7+c_3^7=2\,c_1^6.}\label{triangle_rule_example}\\[-30pt]\nonumber
\end{align}
While not visually manifest, it is not hard to check that shrinking each highlighted triangle/edge in the first line of (\ref{triangle_rule_example}) results in graphs isomorphic to the one shown in the second line. And indeed, the coefficients of the six and seven loop correlators (obtained independently) satisfy this identity: $\{c_1^7,c^7_2,c^7_3,c^6_1\}\!=\!\{\!\pl1,\!\pl1,\!\mi1,\!\pl1\}$. (The coefficient of 2 appearing in front of $c_2^7$ results from the fact that the symmetry factor of the initial graph is 1, while its image under $\mathcal{T}$ has a symmetry factor of 2.)

\vspace{-6pt}\subsubsection*{Proof and Origins of the Triangle Rule}\label{subsubsec:proof_of_triangle_rule}\vspace{-4pt}
The triangle rule arises from a reformulation of the Euclidean short distance limit of correlation functions discussed in \mbox{ref.\ \cite{Eden:2012tu,Eden:2012fe}}. In the Euclidean short distance limit  $x_2\!\rightarrow\!x_1$, the operator product expansion dictates that the leading divergence of the logarithm of  the correlation function is proportional to the one loop divergence. More precisely,
\vspace{-5pt}\eq{\lim_{x_2\rightarrow x_1}\!\log\!\Big(1+\sum_{\ell\geq1}a^\ell\,F^{(\ell)}\Big)=\gamma(a)\!\lim_{x_2\rightarrow x_1}\!F^{(1)}+\ldots,\label{konishi_log_relation}\vspace{-7pt}}
where `$a$' refers to the coupling, $F$ is defined by,
\vspace{-5pt}\eq{F^{(\ell)}\equiv3\,\frac{\mathcal{G}^{(\ell)}_4(x_1,x_2,x_3,x_4)}{\mathcal{G}^{(0)}_4(x_1,x_2,x_3,x_4)}\,,\label{definition_of_F}\vspace{-5pt}}
and where the dots in (\ref{konishi_log_relation}) refer to subleading terms in this limit. The proportionality constant $\gamma(a)$ here is the anomalous dimension of the Konishi operator, and the factor 3 in (\ref{definition_of_F}) also has a physical origin---ultimately arising from the tree-level three-point function of two stress-energy multiplets and the Konishi multiplet.\footnote{See \mbox{refs.\ \cite{Eden:2012tu,Eden:2012fe}} for details. There, the double coincidence limit was taken $x_2\!\rightarrow\!x_1$, $x_4\!\rightarrow\!x_3$, but due to conformal invariance this is in fact equivalent to the single coincidence limit we consider.}

The important point for us from (\ref{konishi_log_relation}) is that the logarithm of the correlator has the same divergence as the one loop correlator, whereas the correlator itself at $\ell$ loops diverges as the $\ell^{\text{th}}$ power of the one loop correlator $\lim_{x_2\rightarrow x_1}\!\big(\mathcal{G}_4^{(\ell)}\big)\!\sim\!\log^\ell\!\left(\x{1}{2}\right)$. At the integrand level this divergence arises from loop integration variables approaching $x_2\!=\!x_1$. The only way for a loop integral of this form---with symmetrized integration variables---to be reduced to a single log divergence is if the integrand had reduced divergence in the simultaneous limit $x_5, x_2\!\rightarrow\!x_1$, where we recall that $x_5$ is one of the loop integration variables.\footnote{The weaker requirement that the integrand only had a reduced divergence in the limit where two integration variables both approach $x_1\!=\!x_2$ would result in a divergence of at most $\log^2$, etc.}

More precisely then, defining the relevant perturbative logarithm of the correlation function as ${g}^{(\ell)}$:
\vspace{-4pt}\eq{\sum_{\ell\geq1}a^\ell g^{(\ell)}\equiv\log\!\Big(1+\sum_{\ell\geq1}a^\ell\,F^{(\ell)}\Big),\label{definition_of_g}\vspace{-2pt}}
then at the integrand-level (\ref{konishi_log_relation}) implies:\footnote{In this section we are using the same notation for both integrated functions and integrands.}
\vspace{-4pt}\eq{\lim_{x_5,x_2\rightarrow x_1}\left(\frac{g^{(\ell)}(x_1, \dots, x_{4+\ell})}{g^{(1)}(x_1,\dots,x_{5})}\right)=0,\qquad\ell\!>\!1\,.\label{eq:3b}\vspace{-3pt}}
This equation gives a clean integrand-level consequence of the reduced divergence; however, it is phrased in terms of the logarithm of the integrand rather than the integrand itself, and this does not translate directly into a graphical rule. However, notice the relation between the $\log$-expansion $g$ and the correlator $F$,
\vspace{-6pt}\eq{g^{(\ell)} = F^{(\ell)} - \frac 1\ell g^{(1)}(x_{5}) F^{(\ell-1)} - \sum_{m=2}^{\ell-1} \frac{m}{\ell} g^{(m)}(x_{5}) F^{(\ell-m)}\, .\label{logarithm_expansion}\vspace{-3pt}}
This formula can be read at the level of the integrand, and we write the dependence of the loop variable $x_{5}$ explicitly, the dependence on all other loop variables is completely symmetrized.\footnote{Note that although not manifest, the loop  variable $x_{5}$ also appears completely symmetrically in the above formula. For example, consider terms of the form $F^{(1)}F^{(\ell-1)}$. One such term arises from the second term in (\ref{logarithm_expansion}), giving $1/\ell \times F^{(1)}(x_{5}) F^{(\ell-1)}$.   Other such terms arise from the sum with $m\!=\!\ell\mi1$, giving ${(\ell\mi1)}/\ell \times F^{(\ell-1)}(x_{5}) F^{(1)}$. We see that the integration variable appears with weight 1 in $F^{(1)}$ and weight $\ell\mi1$ in $F^{(\ell-1)}$---{\it i.e.}\ completely symmetrically.} From equation (\ref{logarithm_expansion}), it is straightforward to see (using an induction argument) that (\ref{eq:3b}) is equivalent to
\vspace{-3pt}\eq{\lim_{x_2,x_{5} \rightarrow x_1} \frac{F^{(\ell)}(x_1,\dots,x_{4+\ell})}{g^{(1)}(x_1,x_2,x_3,x_4,x_{5})}= \frac1\ell\,\lim_{x_2\rightarrow x_1} F^{(\ell-1)}(x_1, \dots,\hat x_5,\dots, x_{4+\ell})\,,\label{eq:6}\vspace{-3pt}}
where the variable $x_5$ is missing in the right-hand side. This is now a direct rewriting of the reduced divergence at the level of integrands and as a relation for the loop level correlator (rather than the more complicated logarithm).

Note that everything in the discussion of this section so far can be transferred straightforwardly onto the soft/collinear divergence constraint; and indeed, a rephrasing of the soft/collinear constraint similar to (\ref{eq:6}) was conjectured in \mbox{ref.\ \cite{Golden:2012hi}}, with the relevant limit being $x_5$ approaching the line joining $x_1$ and $x_2$, $\lim_{x_{5}\rightarrow [x_1,x_2]}$.

Now inputting the one loop correlator, $\lim_{x_2,x_{5}\rightarrow x_1}g^{(1)}(x_1,\dots,x_{5})= 6/(\x{1}{5}\x{2}{5})$, and rewriting this in terms of $\mathcal{F}^{(\ell)}$, (\ref{eq:6}) becomes simply
\vspace{-3pt}\eq{\lim_{x_2,x_{5}\rightarrow x_1}(\x{1}{2} \x{1}{5}\x{2}{5})\times {{\mathcal F}^{(\ell)}(x_1, \dots, x_{4+\ell})}= 6\lim_{x_2\rightarrow x_1} (\x{1}{2})\times {\mathcal F}^{(\ell-1)}(x_1, \dots, x_{3+\ell})\, .\label{eq:7}\vspace{-3pt}}

The final step in this rephrasing of the coincidence limit  is to view (\ref{eq:7}) graphically. Clearly the limit on the left-hand side will only be non-zero if the corresponding term in the labelled \mbox{$f$-graph} contains the triangle with vertices $x_1 ,x_2, x_5$. The limit then deletes this triangle and shrinks it to a point. On the right-hand side, we similarly choose terms in the labelled \mbox{$f$-graphs} containing the edge $x_1\!\!\leftrightarrow\!x_2$, delete this edge and then shrink to a point. The equation has to hold graphically and we no longer need to consider explicit labels. Simply shrink all inequivalent (up to automorphisms) triangles of the linear sum of graphs on the left-hand side and equate it to the result of shrinking all inequivalent (again, up to automorphisms) edges of the linear sum of graphs on the right-hand side. The different (non-isomorphic) shrunk graphs are independent, and thus for each shrunk graph we obtain an equation relating $\ell$ loop coefficients to $(\ell\mi1)$ loop coefficients. There are six different labelings of the triangle and two different labelings of the edge which all reduce to the same expression in this limit, thus the factor of 6 in the algebraic expression (\ref{eq:7}) becomes the factor of 2 in the equivalent graphical version (\ref{algebraic_but_figurative_triangle_rule}).

\newpage
\vspace{-6pt}\subsection{The Pentagon Rule: Equivalence of One Loop Pentagons}\label{subsec:pentagon_rule}\vspace{-6pt}
Let us now describe the pentagon rule. It is perhaps the hardest to describe (and derive), but it ultimately turns out to imply much simpler relations among coefficients than the triangle rule. In particular, the pentagon rule will always imply that the sum of some subset of coefficients $\{c_i^\ell\}$ vanishes---with no relative factors between terms in the sum. Let us first describe operationally how these identities are found graphically, and then describe how this rule can be deduced from considerations of 5-point light-like limits according to (\ref{f_to_5pt_amp_map2}). 

Graphically, each pentagon rule identity involves a relation between $f$-graphs involving the following topologies:
\vspace{-4pt}\eq{\fig{-34.75pt}{1}{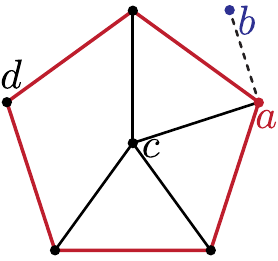}\;\bigger{\Rightarrow}\left\{\!\!\fig{-34.75pt}{1}{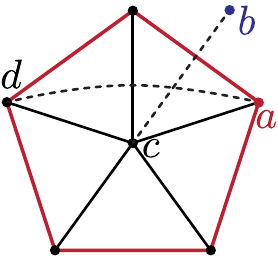}\right\}\label{topologies_of_the_pentagon_rule}}
Each pentagon rule identity involves an $f$-graph with a face with the topology on the left-hand side of the figure above, (\ref{topologies_of_the_pentagon_rule}). This sub-graph is easily identified as having the structure of $\mathcal{M}_{\text{even}}^{(1)}$---see equation (\ref{five_point_one_loop_terms}). (This is merely suggestive: we will soon see that it is the role these graphs play in $\mathcal{M}_{\text{odd}}^{(1)}$ that is critical.) Importantly, these $f$-graphs may involve any number of numerators of the form $\x{{\color{hred}a}}{{\color{hblue}b}}$---including some that are `implicit': any points ${\color{hblue}x_b}$ separated from ${\color{hred}x_a}$ by a face (not connected by an edge), because for such points ${\color{hblue}x_b}$, multiplication by $\x{{\color{hred}a}}{{\color{hblue}b}}/\x{{\color{hred}a}}{{\color{hblue}b}}$ would not affect planarity of the factors in the denominator. The graphs on the right-hand side of (\ref{topologies_of_the_pentagon_rule}), then, are the collection of those \mbox{$f$-graphs} obtained from that on the left-hand side by multiplication by a simple cross-ratio:
\eq{f_i^{(\ell)}({\color{hred}x_{a}},{\color{hblue}x}_{{\color{hblue}b}},x_c,x_d)\mapsto f_{i'}^{(\ell)}\equiv f_i^{(\ell)}\frac{\x{{\color{hred}a}}{d}\x{{\color{hblue}b}}{c}}{\x{{\color{hred}a}}{{\color{hblue}b}}\x{c}{d}}.\label{cross_ratio_relation_for_pentagon_rule}}
There is one final restriction that must be mentioned. The generators of pentagon rule identities---$f$-graphs including subgraphs with the topology shown on the left-hand side of (\ref{topologies_of_the_pentagon_rule})---must not involve any numerators connecting points on the pentagon {\it other than} between ${\color{hred}x_a}$ and $x_d$ (arbitrary powers of $\x{{\color{hred}a}}{d}$ are allowed). 

While the requirements for the graphs that participate in pentagon rule identities may seem stringent, each is important---as we will see when we describe the rule's proof. But the identities that result are very powerful: they always take the form that the sum of the coefficients of  the graphs involved (both the initial graph, and all its images in (\ref{topologies_of_the_pentagon_rule})) must vanish. 

Let us illustrate these relations with a concrete example from seven loops. Below, we have drawn an $f$-graph on the left, highlighting in blue the three points $\{{\color{hblue}x_b}\}$ that satisfy requirements described above; and on the right we have drawn the three $f$-graphs related to the initial graph according to (\ref{cross_ratio_relation_for_pentagon_rule}):
\vspace{-6pt}\eq{\hspace{-234pt}\fig{-54.75pt}{1}{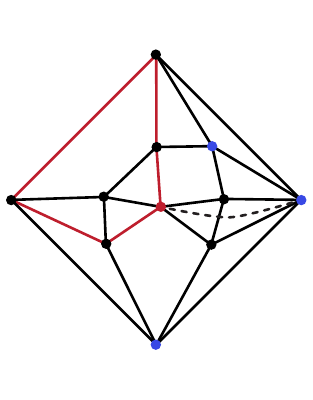}\bigger{\Rightarrow}\!\!\left\{\rule{0pt}{47.5pt}\right.\!\!\hspace{-7.5pt}\fig{-54.75pt}{1}{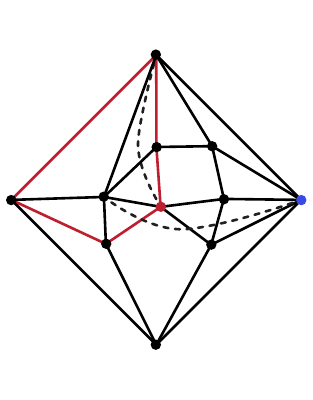},\hspace{-5pt}\fig{-54.75pt}{1}{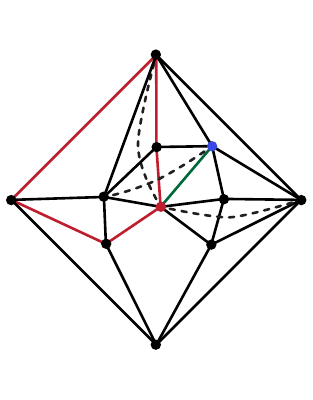},\hspace{-5pt}\fig{-54.75pt}{1}{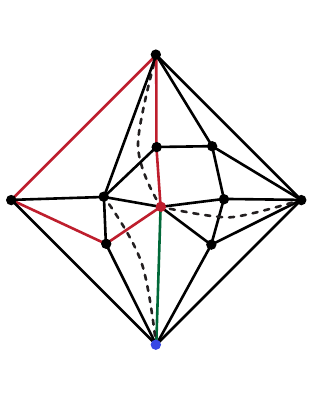}\hspace{-7.5pt}\left.\rule{0pt}{47.5pt}\right\}\hspace{-200pt}\label{seven_loop_pentagon_rule_example}\vspace{-6pt}}
Notice that two of the three points ${\color{hblue}x_b}$ are `implicit' in the manner described above. Labeling the coefficients of the $f$-graphs in (\ref{seven_loop_pentagon_rule_example}) from left to right as \mbox{$\{c_1^{7},c_2^7,c_3^7,c_4^7\}$}, the pentagon rule would imply that \mbox{$c_1^7\pl c_2^7\pl c_3^7\pl c_4^7\!=\!0$.} And indeed, these coefficients of terms in the seven loop correlator turn out to be: \mbox{$\{c_1^{7},c_2^7,c_3^7,c_4^7\}\!=\!\{0,0,\!\pl1,\!\mi1\},$} which do satisfy this identity. 

As usual, there are no symmetry factors to consider; but it is important that only {\it distinct} images are included in the set on the right-hand side of (\ref{topologies_of_the_pentagon_rule}). As will be discussed in \mbox{section \ref{sec:results}}, the pentagon rule is strong enough to fix all coefficients but one not already fixed by the square rule through seven loops. 

\vspace{-6pt}\subsubsection*{Proof of the Pentagon Rule}\label{subsubsec:proof_of_pentagon_rule}\vspace{-4pt}
The pentagon rule~(\ref{topologies_of_the_pentagon_rule}) arises from examining the 5-point light-like limit of the correlator and its relation to the five-particle amplitude (just as the square rule arises from the 4-point light-like limit and its relation to the four-particle amplitude explained in \mbox{section \ref{subsec:square_rule}}). As described in \mbox{section \ref{subsec:higher_point_amplitude_extraction}}, in the pentagonal light-like limit the correlator is directly related to the five-particle amplitude as in (\ref{f_to_5pt_amp_map2}).

In particular let us focus on the terms involving one loop amplitudes in (\ref{f_to_5pt_amp_map2}): ${\mathcal F}^{(\ell+1)}$ contains the terms,
\vspace{-5pt}\eq{\hspace{-75pt}\frac{1}{\xi^{(5)}}\left(\mathcal{M}_{\text{even}}^{(1)}\mathcal{M}_{\text{even}}^{(\ell-1)}+\epsilon_{123456}\epsilon_{12345(m+6)}\widehat{\mathcal{M}}_{\text{odd}}^{(1)}\widehat{\mathcal{M}}_{\text{odd}}^{(\ell-1)}\right)\,.\label{eq:24}\hspace{-40pt}\vspace{-5pt}}
Indeed any term in the correlator which graphically has a plane embedding with the topology of a 5-cycle whose `inside'  contains a single vertex and whose `outside' contains $\ell\mi1$ vertices has to arise from the above terms \cite{Ambrosio:2013pba}.

Inserting the one loop expressions~\eqref{five_point_one_loop_terms} and the algebraic identity (valid only in the pentagonal light-like limit),
\vspace{-0pt}\eq{\begin{split}&\hspace{-20pt}\phantom{=\,}\frac{\epsilon_{123456}\,\epsilon_{123457}}{\x{1}{2}\x{2}{3}\x{3}{4}\x{4}{5}\x{1}{5}}\\&\hspace{-20pt}=2\,\x{6}{7}+\left[\frac{\x{1}{6}\x{2}{7}\x{3}{5}+\x{1}{7}\x{2}{6}\x{3}{5}}{\x{1}{3}\x{2}{5}}-\frac{\x{1}{7}\x{3}{6}+\x{1}{6}\x{3}{7}}{\x{1}{3}}-\frac{\x{1}{6}\x{1}{7}\x{2}{4}\x{3}{5}}{\x{1}{3}\x{1}{4}\x{2}{5}}+\text{cyclic}\right]\,,\hspace{-24pt}\end{split}\label{eq:19}\vspace{-0pt}}
then \eqref{eq:24} becomes the following contribution to ${\mathcal F}^{(\ell+1)}$
\vspace{-5pt}\eq{\begin{split}&\hspace{-30pt}\frac{1}{\x{1}{2}\x{2}{3}\x{3}{4}\x{4}{5}\x{1}{5}}\Bigg(2\,\frac{\x{6}{7}}{\x{1}{6}\x{2}{6}\x{3}{6}\x{4}{6}\x{5}{6}}\hat{\mathcal{M}}^{(\ell-1)}_{\text{odd}}+\Bigg\{\frac{1}{\x{1}{6}\x{2}{6}\x{3}{6}\x{4}{6}}\Bigg[\frac{1}{\x{1}{4}\x{2}{5}\x{3}{5}}\mathcal{M}_{\text{even}}^{(\ell-1)}\hspace{-40pt}\\&\hspace{-30pt}+\Bigg(\frac{\x{1}{7}\x{2}{4}}{\x{1}{4}\x{2}{5}}+\frac{\x{4}{7}\x{1}{3}}{\x{1}{4}\x{3}{5}}-\frac{\x{3}{7}}{\x{3}{5}}-\frac{\x{2}{7}}{\x{2}{5}}-\frac{\x{5}{7}\x{1}{3}\x{2}{4}}{\x{2}{5}\x{3}{5}\x{1}{4}}\Bigg)\hat{\mathcal{M}}_{\text{odd}}^{(\ell-1)}\Bigg]+\text{cyclic}\Bigg\}\Bigg)\,.\hspace{-40pt}\end{split}\label{eq:12}\vspace{-5pt}}

We wish to now consider all terms in ${\mathcal F}^{(\ell+1)}$ containing the structure occurring in the pentagon rule, namely a `pentawheel' with a spoke missing,
\vspace{-5pt}\eq{\fig{-34.75pt}{1}{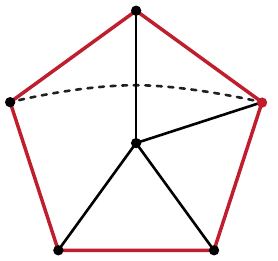}\label{eq:23}\vspace{-5pt}}
with numerators (if present at all within this subgraph) allowed {\it only} between the vertex with the missing spoke and the marked point (as shown). A term in ${\mathcal F}^{(\ell+1)}$ containing this subgraph inevitably contributes to the pentagonal light-like limit and by its topology it has to arise from the ${\mathcal M}^{(1)} \times {\mathcal M}^{(\ell-1)}$ terms, {\it i.e.}\ somewhere in~\eqref{eq:12}. We now proceed to investigate all seven terms in~\eqref{eq:12} to show that this structure of interest can only arise from the fifth and sixth terms.  

We start with the second term of \eqref{eq:12}
\vspace{-5pt}\eq{\frac{1}{\x{1}{2}\x{2}{3}\x{3}{4}\x{4}{5}\x{5}{1}}\,\frac{1}{\x{1}{6}\x{2}{6}\x{3}{6}\x{4}{6}}\,\frac{1}{\x{1}{4}\x{2}{5}\x{3}{5}}{\mathcal M}_{\text{even}}^{(\ell-1)}\,,\label{eq:3}\vspace{-5pt}}
arising from the even part of the amplitude, which is the most subtle one. Graphically, this term can be displayed as:
\vspace{-5pt}\eq{\fig{-34.75pt}{1}{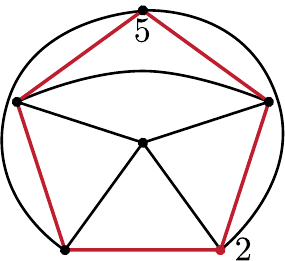}\label{pentagon_proof_figure_2}\vspace{-5pt}}
In order for this to yield the structure \eqref{eq:23} in a planar \mbox{$f$-graph}, the amplitude ${\mathcal M}_{\text{even}}^{(\ell-1)}$ must either contain a numerator $\x{1}{4}$ (to cancel the corresponding propagator above) or alternatively it must contain the numerator terms $\x{2}{5}$ and $\x{3}{5}$ in order to allow the edge $\x{1}{4}$ to be drawn  outside the pentagon without any edge crossing. Analyzing these different possibilities one concludes that this requires all three numerators $\x{1}{4}\x{2}{5}\x{3}{5}$ to be present in a term of ${\mathcal M}_{\text{even}}^{(\ell-1)}$. Now using the amplitude/correlator duality again in a different way note that such a contribution to ${\mathcal M}_{\text{even}}^{(\ell-1)}$ must also contribute to the lower-loop correlator ${\mathcal F}^{(\ell)}$ through~(\ref{f_to_5pt_amp_map2})
\vspace{-5pt}\eq{\lim_{\substack{\text{5-point}\\\text{light-like}}}\!\!\left(\xi^{(5)}\mathcal{F}^{(\ell-1)}\right)={\mathcal M}_\text{even}^{(\ell-1)} + \ldots\,.\label{eq:28}\vspace{-5pt}}
So a term in  ${\mathcal M}_{\text{even}}^{(\ell-1)}$ with numerators $\x{1}{4}\x{2}{5}\x{3}{5}$ contributes a term with topology,
\vspace{-5pt}\eq{\fig{-34.75pt}{1}{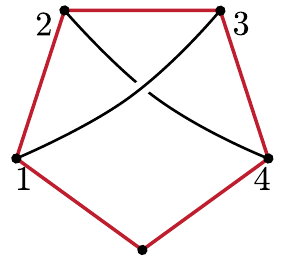}\label{pentagon_proof_figure_3}\vspace{-5pt}}
(Here the numerators $\x{1}{4}\x{2}{5}\x{3}{5}$ cancel three of the denominators of $1/\xi^{(5)}$, but they leave the pentagon and two further edges attached to the pentagon as shown.)

We see that this term can never be planar (this term in ${\mathcal M}_{\text{even}}^{(\ell-1)}$ has to be attached to all five external legs by conformal invariance so one cannot pull one of the offending edges outside the pentagon) {\it unless} there is a further numerator term, either $\x{2}{4}$ or $\x{1}{3}$ to cancel one of these edges. But in this case inserting this back into~\eqref{eq:3} we obtain the required structure~\eqref{eq:23} but with this further numerator which is of the type explicitly disallowed from our rule.

Having ruled out the second term, we consider the other terms of~\eqref{eq:12}.  The first term can clearly never give a pentawheel with a spoke missing. The contribution of the third term of~\eqref{eq:12} has the diagrammatic form:
\vspace{-5pt}\eq{\fig{-34.75pt}{1}{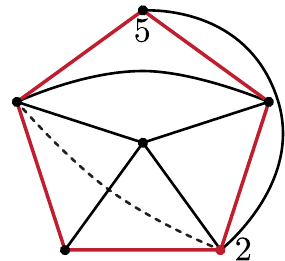}\label{pentagon_proof_figure_4}\vspace{-2.5pt}}
and so could potentially give a contribution of the form of a pentawheel with a spoke missing if $\hat{\mathcal M}_\text{odd}^{(\ell-1)}$ has a numerator $\x{1}{4}$ to cancel the corresponding edge. However in any case such a term would also contain the numerator $\x{2}{4}$ which we disallow in~\eqref{eq:23}. The third and last terms are similarly ruled out as a source for the structure in question. So we conclude that the fifth and sixth terms are the only ones which can yield the structure we focus on in the pentagon rule.

Given this important fact, we are now in a position to understand the origin of the pentagon rule. Every occurrence of the structure~\eqref{eq:23} arises from the fifth or sixth terms in~\eqref{eq:12}, namely from $\x{3}{7}/\x{3}{5}\times \hat {\mathcal M}_{\text{odd}}^{(\ell-1)}$ (where $x_3$ is the marked point of the pentagon). But we also know \cite{Ambrosio:2013pba} that $\hat {\mathcal M}_{\text{odd}}^{(\ell-1)}$ is in direct one-to-one correspondence with pentawheel structures of $f^{(\ell+1)}$ (the first term in~\eqref{eq:12}). Thus there is a direct link between the pentawheel structures and the  structure~\eqref{eq:23} and this link appears with a sign due to the sign difference between the first and fifth/sixth terms in~\eqref{eq:12}. To get from the first term of~\eqref{eq:12} to the fifth term, one multiplies by $\x{3}{7}\x{5}{6}/(\x{3}{5}\x{6}{7})$---that is, deleting the two edges, $\x{3}{7}$ and $\x{5}{6}$, and deleting the two numerator lines $\x{6}{7},\x{3}{5}$. This is precisely the operation involved in the five-point rule described in more detail above (see~(\ref{cross_ratio_relation_for_pentagon_rule})).

\newpage
\vspace{-6pt}\section{Bootstrapping Amplitudes/Correlators to Many Loops}\label{sec:results}\vspace{-6pt}
In this section, we survey the relative strengths of the three rules described in the pervious section, and then some of the more noteworthy aspects of the forms found for the correlator through ten loops. Before we begin, however, it is worth emphasizing that the three rules we have used are only three among many which follow from the way in which lower loop (and higher point) amplitudes are encoded in the correlator $\mathcal{F}^{(\ell)}$ via equations (\ref{f_to_4pt_amp_map_with_series_expansion}) and (\ref{f_to_npt_amp_map}). The triangle, square, and pentagon rules merely represent those we implemented first, and which proved sufficient through ten loops. And finally, it is worth mentioning that we expect the soft-collinear bootstrap criterion to continue to prove sufficient to fix all coefficients at all loops, even if using this tool has proven computationally out of reach beyond eight loops. (If it were to be translated into a purely graphical rule, it may prove extraordinarily powerful.)

~\\[-34pt]\paragraph{The Square Rule:}~\\
\indent As described in the previous section, the square rule is undoubtedly the most powerful of the three, and results in the simplest possible relations between coefficients---namely, that certain $\ell$ loop coefficients are identical to particular $(\ell\mi1)$ loop coefficients. As illustrated in \mbox{Table \ref{square_rule_strength_table}}, the square rule is strong enough to fix $\sim\!95$\% of the $22,\!097,\!035$ $f$-graphs coefficients at eleven loops. The role of the triangle and pentagon rules, therefore, can be seen as tools to fix the coefficients not already fixed by the square rule. 

~\\[-34pt]\paragraph{The Triangle Rule:}~\\
\indent Similar to the square rule, the triangle rule is strong enough to fix all coefficients through three loops, but will leave one free coefficient at four loops. Conveniently, the relations required by the triangle rule are not the same as those of the square rule, and so the combination of the two fix everything. In fact, the square and triangle rule together immediately fix all correlation functions through seven loops, and all but 22 of the $2,\!709$ eight loop coefficients. (This fact was known when the eight loop correlator was found in \mbox{ref.\ \cite{Bourjaily:2015bpz}}, which is why we alluded to these new rules in the conclusions of that Letter.)

Interestingly, applying the triangle and square rules to nine loops fixes all but 3 of the $43,\!017$ {\it new coefficients}, including 20 of those not already fixed at eight loops. (To be clear, this means that, without any further input, there would be a total of $3\pl2$ unfixed coefficients at nine loops.) Motivated by this, we implemented the triangle and square rules at ten loops, and found that these rules sufficed to determine eight and nine loop correlators uniquely. At ten loops, we found the complete system of equations following from the two rules to fix all but $1,\!570$ of the coefficients of the $900,\!145$ $f$-graphs. 

These facts are summarized in \mbox{Table \ref{square_and_triangle_rules_strength_table}}. Notice that the number of unknowns quoted in that table for $\ell$ loops are the number of coefficients given the lower loop correlator. If the coefficients at lower loops were not assumed, then there would be $5$ unknowns at nine loops rather than 3; but the number quoted for ten loops would be the same---because all lower loop coefficients are fixed by the ten loop relations. 

\begin{table}[t]\caption{Statistics of coefficients fixed by the square \& triangle rules through $\ell\!=\!10$ loops.\label{square_and_triangle_rules_strength_table}}\vspace{-10pt}$$\fwbox{0pt}{\begin{array}{|r|r|r|r|r|r|r|r|r|r|}\cline{2-10}\multicolumn{1}{r}{\ell\!=}&\multicolumn{1}{|c|}{2}&\multicolumn{1}{c|}{3}&\multicolumn{1}{c|}{4}&\multicolumn{1}{c|}{5}&\multicolumn{1}{c|}{6}&\multicolumn{1}{c|}{7}&\multicolumn{1}{c|}{8}&\multicolumn{1}{c|}{9}&\multicolumn{1}{c|}{10}\\\hline\text{number of $f$-graph coefficients:}&\,1\,&\,1\,&\,3\,&\,7\,&\,36\,&\,220\,&\,2,\!709\,&\,43,\!017\,&\,900,\!145\,\\\hline\text{unknowns remaining after square rule:}&\,\,0\,&\,\,0\,&\,\,1\,&1\,&5\,&22\,&293\,&2,\!900\,&52,\!475\,\\\hline\text{unknowns after square \& triangle rules:}&\,\,0\,&\,\,0\,&\,\,0\,&\,\,0\,&\,\,0\,&\,\,0\,&22\,&3\,&1,\!570\,\\\hline\end{array}}$$\vspace{-18pt}\vspace{-0pt}\end{table}

~\\[-34pt]\paragraph{The Pentagon Rule:}~\\
\indent The pentagon rule is not quite as strong as the others, but the relations implied are much simpler to implement. In fact, there are no instances of $f$-graphs for which the pentagon rule applies until four loops, when it implies a single linear relation among the three coefficients. This relation, when combined with the square rule fixes the four loop correlator, and the same is true for five loops. However at six loops, the two rules combined leave 1 (of the $36$) $f$-graph coefficients undetermined. The reason for this is simple: there exists an $f$-graph at six loops which neither contributes to $\mathcal{A}^{(5)}_4\mathcal{A}^{(1)}_4$ nor to $\mathcal{M}_5^{(4)}\overline{\mathcal{M}}_5^{(1)}$. This is easily seen by inspection of the $f$-graph in question: 
\vspace{-5pt}\eq{\fig{-54.75pt}{1}{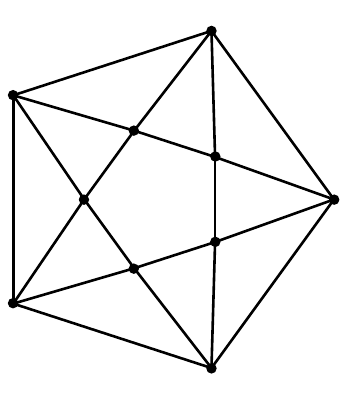}\label{six_loop_prism_graph}\vspace{-5pt}}
We will have more to say about this graph and its coefficient below. There is one graph at seven loops related to (\ref{six_loop_prism_graph}) by the square rule that is also left undetermined, but all other coefficients (219 of the 220) are fixed by the combination of the square and pentagon rules. 

The number of coefficients fixed by the square and pentagon rules through nine loops is summarized in \mbox{Table \ref{square_and_pentagon_rules_strength_table}}. As before, only the number of {\it new} coefficients are quoted---assuming that the lower loop coefficients are known. 

\begin{table}[b]\vspace{-40pt}$$\fwbox{0pt}{\begin{array}{|r|r|r|r|r|r|r|r|r|}\cline{2-9}\multicolumn{1}{r}{\ell\!=}&\multicolumn{1}{|c|}{2}&\multicolumn{1}{c|}{3}&\multicolumn{1}{c|}{4}&\multicolumn{1}{c|}{5}&\multicolumn{1}{c|}{6}&\multicolumn{1}{c|}{7}&\multicolumn{1}{c|}{8}&\multicolumn{1}{c|}{9}\\\hline\text{number of $f$-graph coefficients:}&\,1\,&\,1\,&\,3\,&\,7\,&\,36\,&\,220\,&\,2,\!709\,&\,43,\!017\,\\\hline\text{unknowns remaining after square rule:}&\,\,0\,&\,\,0\,&\,\,1\,&\,\,1\,&\,\,5\,&22\,&293\,&2,\!900\,\\\hline\text{unknowns after square \& pentagon rules:}&\,\,0\,&\,\,0\,&\,\,0\,&\,\,0\,&\,\,\,1\,&\,\,\,0\,&\,\,\,17\,&\,\,\,64\,\\\hline\end{array}}$$\vspace{-18pt}\caption{Statistics of coefficients fixed by the square \& pentagon rules through $\ell\!=\!9$ loops.\label{square_and_pentagon_rules_strength_table}}\vspace{-30pt}\end{table}

\newpage
\vspace{-0pt}\subsection{Aspects of Correlators and Amplitudes at High Loop-Orders}\label{subsec:statistical_tour}\vspace{-0pt}
While no two of the three rules alone prove sufficient to determine the ten loop correlation function, the three in combination fix all coefficients uniquely---without any outside information about lower loops. As such, the reproduction of the eight (and lower) loop functions found in \mbox{ref.\ \cite{Bourjaily:2015bpz}} can be viewed as an independent check on the code being employed. Moreover, because the three rules each impose mutually overlapping (and individually over constrained) constraints on the coefficients, the existence of any solution is a source of considerable confidence in our results. 

One striking aspect of the correlation function exposed only at high loop-order is that the (increasingly vast) majority of coefficients are zero: while all possible $f$-graphs contribute through five loops, only 26 of the 36 graphs at six loops do; by ten loops, $85\%$ of the coefficients vanish. (At eleven loops, {\it at least} $19,\!388,\!448$ coefficients vanish ($88\%$) due to the square rule alone.) This pattern is illustrated in \mbox{Table \ref{correlator_contributions_table}}, where we count all contributions---both for $f$-graphs, and planar DCI integrands. 

The two principle novelties discovered for the eight loop correlator \cite{Bourjaily:2015bpz} also persist to higher loops. Specifically, we refer to the fact that there are contributions to the amplitude that are finite (upon integration) even on-shell, and contributions to the correlator that are (individually) divergent even off-shell. The meaning of the finite integrals remains unclear (although they would have prevented the use of the soft-collinear bootstrap without grouping terms according to $f$-graphs); but the existence of divergent contributions imposes an important constraint on the result: because the correlator is strictly finite off-shell, all such divergences must cancel in combination. (Moreover, these contributions impose an interesting technical obstruction to evaluation, as they cannot be easily regulated in four dimensions---such as by going to the Higgs branch of the theory \cite{Alday:2009zm}.)
 
\begin{table}[b]\vspace{-15pt}$$\hspace{1.5pt}\begin{array}{|@{$\,$}c@{$\,$}|@{$\,$}r@{$\,$}|@{$\,$}r@{$\,$}|@{$\,\,$}r@{$\,\,$}|@{$\;\;\;\;\;\;\;$}|@{$\,$}r@{$\,$}|@{$\,$}r@{$\,$}|@{$\,\,$}r@{$\,\,$}|}\multicolumn{1}{@{$\,$}c@{$\,$}}{\begin{array}{@{}l@{}}\\[-4pt]\text{$\ell\,$}\end{array}}&\multicolumn{1}{@{$\,$}c@{$\,$}}{\!\begin{array}{@{}c@{}}\text{number of}\\[-4pt]\text{$f$-graphs}\end{array}}\,&\multicolumn{1}{@{$\,$}c@{$\,$}}{\begin{array}{@{}c@{}}\text{no.\ of $f$-graph}\\[-4pt]\text{contributions}\end{array}}\,\,&\multicolumn{1}{@{$\,$}c@{$\,$}}{\begin{array}{@{}c@{$\,\,\,\,\;\;\;\;\;$}}\text{}\\[-4pt]\text{\!\!(\%)}\end{array}}&\multicolumn{1}{@{$\,$}c@{$\,$}}{\begin{array}{@{}c@{}}\text{number of}\\[-4pt]\text{DCI integrands}\end{array}}\,&\multicolumn{1}{@{$\,$}c@{$\,$}}{\begin{array}{@{}c@{}}\text{no.\ of integrand}\\[-4pt]\text{contributions}\end{array}}\,&\multicolumn{1}{@{$\,$}c@{$\,$}}{\begin{array}{@{}c@{}}\text{}\\[-4pt]\text{(\%)}\end{array}}\\[-0pt]\hline1&1&1&100&1&1&100\\\hline2&1&1&100&1&1&100\\\hline3&1&1&100&2&2&100\\\hline4&3&3&100&8&8&100\\\hline5&7&7&100&34&34&100\\\hline6&36&26&72&284&229&81\\\hline7&220&127&58&3,\!239&1,\!873&58\\\hline8&2,\!709&1,\!060&39&52,\!033&19,\!949&38\\\hline9&43,\!017&10,\!525&24&1,\!025,\!970&247,\!856&24\\\hline10&900,\!145&136,\!433&15&24,\!081,\!425&3,\!586,\!145&15\\\hline\end{array}\vspace{-16pt}$$\vspace{-6pt}\caption{Statistics of $f$-graph and DCI integrand {\it contributions} through $\ell\!=\!10$ loops.\label{correlator_contributions_table}}\vspace{-24pt}\end{table}
\newpage 

At eight loops there are exactly 4 $f$-graphs which lead to finite DCI integrands, and all 4 have non-vanishing coefficients. At nine loops there are 45, of which 33 contribute; at ten loops there are $1,\!287$, of which $570$ contribute. For the individually divergent contributions, their number and complexity grow considerably beyond eight loops. The first appearance of such divergences happened at eight loops---with terms that had a so-called `$k\!=\!5$' divergence (see \cite{Bourjaily:2015bpz} for details). Of the 662 $f$-graphs with a $k\!=\!5$ divergence at eight loops, only 60 contributed. At nine loops there are $15,\!781$, of which $961$ contribute; at ten loops, there are $424,\!348$, of which $21,\!322$ contribute. Notice that terms with these divergences grow proportionally in number---and even start to have the feel of being ubiquitous asymptotically. We have not enumerated all the divergent contributions for $k\!>\!5$, but essentially all categories of such divergences exist and contribute to the correlator. (For example, there are $971$ contributions at ten loops with (the simplest category of) a $k\!=\!7$ divergence.)

While the coefficients of $f$-graphs are encouragingly simple at low loop-orders, the variety of possible coefficients seems to grow considerably at higher orders. The distribution of these coefficients is given in \mbox{Table \ref{coefficient_statistics_table}}. While all coefficients through five loops were $\pm\!1$, those at higher loops include many novelties. (Of course, the increasing dominance of zeros among the coefficients is still rather encouraging.)

Interestingly, it is clear from \mbox{Table \ref{coefficient_statistics_table}} that new coefficients (up to signs) only appear at even loop-orders. The first term with coefficient $\!\mi1$ occurs at four loops, and the first appearance of $\!\pl2$ at six loops. At eight loops, we saw the first instances of $\pm\frac{1}{2}$, $\pm\frac{3}{2}$, and also $\!\mi5$. And there are many novel coefficients that first appear at ten loops. 
\begin{table}[t]$$\hspace{-1.2pt}\begin{array}{|c|r|r|r|r|r|r|r|r|r|r|r|r|r|r|}\multicolumn{1}{c}{}&\multicolumn{14}{c}{\text{number of $f$-graphs at $\ell$ loops having coefficient:}}\\\cline{2-15}\multicolumn{1}{c}{\ell}&\multicolumn{1}{|c|}{\pm1\phantom{}}&\multicolumn{1}{c|}{0}&\multicolumn{1}{c|}{\pm2}&\multicolumn{1}{c|}{\pm1/2}&\multicolumn{1}{c|}{\pm3/2}&\multicolumn{1}{c|}{\pm5}&\multicolumn{1}{c|}{\pm1/4}&\multicolumn{1}{c|}{\pm3/4}&\multicolumn{1}{c|}{\pm5/4}&\multicolumn{1}{c|}{+7/4}&\multicolumn{1}{c|}{\pm9/4}&\multicolumn{1}{c|}{\pm5/2}&\multicolumn{1}{c|}{+4}&\multicolumn{1}{c|}{+14}\\\hline1&1&{\color{dim}0}&{\color{dim}0}&{\color{dim}0}&{\color{dim}0}&{\color{dim}0}&{\color{dim}0}&{\color{dim}0}&{\color{dim}0}&{\color{dim}0}&{\color{dim}0}&{\color{dim}0}&{\color{dim}0}&{\color{dim}0}\\\hline2&1&{\color{dim}0}&{\color{dim}0}&{\color{dim}0}&{\color{dim}0}&{\color{dim}0}&{\color{dim}0}&{\color{dim}0}&{\color{dim}0}&{\color{dim}0}&{\color{dim}0}&{\color{dim}0}&{\color{dim}0}&{\color{dim}0}\\\hline3&1&{\color{dim}0}&{\color{dim}0}&{\color{dim}0}&{\color{dim}0}&{\color{dim}0}&{\color{dim}0}&{\color{dim}0}&{\color{dim}0}&{\color{dim}0}&{\color{dim}0}&{\color{dim}0}&{\color{dim}0}&{\color{dim}0}\\\hline4&3&{\color{dim}0}&{\color{dim}0}&{\color{dim}0}&{\color{dim}0}&{\color{dim}0}&{\color{dim}0}&{\color{dim}0}&{\color{dim}0}&{\color{dim}0}&{\color{dim}0}&{\color{dim}0}&{\color{dim}0}&{\color{dim}0}\\\hline5&7&{\color{dim}0}&{\color{dim}0}&{\color{dim}0}&{\color{dim}0}&{\color{dim}0}&{\color{dim}0}&{\color{dim}0}&{\color{dim}0}&{\color{dim}0}&{\color{dim}0}&{\color{dim}0}&{\color{dim}0}&{\color{dim}0}\\\hline6&25&10&1&{\color{dim}0}&{\color{dim}0}&{\color{dim}0}&{\color{dim}0}&{\color{dim}0}&{\color{dim}0}&{\color{dim}0}&{\color{dim}0}&{\color{dim}0}&{\color{dim}0}&{\color{dim}0}\\\hline7&126&93&1&{\color{dim}0}&{\color{dim}0}&{\color{dim}0}&{\color{dim}0}&{\color{dim}0}&{\color{dim}0}&{\color{dim}0}&{\color{dim}0}&{\color{dim}0}&{\color{dim}0}&{\color{dim}0}\\\hline8&906&1,\!649&9&141&3&1&{\color{dim}0}&{\color{dim}0}&{\color{dim}0}&{\color{dim}0}&{\color{dim}0}&{\color{dim}0}&{\color{dim}0}&{\color{dim}0}\\\hline9&7,\!919&32,\!492&54&2,\!529&22&1&{\color{dim}0}&{\color{dim}0}&{\color{dim}0}&{\color{dim}0}&{\color{dim}0}&{\color{dim}0}&{\color{dim}0}&{\color{dim}0}\\\hline10&78,\!949&763,\!712&490&50,\!633&329&9&5,\!431&559&18&5&4&4&1&1\\\hline\end{array}$$\vspace{-24pt}\caption{Statistics of $f$-graph coefficients in the expansion of $\mathcal{F}^{(\ell)}$ through $\ell\!=\!10$ loops.\label{coefficient_statistics_table}}\vspace{-10pt}\end{table}

While most of the `new' coefficients occur with sufficient multiplicity to require further consideration (more than warranted here), there is at least one class of contributions which seems predictably novel. Consider the following six, eight, and ten loop $f$-graphs: 
\vspace{-10pt}\eq{\fig{-54.75pt}{1}{six_loop_prism_graph}\quad\fig{-54.75pt}{1}{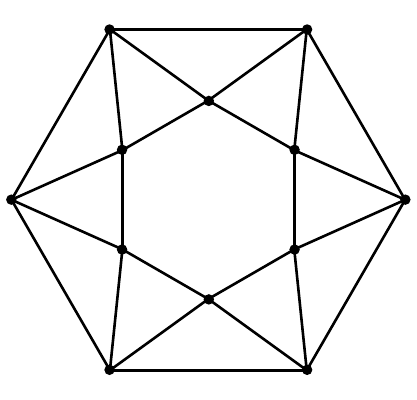}\quad\fig{-54.75pt}{1}{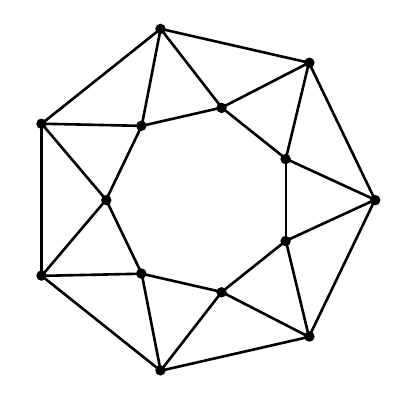}\label{prism_graph_figures}\vspace{-10pt}}
These graphs all have the topology of a $(\ell/2\pl2)$-gon anti-prism, and all represent contributions with unique (and always exceptional) coefficients. In particular, these graphs contribute to the correlator with coefficients $\!\pl2$, $\!\mi5$ and $\!\!\pl14$, respectively. (Notice also that the four loop $f$-graph $f_3^{(4)}$ shown in (\ref{one_through_four_loop_f_graphs}) is an anti-prism of this type---and is the first term having contribution $\!\mi1$---as is the only two loop $f$-graph (the octahedron), which also follows this pattern.) Each of the $f$-graphs in (\ref{prism_graph_figures}) contribute a unique DCI integrand to the $\ell$ loop amplitude,
\vspace{-12pt}\eq{\fig{-54.75pt}{1}{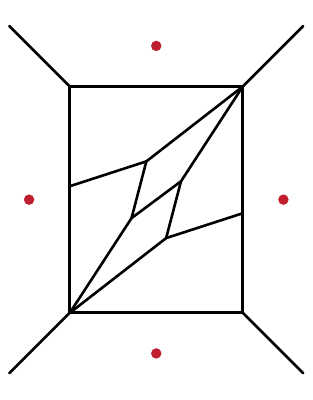}\qquad\fig{-54.75pt}{1}{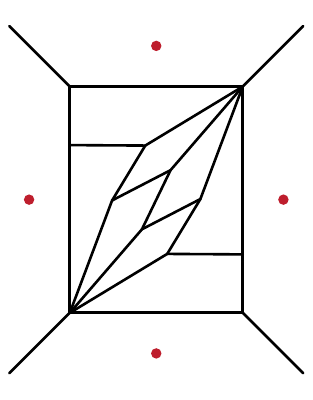}\qquad\fig{-54.75pt}{1}{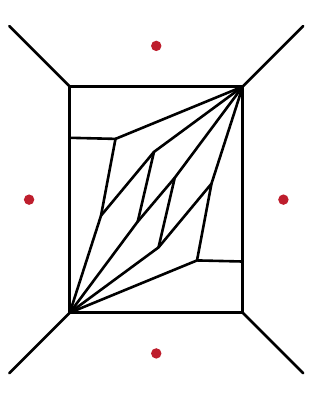}\label{prism_dci_integrands}\vspace{-12pt}}
with each drawn in momentum space as Feynman graphs for the sake of intuition. From these, a clear pattern emerges---leading us to make a rather speculative guess for the coefficients of these terms. It seems plausible that the coefficients of anti-prism graphs are given by the Catalan numbers---leading us to predict that the coefficient of the octagonal anti-prism $f$-graph at twelve loops, for example, will be $\!\mi42$. Testing this conjecture---let alone proving it---however, must await further work. 

The only other term that contributes at ten loops with a unique coefficient is the following, which has coefficient $\!\pl4$: 
\vspace{-10pt}\eq{\fig{-54.75pt}{1}{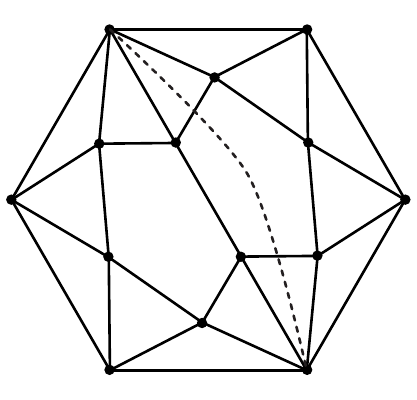}\;\;\bigger{\supset}\fig{-54.75pt}{1}{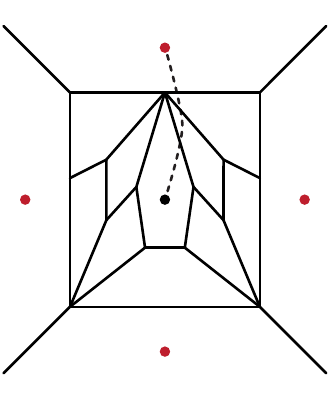}\label{ten_loop_coefficient_4_graph},\ldots\vspace{-10pt}}

We hope that the explicit form of the correlation functions provided at  \href{http://goo.gl/JH0yEc}{http://goo.gl/JH0yEc} (see \mbox{Appendix \ref{appendix:mathematica_and_explicit_results}}) will provide sufficient data for other researchers to find new patterns within the structure of coefficients. 

\newpage
\vspace{-6pt}\section{Conclusions and Future Directions}\label{sec:conclusions}\vspace{-6pt}
In this work, we have described a small set of simple, graphical rules which prove to be extremely efficient in fixing the possible contributions to the $\ell$ loop four-point correlation function in planar maximally supersymmetric $(\mathcal{N}\!=\!4)$ Yang-Mills theory (SYM). And we have described the form that results when used to fix the correlation function through ten loop-order. While clearly this is merely the simplest non-trivial observable in (arguably) the simplest four-dimensional quantum field theory, it exemplifies many of the features (and possible tools) we expect will be applicable to more general quantum field theories. And even within the limited scope of planar SYM, this single function contains important information about higher-point amplitudes. 

It is important to reiterate that the rules we have described are merely necessary conditions---and not obviously sufficient to all orders. But these three rules are merely three among many that follow from the consistency of the amplitude/correlator duality. Even without extension beyond ten loops, it would be worthwhile (and very interesting) to explore the strengths of the various natural generalizations of the rules we have described. 

Another important open direction would be to explore the systematic extraction of higher-point (lower loop) amplitudes from the four-point correlator. This has proven exceptionally direct and straight-forward for five-point amplitudes, but further work should be done to better understand the systematics (and potential difficulties) of this procedure for higher multiplicity. (Even six-particle amplitude extraction remains largely unexplored.)

Finally, it is natural to wonder how far this programme can be extended beyond ten loops. Although the use of graphical rules essentially eliminates the challenges of setting up the linear algebra problem to be solved, solving the system of equations that result (with millions of unknowns) rapidly becomes rather non-trivial. However, such problems of linear algebra (involving (very) large systems of equations) arise in many areas of physics and computer science, and there is reason to expect that they may be surmounted through the use of programmes such as that described in \mbox{ref.\ \cite{vonManteuffel:2014ixa}} (an impressive implementation of Laporta's algorithm). At present, it is unclear where the next computational bottle-neck will be, but it is worth pushing these tools as far as they can go---certainly to eleven loops, and possibly even twelve.

\newpage
\vspace{-0pt}\section*{Acknowledgements}\vspace{-6pt}
The authors gratefully acknowledge helpful discussions with Zvi Bern, Simon Caron-Huot, JJ Carrasco, Dmitry Chicherin, Burkhard Eden, Gregory Korchemsky, Emery Sokatchev, and Marcus Spradlin. This work was supported in part by the Harvard Society of Fellows, a grant from the Harvard Milton Fund, by the Danish National Research Foundation (DNRF91), and by a MOBILEX research grant from the Danish Council for Independent Research (JLB); by an STFC studentship (VVT); and by an STFC Consolidated Grant ST/L000407/1 and the Marie Curie network GATIS (gatis.desy.eu) of the European Union's Seventh Framework Programme FP7/2007-2013 under REA Grant Agreement No.\ 317089 (PH). PH would also like to acknowledge the hospitality of Laboratoire dÕAnnecy-le-Vieux de Physique Th\'eorique, UMR 5108, where this work was completed.

\vspace{20pt}\appendix\section{Obtaining and Using the Explicit Results in {\sc Mathematica}}\label{appendix:mathematica_and_explicit_results}\vspace{-6pt}
Our full results, including all contributions to the amplitude and correlator $\mathcal{F}^{(\ell)}$ through ten loops, have been made available at the site  \href{http://goo.gl/JH0yEc}{http://goo.gl/JH0yEc}. These can be obtained by downloading the compressed file {\tt multiloop\rule[-0.75pt]{7.5pt}{0.5pt}data.zip}, or by downloading each data file individually (which are encoded somewhat esoterically). These files include a {\sc Mathematica} package, {\tt consolidated\rule[-0.75pt]{7.5pt}{0.5pt}multiloop\rule[-0.75pt]{7.5pt}{0.5pt}data.m}, and a notebook {\tt multiloop\rule[-0.75pt]{7.5pt}{0.5pt}demo.nb}. 

The demonstration notebook illustrates the principle data defined in the package, and examples of how these functions are represented. Also included in the package are several general-purpose functions that may be useful to the reader---for example, a functions that compute symmetry factors and check if two functions are isomorphic (as graphs). Principle among the data included in this package are the list of all $f$-graphs at $\ell$ loops with non-vanishing coefficients for $\ell\!=\!1,\ldots,10$, and the corresponding coefficients. Also included is a list of all $\ell$ loop DCI integrands obtained from each $f$-graph in the light-like limit. 

Importantly, we have only included terms with non-vanishing coefficients---in order to reduce the file size of the data. The complete list of $f$-graphs at each loop order can be obtained by contacting the authors. 

\newpage
\providecommand{\href}[2]{#2}\begingroup\raggedright\endgroup

\end{document}

\pdfoutput=1

\documentclass[12pt,a4paper,hyperref]{jhepLike}
\let\ifpdf\relax
\usepackage{ifpdf,color}
\let\normalcolor\relax
\usepackage{mathtools,cite}
\widowpenalty=500\clubpenalty=1000
\unitlength=1mm\textheight=8.7in

\newcommand{\eq}[1]{\vspace{-0.5pt}\begin{equation}#1\vspace{-0.5pt}\end{equation}}
\newcommand{\fwbox}[2]{\text{\makebox[#1][c]{$\hspace{-150pt}\displaystyle#2\hspace{-150pt}$}}}
\newcommand{\fwboxL}[2]{\text{\makebox[#1][l]{$#2$}}}
\newcommand{\fwboxR}[2]{\text{\makebox[#1][r]{$#2$}}}
\newcommand{\fig}[3]{\raisebox{#1}{\ \includegraphics[scale=#2]{#3}}}
\newcommand{\mi}{\raisebox{0.75pt}{\scalebox{0.75}{$\,-\,$}}}
\newcommand{\pl}{\raisebox{0.75pt}{\scalebox{0.75}{$\,+\,$}}}
\renewcommand{\phi}{\varphi}
\newcommand{\ab}[1]{\langle\hspace{-0.5pt}#1\hspace{-0.5pt}\rangle}
\newcommand{\bigger}[1]{\raisebox{-2.25pt}{\scalebox{1.75}{$#1$}}}
\newcommand{\x}[2]{x_{#1\hspace{0.5pt}#2}^2}
\renewcommand{\hat}{\widehat}
\definecolor{dim}{rgb}{0.75,0.75,0.75}
\definecolor{hblue}{rgb}{0.18,0.19,0.572}
\definecolor{hred}{rgb}{0.745,0.118,0.176}

%
%
%
%

\thispagestyle{empty}
\title{{\LARGE \mbox{Amplitudes and Correlators to Ten Loops}}\\ {\LARGE\mbox{Using Simple, Graphical Bootstraps}}}
\author{{\normalsize \mbox{Jacob~L.~Bourjaily$^1$, Paul~Heslop$^2$, Vuong-Viet~Tran\mbox{$^{2}$}}}\\
\mbox{{\mbox{$^1$}\ Niels Bohr International Academy and Discovery Center, University of Copenhagen}}\\
\mbox{{\mbox{$^{\phantom{1}}$}\ Blegdamsvej 17, DK-2100 Copenhagen \O, Denmark}}\\
\mbox{{\mbox{$^2$}\ Centre for Particle Theory, Department of Mathematical Sciences, Durham University,}}\\
\mbox{{\mbox{$^{\phantom{2}}$}\ South Road, Durham DH1 3LE, United Kingdom}}\\\vspace{-10pt}\\
\mbox{\hspace{-13pt}{{\it E-mails:} {\tt bourjaily@nbi.ku.dk, paul.heslop@durham.ac.uk,}}}\\\vspace{-15pt}\\\mbox{\hspace{28pt}{ {\tt vuong-viet.tran@durham.ac.uk}}}\vspace{-10pt}
}
\keywords{scattering amplitudes, correlation functions, supersymmetry}
\date{\today}
\abstract{%
We introduce two new graphical-level relations among possible contributions to the four-point correlation function and scattering amplitude in planar, maximally supersymmetric Yang-Mills theory. When combined with the rung rule, these prove powerful enough to fully determine both functions through ten loops. This then also yields the full five-point amplitude to eight loops and the parity-even part to nine loops. We derive these rules, illustrate their applications, and compare their relative strengths for fixing coefficients. We survey some of the features of the previously unknown nine and ten loop expressions, and provide explicit formulae for each as part of this work's submission files to the {\tt arXiv}. 
}
\preprint{DCPT-16/31}

\begin{document}
\vspace{-27.8pt}\section{Introduction}\label{sec:introduction}\vspace{-6pt}
Among all four-dimensional quantum field theories lies a unique example singled out for its remarkable symmetry and mathematical structure as well as its key role in the AdS/CFT correspondence. This is maximally supersymmetric ($\mathcal{N}\!=\!4$) Yang-Mills theory (SYM) in the planar limit \cite{ArkaniHamed:2008gz}. It has been the subject of great interest over recent years, and the source of many remarkable discoveries that may extend to much more general quantum field theories. These features include a connection to Grassmannian geometry \cite{ArkaniHamed:2009dn,ArkaniHamed:2009sx,ArkaniHamed:2009dg,ArkaniHamed:2012nw,ArkaniHamed:book}, extra simplicity for planar theories' loop integrands \cite{ArkaniHamed:2010gh,Bourjaily:2011hi,Bourjaily:2015jna}, the existence of all-loop recursion relations \cite{ArkaniHamed:2010kv}, and the existence of unanticipated symmetries \cite{Drummond:2007cf,Drummond:2008vq,Brandhuber:2008pf,Drummond:2009fd} and related dualities between observables in the theory \cite{Alday:2007hr,Drummond:2007aua,Brandhuber:2007yx,Alday:2010zy,Eden:2010zz,Mason:2010yk,CaronHuot:2010ek,Eden:2011yp,Adamo:2011dq,Eden:2011ku}. Of these, the duality between scattering amplitudes and correlation functions, will play a fundamental role throughout this work. 

Much of this progress has been fueled through concrete theoretical data: heroic efforts of computation are made to determine observables (with more states, and at higher orders of perturbation); and this data leads to the discovery of new patterns and structures that allow these efforts to be extended even further. This virtuous cycle---even when applied merely to the `simplest' quantum field theory---has taught us a great deal about the structure of field theory in general, and represents an extremely fruitful way to improve our ability to make predictions for experiments. 

In this paper, we greatly extend the reach of this theoretical data by computing a particular observable in this simple theory to {\it ten} loops---mere months after eight loops was first determined. This is made possible through the use of powerful new {\it graphical} rules described in this work. The observable in question is the four-point correlation function among scalars---the simplest operator that receives quantum corrections in planar SYM. This correlation function is closely related to the four-particle scattering amplitude, obtained (in essence) from the correlator by taking the external states to be on-shell as reviewed below. But the information contained in this single function is vastly more general: it contains information about all scattering amplitudes in the theory---including those involving more external states (at lower loop-orders). As such, our determination of the four-point correlator at ten loops immediately provides information about the five-point amplitude at nine loops, the six-point amplitude at eight loops, etc.\ \cite{Ambrosio:2013pba}. 

Before we review this correspondence and describe the rules used to obtain the ten loop correlator, it is worth taking a moment to reflect on the history of our knowledge about it. Considered as an amplitude, it has been the subject of much interest for a long time. The tree-level amplitude was first expressed in supersymmetric form by Nair in \mbox{ref.\ \cite{Nair:1988bq}}. It was computed using unitarity to two loops in 1997 \cite{Bern:1997nh} (see also \cite{Anastasiou:2003kj}), to three loops in 2005 \cite{Bern:2005iz}, to five loops in 2007---first at four loops \cite{Bern:2006ew}, and five quickly thereafter \cite{Bern:2007ct}---and to six loops around 2009 \cite{Bern:2012di} (although published later). The extension to seven loops required significant new technology. This came from the discovery of the soft-collinear bootstrap in 2011 \cite{Bourjaily:2011hi}. Although not known at the time, the soft-collinear bootstrap method (as described in \mbox{ref.\ \cite{Bourjaily:2011hi}}), would have failed beyond seven loops; but luckily, the missing ingredient would be supplied by the duality between amplitudes and correlation functions discovered in \cite{Eden:2010zz,Alday:2010zy} and elaborated in \cite{Eden:2010ce,Eden:2011yp,Eden:2011ku,Adamo:2011dq}. The determination of the four-point correlator in planar SYM followed a somewhat less linear trajectory. One and two loops were obtained soon after (and motivated by) the AdS/CFT correspondence between 1998 and 2000 \cite{GonzalezRey:1998tk,Eden:1998hh,Eden:1999kh,Eden:2000mv,Bianchi:2000hn}. But despite a great deal of effort by a number of groups, the three loop result had to wait over 10 years until 2011---at which time the four, five, and six loop results were found in quick succession \cite{Eden:2011we,Eden:2012tu,Ambrosio:2013pba,Drummond:2013nda}; seven loops was reached in 2013 \cite{Ambrosio:2013pba}. 

The breakthrough for the correlator, enabling this rapid development, was the discovery of a hidden symmetry \cite{Eden:2011we,Eden:2012tu}. On the amplitude side, the extension of the above methods to eight loops also required the exploitation of this symmetry via the duality between amplitudes and correlators. This hidden symmetry (reviewed below) greatly simplifies the work required to extend the soft-collinear bootstrap, making it possible to determine the eight loop functions in 2015 \cite{Bourjaily:2015bpz}. 

While the eight loop amplitude and correlator were determined (the `hard way',) using just the soft-collinear bootstrap and hidden symmetry, we had already started exploring alternative methods to find these functions which seemed quite promising. These were mentioned in the conclusions of \mbox{ref.\ \cite{Bourjaily:2015bpz}}---the details of which we describe in this note. This new approach, based not on algebraic relations but graphical ones, has allowed for a watershed of new theoretical data similar to that of 2007: within a few short months, we were able to fully determine both the nine and ten loop correlation functions. The reason for this great advance---the (computational) advantages of graphical rules---will be discussed at the end of this introduction. 

Our work is organized as follows. In \mbox{section \ref{sec:review_of_duality}} we review the representation of amplitudes and correlation functions, and the duality between them. This will include a summary of the notation and conventions used throughout this paper, and also a description of the way that the terms involved are represented both algebraically and graphically. We elaborate on how the plane embedding of the terms that contribute to the correlator (viewed as graphs) allow for the direct extraction of amplitudes at corresponding (and lower) loop-orders---including amplitudes involving more than four external states---in \mbox{section \ref{subsec:amplitude_extraction}}. The three graphical rules sufficient to fix all possible contributions (at least through ten loops) are described in \mbox{section \ref{sec:graphical_bootstraps}}. We will refer to these as the triangle, square, and pentagon rules. 

The triangle and the square rules relate terms at different loop orders, while the pentagon rule relates terms at a given loop-order. While the square rule is merely the graphical manifestation of the so-called `rung' rule \cite{Bern:1997nh,Eden:2012tu} (generalized by the hidden symmetry of the correlator), the triangle and pentagon rules are new. We provide illustrations of each and proofs of their validity in \mbox{section \ref{sec:graphical_bootstraps}}. These rules have varying levels of strength. While the square rule is well-known to be insufficient to determine the amplitude or correlator at all orders (and the same is true for the pentagon rule), we expect that the combination of the square and triangle rules {do} prove sufficient---but only after their consequences at higher loop-orders are taken also into account. (For example, the pentagon rule was not required for us to determine the nine loop correlator---but the constraints that follow from the square and triangle rules at ten loops were necessary.) In \mbox{section \ref{sec:results}} we describe the varying strengths of each of these rules, and summarize the expressions found for the correlation function and amplitude through ten loops in \mbox{section \ref{subsec:statistical_tour}}. The explicit expressions for the ten loop correlator and amplitude have been included as part of this work's submission files to the {\tt arXiv}. Details on how these can be obtained, how the data has been encoded, and the functionality provided (as part of a bare-bones {\sc Mathematica} package) are described in \mbox{Appendix \ref{appendix:mathematica_and_explicit_results}}.

Before we begin, however, it seems appropriate to first describe what accounts for the advance---from eight to ten loops---in such a short interval of time. This turns out to be entirely a consequence of the computational power of working with graphical objects over algebraic expressions. The superiority of a graphical framework may not be manifest to all readers, and so it is worth describing why this is the case---and why a direct extension of the soft-collinear bootstrap beyond eight loops (implemented algebraically) does not seem within the reach of existing resources.

\paragraph{Why {Graphical} Rules?}~\\
\indent It is worth taking a moment to describe the incredible advantages of {\it graphical} methods over analytic or algebraic ones. The integrands of planar amplitudes or correlators can only meaningfully be defined if the labels of the internal loop momenta are fully symmetrized. Only then do they become well-defined, rational functions. But this means that, considered as algebraic functions, even {\it evaluation} of an integrand requires summing over all the permuted relabelings of the loop momenta (not to mention any cyclic or dihedral symmetrization of the legs that is also required). Thus, any analysis that makes use of evaluation will be rendered computationally intractable beyond some loop-order by the simple factorial growth in the time required by symmetrized evaluation. 

This is the case for the soft-collinear bootstrap as implemented in \mbox{ref.\ \cite{Bourjaily:2015bpz}}. At eight loops, the system of equations required to find the coefficients is a relatively straight-forward problem in linear algebra; and solving this system of equations is well within the limits of a typical laptop computer. However, {\it setting up} this linear algebra problem requires the evaluation of many terms---each at a sufficient number of points in loop-momentum space. And even with considerable ingenuity (and access to dozens of CPUs), these evaluations required more than two weeks to complete. Extending this method to nine loops would cost an additional factor of 9 from the combinatorics, and also a factor of 15 from the growth in the number of unknowns. This seems well beyond the reach of present-day computational resources. 

However, when the terms involved in the representation of an amplitude or correlator are considered more abstractly as {\it graphs}, the symmetrization required by evaluation becomes irrelevant: relabeling the vertices of a graph clearly leaves the {\it graph} unchanged. And it turns out that graphs can be compared with remarkable efficiency. Indeed, {\sc Mathematica} has built-in (and impressive) functionality for checking if two graphs are isomorphic (providing all isomorphisms that may exist). This means that relations among terms, when expressed as identities among graphs, can be implemented well beyond the limits faced for any method requiring evaluation.

We do not yet know of how the soft-collinear bootstrap can be translated as a graphical rule. And this prevents its extension beyond eight loops---at least at any time in the near future. However, the graphical rules we describe here prove sufficient to uniquely fix the amplitude and correlator through at least ten loops, and reproduce the eight loop answer in minutes rather than weeks. The extension of these ideas---perhaps amended by a broader set of analogous rules---to higher loops seems plausible using existing computational resources. Details of what challenges we expect in going to higher orders will be described in the conclusions.

\newpage
\vspace{-6pt}\section{Review of Amplitude/Correlator Duality}\label{sec:review_of_duality}\vspace{-6pt}
Let us briefly review the functional forms of the four-particle amplitude and correlator in planar maximally supersymmetric ($\mathcal{N}\!=\!4$) Yang-Mills theory (SYM), the duality that exists between these observables, and how each can be represented analytically as well graphically at each loop-order. This will serve as a casual review for readers already familiar with the subject; but for those less familiar, we will take care to be explicit about the (many, often implicit) conventions. 

The most fundamental objects of interest in any conformal field theory are gauge-invariant operators and their correlation functions. Perhaps the simplest operator in planar SYM is $\mathcal{O}(x)\!\equiv\!\mathrm{Tr}(\phi(x)^2)$, where $\phi$ is one of the six scalars of the theory and the trace is taken over gauge group indices (in the adjoint representation). This is a very special operator: it is related by (dual) superconformal symmetry to both the stress-energy tensor and the on-shell Lagrangian, is dual to supergravity states on AdS$_5$, protected from renormalization, and annihilated by half of the supercharges of the theory. Moreover, its two- and three-point correlation functions are protected from perturbative corrections. 

The four-point correlator involving $\mathcal{O}(x)$,
\eq{\mathcal{G}_4(x_1,x_2,x_3,x_4)\equiv\langle\mathcal{O}(x_1)\overline{\mathcal{O}}(x_2)\mathcal{O}(x_3)\overline{\mathcal{O}}(x_4)\rangle,\label{definition_of_correlator}}
is therefore the first non-trivial observable of interest in the theory. This correlator, computed perturbatively in loop-order and divided by the tree-level correlator is related to the four-particle amplitude (also divided by the tree) in a simple way \cite{Alday:2010zy,Eden:2010zz}:
\eq{\lim_{\substack{\text{4-point}\\\text{light-like}}}\left(\frac{\mathcal{G}_4^{}(x_1,x_2,x_3,x_4)}{\mathcal{G}_4^{(0)}\!(x_1,x_2,x_3,x_4)}\right)=\mathcal{A}^{}_4(x_1,x_2,x_3,x_4)^2,\label{correlator_amplitude_relation}}
where the amplitude is represented in dual-momentum coordinates, \mbox{$p_a\!\equiv\!x_{a+1}\mi x_a$}, and the light-like limit corresponds to taking the four (otherwise generic) points $x_a\!\in\!\mathbb{R}^{3,1}$ to be light-like separated: defining $x_{ab}\!\equiv\!x_b\mi x_a$, this corresponds to the limit where $\x{1}{2}\!=\!\x{2}{3}\!=\!\x{3}{4}\!=\!\x{1}{4}\!=\!0$. Importantly, while the correlator is generally finite upon integration, the limit taken in (\ref{correlator_amplitude_relation}) is divergent; however, the correspondence exists at the level of the loop {\it integrand}---both of which can be uniquely defined in any (planar) quantum field theory upon symmetrization in (dual) loop-momentum space. 

As a loop integrand, both sides of the identity (\ref{correlator_amplitude_relation}) are rational functions in $(4\pl\ell)$ points in $x$-space---to be integrated over the $\ell$ additional points, which we will (suggestively) denote as $x_{4+1},\ldots,x_{4+\ell}$. While the external points $x_1,\ldots,x_4$ would seem to stand on rather different footing relative to the loop momenta, it was noticed in \mbox{ref.\ \cite{Eden:2011we}} that this distinction disappears completely if one considers instead the function (appropriate for the component of the supercorrelator in (\ref{definition_of_correlator})),
\eq{\mathcal{F}^{(\ell)}(x_1,\ldots,x_4,x_5,\ldots,x_{4+\ell})\equiv\frac{1}{2}\left(\frac{G_4^{(\ell)}(x_1,x_2,x_3,x_4)}{G_4^{(0)}\!(x_1,x_2,x_3,x_4)}\right)/\xi^{(4)},\label{definition_of_f}}
where $\xi^{(4)}$ is defined to be $\x{1}{2}\x{2}{3}\x{3}{4}\x{1}{4}(\x{1}{3}\x{2}{4})^2$. As the attentive reader may infer, we will later have use to generalize this---yielding $\xi^{(4)}$ as a particular instance of,
\eq{\xi^{(n)}\equiv\prod_{a=1}^n\x{a}{a+1}\x{a}{a+2},\label{definition_of_general_xi}}
where cyclic ordering on $n$ points $x_a$ is understood (as well as the symmetry \mbox{$\x{a}{b}\!=\!\x{b}{a}$}). With this slight modification, it was discovered in \mbox{ref.\ \cite{Eden:2011we}} that the function $\mathcal{F}^{(\ell)}$ is fully {\it permutation invariant in its arguments}. This hidden symmetry is quite remarkable, and is responsible for a dramatic simplification in the representation of both the amplitude and the correlator. Because of the close connection between $\mathcal{F}^{(\ell)}$ and the correlation function defined via (\ref{definition_of_f}), we will frequently refer to $\mathcal{F}^{(\ell)}$ as `the $\ell$ loop correlation function' throughout the rest of this work; we hope this slight abuse of language will not lead to any confusion to the reader.

\vspace{-6pt}\subsection{$f$-Graphs: Their Analytic and Graphical Representations}\label{subsec:fgraphs_and_conventions}\vspace{-0pt}

Considering the full symmetry of $\mathcal{F}^{(\ell)}$ among its $(4\pl\ell)$ arguments, we are led to think of the possible contributions more as graphs than algebraic expressions. Conformality requires that any such contribution must be weight $\!\mi4$ in each of its arguments; locality ensures that only factors of the form $\x{a}{b}$ can appear in the denominator; analyticity requires that there are at most single poles in these factors (for the amplitude---for the correlator, analysis of OPE limits); and finally, planarity informs us that these factors must form a plane graph. The denominator of any possible contribution, therefore, can be encoded as a plane graph with edges $a\!\leftrightarrow\!b$ for each factor $\x{a}{b}$. (Because $\x{a}{b}\!\!=\!\x{b}{a}$, these graphs are naturally {\it undirected}.) 

We are therefore interested in plane graphs involving $(4\pl\ell)$ points, with valency at least 4 in each vertex. Excess conformal weight from vertices with higher valency can be absorbed by factors in the numerator. Conveniently, it is not hard to enumerate all such plane graphs---one can use the program {\tt CaGe} \cite{CaGe}, for example. Decorating each of these plane graphs with all inequivalent numerators capable of rending the net conformal weight of every vertex to be $\!\mi4$ results in the space of so-called `$f$-graphs'. The enumeration of the possible $f$-graph contributions that result from this exercise (through eleven loop-order) is given in \mbox{Table \ref{f_graph_statistics_table}}. Also in the Table, we have listed the number of (graph-inequivalent) planar, (dual-)conformally invariant (`DCI') integrands that exist. (The way in which these contributions to the four-particle amplitude are obtainable from each $f$-graph is described below.)

\newpage
\begin{table}[t]$\hspace{1.5pt}\begin{array}{|@{$\,$}c@{$\,$}|@{$\,$}r@{$\,$}|@{$\,$}r@{$\,$}|@{$\,$}r@{$\,$}|@{$\,$}r@{$\,$}|}\multicolumn{1}{@{$\,$}c@{$\,$}}{\begin{array}{@{}l@{}}\\[-4pt]\text{$\ell\,$}\end{array}}&\multicolumn{1}{@{$\,$}c@{$\,$}}{\!\begin{array}{@{}c@{}}\text{number of}\\[-4pt]\text{plane graphs}\end{array}}\,&\multicolumn{1}{@{$\,$}c@{$\,$}}{\begin{array}{@{}c@{}}\text{number of graphs}\\[-4pt]\text{admitting decoration}\end{array}}\,&\multicolumn{1}{@{$\,$}c@{$\,$}}{\begin{array}{@{}c@{}}\text{number of decorated}\\[-4pt]\text{plane graphs ($f$-graphs)}\end{array}}\,&\multicolumn{1}{@{$\,$}c@{$\,$}}{\begin{array}{@{}c@{}}\text{number of planar}\\[-4pt]\text{DCI integrands}\end{array}}\,\\[-0pt]\hline1&0&0&0&1\\\hline2&1&1&1&1\\[-0pt]\hline3&1&1&1&2\\\hline4&4&3&3&8\\\hline5&14&7&7&34\\\hline6&69&31&36&284\\\hline7&446&164&220&3,\!239\\\hline8&3,\!763&1,\!432&2,\!709&52,\!033\\\hline9&34,\!662&13,\!972&43,\!017&1,\!025,\!970\\\hline10&342,\!832&153,\!252&900,\!145&24,\!081,\!425\\\hline11&3,\!483,\!075&1,\!727,\!655&22,\!097,\!035&651,\!278,\!237\\\hline\end{array}$\vspace{-6pt}\caption{Statistics of plane graphs, $f$-graphs, and DCI integrands through $\ell\!=\!11$ loops.\label{f_graph_statistics_table}}\vspace{-10pt}\end{table}
\noindent(To be clear, \mbox{Table \ref{f_graph_statistics_table}} counts the number of {\it plane} graphs---that is, graphs with a fixed plane embedding. The distinction here is only relevant for graphs that are not 3-vertex connected---which are the only planar graphs that admit multiple plane embeddings. We have found that no such graphs contribute to the amplitude or correlator through ten loops---and we strongly expect their absence can be proven. However, because the graphical rules we describe are sensitive to the plane embedding, we have been careful about this distinction in our analysis---without presumptions on their irrelevance.)

When representing an $f$-graph graphically, we use solid lines to represent every factor in the denominator, and dashed lines (with multiplicity) to indicate the factors that appear in the numerator. For example, the possible $f$-graphs through four loops are as follows:
\eq{\begin{array}{rc@{$\;\;\;\;\;$}rc@{$\;\;\;\;\;$}rc}\\[-40pt]f^{(1)}_1\equiv&\fwbox{75pt}{\fig{-54.75pt}{1}{one_loop_f_graph_1}}&f^{(2)}_1\equiv&\fwbox{75pt}{\fig{-54.75pt}{1}{two_loop_f_graph_1}}&f^{(3)}_1\equiv&\fwbox{75pt}{\fig{-54.75pt}{1}{three_loop_f_graph_1}}\\[-26pt]f^{(4)}_1\equiv&\fwbox{75pt}{\fig{-54.75pt}{1}{four_loop_f_graph_1}}&f^{(4)}_2\equiv&\fwbox{75pt}{\fig{-54.75pt}{1}{four_loop_f_graph_2}}&f^{(4)}_3\equiv&\fwbox{75pt}{\fig{-54.75pt}{1}{four_loop_f_graph_3}}\\[-35pt]\end{array}\vspace{20pt}\label{one_through_four_loop_f_graphs}}
In terms of these, the loop-level correlators $\mathcal{F}^{(\ell)}$ would be expanded according to:
\eq{\mathcal{F}^{(1)}=f^{(1)}_1,\quad \mathcal{F}^{(2)}=f^{(2)}_1,\quad \mathcal{F}^{(3)}=f^{(3)}_1,\quad \mathcal{F}^{(4)}=f^{(4)}_1+f^{(4)}_2-f^{(4)}_3.\label{correlators_through_four_loops}}
(Notice that $f^{(1)}_1$ in (\ref{one_through_four_loop_f_graphs}) is not planar; this is the only exception to the rule; however, it does lead to planar contributions to $\mathcal{G}^{(1)}_4$ and $\mathcal{A}_4^{(1)}$ after multiplication by $\xi^{(4)}$.)

In general, we can always express the $\ell$ loop correlator $\mathcal{F}^{(\ell)}$ in terms of the \mbox{$f$-graphs} $f^{(\ell)}_i$ according to,
\eq{\mathcal{F}^{(\ell)}\equiv\sum_{i}c^{\ell}_i\,f^{(\ell)}_i\,,\label{general_correlator_expansion}\vspace{-5pt}}
where the coefficients $c_i^{\ell}$ (indexed by the complete set of $f$-graphs at $\ell$ loops) are rational numbers---to be determined using principles such as those described below. At eleven loops, for example, there will be $22,\!097,\!035$ coefficients $c_i^{11}$ that must be determined (see \mbox{Table \ref{f_graph_statistics_table}}).

Analytically, these graphs correspond to the product of factors $\x{a}{b}$ in the denominator for each solid line in the figure, and factors $\x{a}{b}$ in the numerator for each dashed line in the figure. This requires, of course, a choice of the labels for the vertices of the graph. For example, 
\vspace{-5pt}\eq{\hspace{-85.5pt}\fig{-54.75pt}{1}{four_loop_f_graph_2_with_labels}\hspace{-8.5pt}\equiv\!\!\frac{\x{1}{6}\x{3}{7}}{\x{1}{2}\x{1}{3}\x{1}{4}\x{1}{5}\x{1}{7}\x{2}{3}\x{2}{7}\x{2}{8}\x{3}{4}\x{3}{6}\x{3}{8}\x{4}{5}\x{4}{6}\x{5}{6}\x{5}{7}\x{6}{7}\x{6}{8}\x{7}{8}}\!.\hspace{-50pt}\vspace{-2pt}}
But any other choice of labels would have corresponded to the same graph, and so we must sum over all the (distinct) relabelings of the function. Of the $8!$ such relabelings, many leave the corresponding function unchanged---resulting (for this example) in 8 copies of each function. Thus, had we chosen to na\"{i}vely sum over all permutations of labels, we would over-count each graph, requiring division by a compensatory `symmetry factor' of 8 in the analytic expression contributing to the amplitude or correlation function. (This symmetry factor is easily computed as the size of the automorphism group of the graph.) However, we prefer not to include such symmetry factors in our expressions, which is why we write the coefficient of this graph in (\ref{correlators_through_four_loops}) as `$\!\pl\!$1' rather than `$\!\pl\!$1/8'. 

And so, to be perhaps overly explicit, we should be clear that this will always be our convention. Contributions to the amplitude or correlator, when converted from graphs to analytic expressions, should be symmetrized and summed; but we will always (implicitly) consider the summation to include only the {\it distinct} terms that result from symmetrization. Hence, no (compensatory) symmetry factors will appear in our coefficients. Had we instead used the convention where $f$-graphs' analytic expressions should be generated by summing over {\it all} terms generated by $\mathfrak{S}_{4+\ell}$, the coefficients of the four loop correlator, for example, would have been $\{\!\pl1/8,\!\pl1/24,\!\mi1/16\}$ instead of $\{\!\pl1,\!\pl1,\!\mi1\}$ as written in (\ref{correlators_through_four_loops}).

\newpage
\vspace{-6pt}\subsection{Four-Particle Amplitude Extraction via Light-Like Limits Along Faces}\label{subsec:amplitude_extraction}\vspace{-0pt}
When the correlation function $\mathcal{F}^{(\ell)}$ is expanded in terms of plane graphs, it is very simple to extract the $\ell$ loop scattering amplitude through the relation (\ref{correlator_amplitude_relation}). To be clear, upon expanding the square of the amplitude in powers of the coupling (and dividing by the tree amplitude), we find that:
\eq{\lim_{\substack{\text{4-point}\\\text{light-like}}}\!\!\Big(\xi^{(4)}\mathcal{F}^{(\ell)}\Big)=\frac{1}{2}\big((\mathcal{A}_4)^2\big)^{(\ell)}=\left(\mathcal{A}_{4}^{(\ell)}+\mathcal{A}_4^{(\ell-1)}\mathcal{A}_4^{(1)}+\mathcal{A}_{4}^{(\ell-2)}\mathcal{A}_4^{(2)}+\ldots\right).\label{f_to_4pt_amp_map_with_series_expansion}}
Before we describe how each term in this expansion can be extracted from the contributions to $\mathcal{F}^{(\ell)}$, let us first discuss which terms survive the light-like limit. Recall from equation (\ref{definition_of_general_xi}) that $\xi^{(4)}$ is proportional to $\x{1}{2}\x{2}{3}\x{3}{4}\x{1}{4}$---each factor of which vanishes in the light-like limit. Because $\xi^{(4)}$ identifies four specific points $x_a$, while $\mathcal{F}^{(\ell)}$ is a permutation-invariant sum of terms, it is clear that these four points can be arbitrarily chosen among the $(4\pl\ell)$ vertices of any $f$-graph; and thus the light-like limit will be non-vanishing iff the graph contains an edge connecting each of the pairs of vertices: $1\!\leftrightarrow\!2$, $2\!\leftrightarrow\!3$, $3\!\leftrightarrow\!4$, $1\!\leftrightarrow\!4$. Thus, terms that survive the light-like limit are those corresponding to a 4-cycle of the (denominator of the) graph. 

Any $n$-cycle of a plane graph divides it into an `interior' and `exterior' according to the plane embedding (viewed on a sphere). And this partition exactly corresponds to that required by the products of amplitudes appearing in (\ref{f_to_4pt_amp_map_with_series_expansion}). We can illustrate this partitioning with the following example of a ten loop $f$-graph (ignoring any factors that appear in the numerator):
\eq{\fig{-54.75pt}{1}{ten_loop_cycles_1}\qquad\fig{-54.75pt}{1}{ten_loop_cycles_2}\qquad\fig{-54.75pt}{1}{ten_loop_cycles_3}\label{example_cycles}}
These three 4-cycles would lead to contributions to $\mathcal{A}_4^{(10)}$, $\mathcal{A}_4^{(9)}\mathcal{A}_4^{(1)}$, and $\mathcal{A}_4^{(5)}\mathcal{A}_4^{(5)}$, respectively. Notice that we have colored the vertices in each of the examples above according to how they are partitioned by the cycle indicated. The fact that the $\ell$ loop correlator $\mathcal{F}^{(\ell)}$ contains within it complete information about lower loops will prove extremely useful to us in the next section. For example, the square (or `rung') rule follows immediately from the requirement that the $\mathcal{A}_4^{(\ell-1)}\mathcal{A}_4^{(1)}$ term in the expansion (\ref{f_to_4pt_amp_map_with_series_expansion}) is correctly reproduced from the representation of $\mathcal{F}^{(\ell)}$ in terms of $f$-graphs. 

The leading term in (\ref{f_to_4pt_amp_map_with_series_expansion}) is arguably the most interesting. As illustrated above, these contributions arise from any 4-cycle of an $f$-graph encompassing no internal vertices. Such cycles correspond to {\it faces} of the graph---either a single square face, or two triangular faces which share an edge. This leads to a direct projection from $f$-graphs into planar `amplitude' graphs that are manifestly dual conformally invariant (`DCI'). Interestingly, the graphs that result from taking the light-like limit along each face of the graph can appear surprisingly different. 

Consider for example the following five loop $f$-graph, which has four non-isomorphic faces, resulting in four rather different DCI integrands:
\vspace{-20pt}\eq{\fwbox{0pt}{\hspace{-225pt}\fwboxR{0pt}{\fig{-54.75pt}{1}{five_loop_f_graph_with_faces}}\fwboxL{0pt}{\hspace{-15pt}\bigger{\Rightarrow}\!\left\{\!\rule{0pt}{40pt}\right.\hspace{-12.5pt}\fig{-54.75pt}{1}{five_loop_planar_projection_2_v2}\hspace{-10pt}\fig{-54.75pt}{1}{five_loop_planar_projection_3_v2}\hspace{-12.5pt}\fig{-54.75pt}{1}{five_loop_planar_projection_4_v2}\hspace{-12.5pt}\fig{-54.75pt}{1}{five_loop_planar_projection_1_v2}\hspace{-5pt}\left.\rule{0pt}{40pt}\right\}}}\label{five_loop_planar_projections_example}\vspace{-20pt}}
Here, we have drawn these graphs in both momentum space and dual-momentum space---with black lines indicating ordinary Feynman propagators (which may be more familiar to many readers), and grey lines indicating the dual graphs (more directly related to the $f$-graph). We have not drawn any dashed lines to indicate factors of $s\!\equiv\!\x{1}{3}$ or $t\!\equiv\!\x{2}{4}$ in numerators that would be uniquely fixed by dual conformal invariance. Notice that one of the faces---the orange one---corresponds to the `outer' four-cycle of the graph as drawn; also, the external points of each planar integrand have been colored according to the face involved. As one further illustration of this correspondence, consider the following seven loop $f$-graph, which similarly leads to four inequivalent DCI integrands (drawn in momentum space):
\vspace{2pt}\eq{\fwbox{0pt}{\hspace{-235pt}\fwboxR{0pt}{\fig{-34.75pt}{1}{seven_loop_f_graph_with_faces}}\fwboxL{0pt}{\hspace{-0pt}\bigger{\Rightarrow}\left\{\rule{0pt}{40pt}\right.\hspace{-10pt}\fig{-34.75pt}{1}{seven_loop_planar_projection_3}\hspace{-5pt}\fig{-34.75pt}{1}{seven_loop_planar_projection_1}\hspace{-2.5pt}\fig{-34.75pt}{1}{seven_loop_planar_projection_2}\hspace{-10.pt}\fig{-34.75pt}{1}{seven_loop_planar_projection_4}\hspace{-10pt}\left.\rule{0pt}{40pt}\right\}}}\label{seven_loop_planar_projections_example}\vspace{-0pt}}

Before moving on, it is worth a brief aside to mention that these projected contributions are to be symmetrized according to the same convention discussed above for $f$-graphs---namely, when considered as analytic expressions, only distinct terms are to be summed. This follows directly from our convention for $f$-graphs and the light-like limit, without any relative symmetry factors required between the coefficients of $f$-graphs and the coefficients of each distinct DCI integrand obtained by taking the light-like limit.

\newpage 
\vspace{-2pt}\subsection{Higher-Point Amplitude Extraction from the Correlator}\label{subsec:higher_point_amplitude_extraction}\vspace{-0pt}
Remarkably enough, although the correlation function $\mathcal{F}^{(\ell)}$ was defined to be closely related to the (actual) four-point correlation function $\mathcal{G}^{(\ell)}_4$ in planar SYM, which accounts for its relation to the four-particle scattering amplitude $\mathcal{A}_4^{(\ell)}$, it turns out that interesting combinations of {\it all} higher-point amplitudes can also be obtained from it \cite{Eden:2011yp,Eden:2011ku,Ambrosio:2013pba}. Perhaps this should not be too surprising, as $\mathcal{F}^{(\ell)}$ is a symmetrical function on $(4\pl\ell)$ points $x_a$; but it is an incredibly powerful observation: it implies that $\mathcal{F}^{(\infty)}$ contains information about {\it all} scattering amplitudes in planar SYM! 

The way in which higher-point, lower-loop amplitudes are encoded in the function $\mathcal{F}^{(\ell)}$ is a consequence of the fully supersymmetric amplitude/correlator duality \cite{Eden:2010zz,Alday:2010zy,Eden:2010ce,Eden:2011yp,Adamo:2011dq,Eden:2011ku} which was unpacked in \mbox{ref.\ \cite{Ambrosio:2013pba}}:
\vspace{-5pt}\eq{\lim_{\substack{\text{n-point}\\\text{light-like}}}\!\!\Big(\xi^{(n)}\mathcal{F}^{(\ell)}\Big)=\frac{1}{2}\sum_{k=0}^{n-4}\mathcal{A}_n^{k}\,\mathcal{A}_n^{n-4-k}/(\mathcal{A}_n^{n-4,(0)}).\label{f_to_npt_amp_map}\vspace{-5pt}}
Here, we have used the notation $\mathcal{A}_n^{k,(\ell)}$ to represent the $\ell$-loop $n$-particle N$^k$MHV amplitude divided by the $n$ particle MHV tree-amplitude. We should point out that division in (\ref{f_to_npt_amp_map}) by the N$^{n-4}$MHV ($\overline{\text{MHV}}$) tree-amplitude is required to absorb the Grassmann $\eta$ weights---resulting in a purely bosonic sum of terms from which all amplitudes can be extracted. 

It is worth mentioning that while for four particles, the $\ell$ loop amplitude can be directly extracted from $\mathcal{F}^{(\ell)}$, and for five-points one can also extract the full amplitude, for higher-point amplitudes it is not yet clear if or how one can obtain full information about amplitudes from the combination on the left-hand side of \eqref{f_to_npt_amp_map}. Elaboration of how this works in detail is beyond the scope of our present work, but because the case of $n\!=\!5$ will play an important role in motivating (and proving) the `pentagon rule' described in the next section, it is worth illustrating at least this case in some detail.

\paragraph{The Pentagonal Light-Like Limit:}~\\
\indent In addition to being the simplest example of how higher-point amplitudes can be extracted from $\mathcal{F}^{(\ell)}$ via (\ref{f_to_npt_amp_map}), the case of five particles will prove quite useful to us in our discussion of the pentagon rule described in the next section. Therefore, let us briefly summarize how this works in practice. 

In the case of five particles, the right-hand side of (\ref{f_to_npt_amp_map}) is simply the product of the MHV and $\overline{\text{MHV}}$ amplitudes---divided by the $\overline{\text{MHV}}$ tree-amplitude (with division by $\mathcal{A}_5^{0,(0)}$ left implicit, as always). Conventionally defining \mbox{$\mathcal{M}_5^{}\!\equiv\!\mathcal{A}_5^{0}/\mathcal{A}_{5}^{0,(0)}$} and \mbox{$\overline{\mathcal{M}}_5^{}\!\equiv\!\mathcal{A}_5^{1}/\mathcal{A}_{5}^{1,(0)}$}, and expanding each in powers of the coupling, the relation (\ref{f_to_npt_amp_map}) becomes more symmetrically expressed as:
\vspace{-6.5pt}\eq{\lim_{\substack{\text{5-point}\\\text{light-like}}}\!\!\Big(\xi^{(5)}\mathcal{F}^{(\ell+1)}\Big)=\sum_{m=0}^{\ell}\mathcal{M}_5^{(m)}\overline{\mathcal{M}}_5^{(\ell-m)}.\label{f_to_5pt_amp_map}\vspace{-0pt}}
Moreover, because the parity-even contributions to the loop integrands $\mathcal{M}_5^{(\ell)}$ and $\overline{\mathcal{M}}_5^{(\ell)}$ are the same, it is further convenient to define:
\eq{\mathcal{M}_{\text{even}}^{(\ell)}\equiv\frac{1}{2}\left(\mathcal{M}_5^{(\ell)}+\overline{\mathcal{M}}_5^{(\ell)}\right)\quad\text{and}\quad\mathcal{M}_{\text{odd}}^{(\ell)}\equiv\frac{1}{2}\left(\mathcal{M}_5^{(\ell)}-\overline{\mathcal{M}}_5^{(\ell)}\right).\label{5pt_even_and_odd_definitions}}

Because any integrand constructed out of factors $\x{a}{b}$ will be manifestly parity-even, it is not entirely obvious how the parity-odd contributions to loop integrands should be represented. Arguably, the most natural way to represent parity-odd contributions is in terms of a six-dimensional  formulation of dual momentum space (essentially the Klein quadric) which was first introduced in this context in \mbox{ref.\ \cite{Mason:2009qx}} following the introduction of momentum twistors in \mbox{ref.\ \cite{Hodges:2009hk}}. Each point $x_a$ is represented by a (six-component) bi-twistor $X_a$. The (dual) conformal group $SO(2,4)$ acts linearly on this six-component object and so  it is natural to define a fully antisymmetric epsilon-tensor, $\epsilon_{abcdef}\!\equiv\!\det\{X_a,\ldots,X_f\}$, in which the parity-odd part of the $\ell$ loop integrand can be represented \cite{Ambrosio:2013pba}:
\eq{\mathcal{M}_{\text{odd}}\equiv i\epsilon_{12345\ell}\,\widehat{\mathcal{M}}_{\text{odd}},\label{definition_of_epsilon_prefactors_for_odd_integrands}}
where $\widehat{\mathcal{M}}_{\text{odd}}$ is a parity-even function, directly expressible in terms of factors $\x{a}{b}$. 

Putting everything together, the expansion (\ref{f_to_5pt_amp_map}) becomes:
\eq{\hspace{-75pt}\lim_{\substack{\text{5-point}\\\text{light-like}}}\!\!\Big(\xi^{(5)}\mathcal{F}^{(\ell+1)}\Big)=\sum_{m=0}^{\ell}\left(\mathcal{M}_{\text{even}}^{(m)}\mathcal{M}_{\text{even}}^{(\ell-m)}+\epsilon_{123456}\epsilon_{12345(m+6)}\widehat{\mathcal{M}}_{\text{odd}}^{(m)}\widehat{\mathcal{M}}_{\text{odd}}^{(\ell-m)}\right).\label{f_to_5pt_amp_map2}\hspace{-40pt}\vspace{-5pt}}

The pentagon rule we derive in the next section amounts to the equality between two different ways to extract the $\ell$-loop 5-particle integrand from $\mathcal{F}^{(\ell+2)}$, by identifying, as part of the contribution, the one loop integrand. As such, it is worthwhile to at least quote these contributions. They are as follows:
\eq{\mathcal{M}_{\text{even}}^{(1)}\equiv\fig{-34.75pt}{1}{five_point_one_loop_even}\qquad\text{and}\qquad\mathcal{M}_{\text{odd}}^{(1)}\equiv\fig{-34.75pt}{1}{five_point_one_loop_odd}\label{five_point_one_loop_terms}}
where the circled vertex in the right-hand figure indicates the last argument of the epsilon-tensor. When converted into analytic expressions, these correspond to:
\eq{\fwbox{0pt}{\fig{-34.75pt}{1}{five_point_one_loop_even}\equiv\frac{\x{1}{3}\x{2}{4}}{\x{1}{6}\x{2}{6}\x{3}{6}\x{4}{6}}+\text{cyclic},\quad\fig{-34.75pt}{1}{five_point_one_loop_odd}\equiv\frac{i\epsilon_{123456}}{\x{1}{6}\x{2}{6}\x{3}{6}\x{4}{6}\x{5}{6}}},\label{five_point_one_loop_terms_analytic}\nonumber}
where the cyclic sum of terms involves only the 5 external vertices. 

\newpage
\vspace{-6pt}\section{(Graphical) Rules For Bootstrapping Amplitudes}\label{sec:graphical_bootstraps}\vspace{-6pt}
As described above, the correlator $\mathcal{F}^{(\ell)}$ can be expanded into a basis of $\ell$ loop \mbox{$f$-graphs} according to (\ref{general_correlator_expansion}). The challenge, then, is to determine the coefficients $c_i^{\ell}$.  We take for granted that the one loop four-particle amplitude integrand may be represented in dual momentum coordinates as:
\vspace{-5pt}\eq{\mathcal{A}_4^{(1)}\equiv\fig{-34.75pt}{1}{four_point_one_loop_amplitude}\equiv\frac{\x{1}{3}\x{2}{4}}{\x{1}{5}\x{2}{5}\x{3}{5}\x{4}{5}},\label{four_point_one_loop_integrand_in_x_space}\vspace{-5pt}}
with which we expect most readers will be familiar. This formula in fact {\it defines} the one loop $f$-graph $f^{(1)}_1$---as there does not exist any planar graph involving five points each having valency at least 4. As such, it is defined so as to ensure that equation (\ref{f_to_4pt_amp_map_with_series_expansion}) holds:
\vspace{-26pt}\eq{f^{(1)}_1\equiv\mathcal{A}_4^{(1)}/\xi^{(4)}\equiv\!\!\fwbox{90pt}{\fig{-54.75pt}{1}{one_loop_f_graph_1}}\equiv\frac{1}{\x{1}{2}\x{1}{3}\x{1}{4}\x{1}{5}\x{2}{3}\x{2}{4}\x{2}{5}\x{3}{4}\x{3}{5}\x{4}{5}}.\label{definition_of_f1}\vspace{-24pt}}
This effectively defines $\mathcal{F}^{(1)}\!\equiv\!f_1^{(1)}$, with a coefficient $c_1^{1}\!\equiv\!\pl1$. Given this seed, we will see that consistency among the products of lower-loop amplitudes in (\ref{f_to_4pt_amp_map_with_series_expansion})---as well as those involving more particles (\ref{f_to_npt_amp_map})---will be strong enough to uniquely determine the coefficients of all $f$-graphs in the expansion for $\mathcal{F}^{(\ell)}$ in terms of lower loop-orders. 

In this section we describe how this can be done in practice through three simple, graphical rules that allow us to `bootstrap' all necessary coefficients through at least ten loops. To be clear, the rules we describe are merely three among many that follow from the self-consistency of equations (\ref{f_to_4pt_amp_map_with_series_expansion}) and (\ref{f_to_npt_amp_map}); they are not obviously the strongest or most effective of such rules; but they are {\it necessary} conditions of any representation of the correlator, and we have found them to be {\it sufficient} to uniquely fix the expansion of $\mathcal{F}^{(\ell)}$ into $f$-graphs, (\ref{general_correlator_expansion}), through at least ten loops. 

Let us briefly describe each of these three rules in qualitative terms, before giving more detail (and derivations) in the following subsections. We refer to these as the `triangle rule', the `square rule', and the `pentagon rule'. Despite the natural ordering suggested by their names, it is perhaps best to start with the square rule---which is simply a generalization of what has long been called the `rung' rule \cite{Bern:1997nh}. 

\paragraph{The Square (or `Rung') Rule:}~\\
\indent The square rule is arguably the most powerful of the three rules, and provides the simplest constraints---directly fixing the coefficients of certain $f$-graphs at $\ell$ loops to be equal to the coefficients of $f$-graphs at $(\ell\mi1)$ loops. 

Roughly speaking, the square rule follows from the requirement that whenever an $f$-graph {\it has} a contribution to $\mathcal{A}_4^{(\ell-1)}\mathcal{A}_4^{(1)}$, this contribution must be correct. It simply reflects the translation of what has long been known as the `rung' rule \cite{Bern:1997nh} into the language of the correlator and $f$-graphs \cite{Eden:2012tu}; however, this translation proves much more powerful than the original, as described in more detail below. As will be seen in the \mbox{section \ref{sec:results}}, for example, the square rule fixes $\sim\!95\%$ of all $f$-graph coefficients at eleven loops---the only coefficients not fixed by the square rule are those of $f$-graphs which do not contribute any terms to $\mathcal{A}_4^{(\ell-1)}\mathcal{A}_4^{(1)}$. 

~\\[-36pt]\paragraph{The Triangle Rule:}~\\
\indent Simply put, the triangle rule states that shrinking triangular faces at $\ell$ loops is equivalent to shrinking edges at $(\ell\mi1)$ loops. By this we mean simply identifying the three vertices of any triangular face of an $f$-graph at $\ell$ loops and identifying two vertices connected by an edge of an $f$-graph at $(\ell\mi1)$ loops, respectively. The result of either operation is never an $f$-graph (as it will not have correct conformal weights, and will often involve vertices connected by more than one edge), but this does not prevent us from implementing the rule graphically. Typically, there are many fewer inequivalent graphs involving shrunken faces/edges, and so the triangle rule typically results in relations involving many $f$-graph coefficients. This makes the equations relatively harder to solve.

As described in more detail below, the triangle rule follows from the Euclidean short distance \cite{Eden:2012tu,Eden:2012fe} limit of correlation functions. We will prove this in the following subsection, and describe more fully its strength in fixing coefficients in \mbox{section \ref{sec:results}}. But it is worth mentioning here that when combined with the square rule, the triangle rule is sufficient to fix $\mathcal{F}^{(\ell)}$ completely through seven loops; and the implications of the triangle rule applied at {\it ten} loops is sufficient to fix $\mathcal{F}^{(\ell)}$ through {\it nine} loops (although the triangle and square rules alone, when imposed at nine loops, would not suffice).

~\\[-36pt]\paragraph{The Pentagon Rule:}~\\
\indent The pentagon rule is the five-particle analog of the square rule---following from the requirement that the $\mathcal{M}^{(\ell-1)}\mathcal{M}^{(1)}$ terms in the expansion (\ref{f_to_5pt_amp_map}) are correct. Unlike the square rule, however, it does not make use of knowing lower-loop five-particle amplitudes; rather, it simply requires that the odd contributions to the amplitude are consistent. We will describe in detail how the pentagon rule is derived below, and give examples of how it fixes coefficients. 

One important aspect of the pentagon rule, however, is that it relates coefficients at a {\it fixed loop-order}. Indeed, as an algebraic constraint, the pentagon rule always becomes the requirement that the sum of some subset of coefficients $c_i^\ell$ is zero (without any relative factors ever required).\\

Before we describe and derive each of these three rules in detail, it is worth mentioning that they lead to mutually overlapping and individually {\it over-constrained} relations on the coefficients of $f$-graphs. As such, the fact that any solution exists to these equations---whether from each individual rule or in combination---strongly implies the correctness of our rules (and the correctness of their implementation in our code). And of course, the results we find are consistent with all known results through eight loops, which have been found using a diversity of other methods. 

\vspace{-0pt}\subsection{The Square (or `Rung') Rule: Removing One Loop Squares}\label{subsec:square_rule}\vspace{-0pt}
Recall from \mbox{section \ref{subsec:amplitude_extraction}} that, upon taking the 4-point light-like limit, an $f$-graph contributes a term to $\mathcal{A}_4^{(\ell-1)}\mathcal{A}_4^{(1)}$ in the expansion (\ref{f_to_4pt_amp_map_with_series_expansion}) if (and only if) there exists a 4-cycle that encloses a single vertex. See, for example, the second illustration given in (\ref{example_cycles}). Because of planarity, the enclosed vertex must have valency exactly 4, and so any such cycle must form a face with the topology:
\vspace{-7pt}\eq{\fig{-34.75pt}{1}{square_rule_face_topology}\label{square_rule_face_toplogy}\vspace{-7pt}}
Whenever an $f$-graph has such a face, it will contribute a term of the form $\mathcal{A}_4^{(\ell-1)}\mathcal{A}_4^{(1)}$ in the light-like limit. If we define the operator $\mathcal{S}(\mathcal{F})$ to be the projection onto such contributions, then the rung rule states that $\mathcal{S}(\mathcal{F}^{(\ell)})/\mathcal{A}_4^{(1)}\!\!=\!\mathcal{A}_4^{(\ell-1)}$. Graphically, division of (\ref{square_rule_face_toplogy}) by the graph for $\mathcal{A}_4^{(1)}$ in (\ref{four_point_one_loop_integrand_in_x_space}) would correspond to the graphical replacement:
\vspace{-0pt}\eq{\hspace{-120pt}\fig{-34.75pt}{1}{square_rule_face_topology}\bigger{\Rightarrow}\left(\fig{-34.75pt}{1}{square_rule_face_topology}\bigger{\times}\fig{-34.75pt}{1}{inverse_one_loop_graph}\right)\bigger{=}\fig{-34.75pt}{1}{image_of_square_after_division}\hspace{-100pt}\label{graphical_square_rule}\vspace{-5pt}}
(Here, we have illustrated division by the graph for $\mathcal{A}_4^{(1)}$---shown in (\ref{four_point_one_loop_integrand_in_x_space})---as multiplication by its inverse.) 

Importantly, the image on the right hand side of (\ref{graphical_square_rule}) resulting from this operation is not always planar! For it to be planar, there must exist a numerator factor connecting any two of the vertices of the square face---to cancel against one or both of the `new' factors in the denominator appearing in (\ref{graphical_square_rule}). When the image is non-planar, however, the graph {\it cannot} contribute to $\mathcal{A}_4^{(\ell-1)}$,\footnote{There is an exception to this conclusion when $\ell\!=\!2$---because $f_1^{(1)}$ is not itself planar.} and thus the coefficient of such an $f$-graph must vanish. For example, consider the following six loop $f$-graph which has a face with the topology (\ref{square_rule_face_toplogy}), and so its contribution to $\mathcal{F}^{(6)}$ would be constrained by the square rule:
\vspace{-14pt}\eq{\fig{-54.75pt}{1}{six_loop_vanishing_by_square_rule_example}\label{six_loop_vanishing_by_square_rule_example}\vspace{-15pt}}
In this case, because there are no numerator factors (indicated by dashed lines) connecting the vertices of the highlighted 4-cycle, its image under (\ref{graphical_square_rule}) would be non-planar, and hence this term cannot appear in $\mathcal{A}_4^{(5)}$. Therefore, the coefficient of this $f$-graph must be zero. (In fact, this reasoning accounts for 8 of the 10 vanishing coefficients that first appear at six loops.) As discussed in \mbox{ref.\ \cite{Bourjaily:2015bpz}}, this immediately implies that there are no possible contributions with `$k\!=\!4$' divergences. 

More typically, however, there is at least one numerator factor in the $\ell$ loop \mbox{$f$-graph} that connects vertices of the one loop square face (\ref{square_rule_face_toplogy}) in order to cancel one or both of the new denominator factors in (\ref{graphical_square_rule}). When this is the case, the image is an $(\ell\mi1)$ loop \mbox{$f$-graph}, and the square rule states that their coefficients are identical. For example, the coefficient of the five loop \mbox{$f$-graph} shown in (\ref{five_loop_planar_projections_example}) is fixed by the square rule to have the same coefficient as $f_3^{(4)}$ shown in (\ref{one_through_four_loop_f_graphs}):
\vspace{-7pt}\eq{\fig{-34.75pt}{1}{five_loop_square_rule_example_1}\,\,\bigger{\Rightarrow}\fig{-34.75pt}{1}{five_loop_square_rule_example_2}\label{five_loop_square_rule_example}\vspace{-7pt}}

In summary, the square rule fixes the coefficient of any $f$-graph that has a face with the topology (\ref{square_rule_face_toplogy}) directly in terms of lower-loop coefficients. And this turns out to constrain the vast majority of possible contributions, as summarized in \mbox{Table \ref{square_rule_strength_table}}. And it is worth emphasizing that the square rule described here is in fact substantially stronger than what has been traditionally called the `rung' rule \cite{Bern:1997nh} for two reasons: first, the square rule unifies collections of planar DCI contributions to amplitudes according to the hidden symmetry of the correlator---allowing us to fix coefficients of even the `non-rung-rule'  integrands such as those appearing in (\ref{five_loop_planar_projections_example}); secondly, the square rule allows us to infer the vanishing of certain coefficients due to the non-existence of lower-loop graphs (due to non-planarity). 

\begin{table}[b]\vspace{-20pt}$$\fwbox{0pt}{\begin{array}{|r|r|r|r|r|r|r|r|r|r|}\cline{2-10}\multicolumn{1}{r}{\ell\!=}&\multicolumn{1}{|c|}{3}&\multicolumn{1}{c|}{4}&\multicolumn{1}{c|}{5}&\multicolumn{1}{c|}{6}&\multicolumn{1}{c|}{7}&\multicolumn{1}{c|}{8}&\multicolumn{1}{c|}{9}&\multicolumn{1}{c|}{10}&\multicolumn{1}{c|}{11}\\\hline\text{number of $f$-graph coefficients:}&1&3&7&36&220&2,\!709&43,\!017&900,\!145&22,\!097,\!035\\\hline\text{number unfixed by square rule:}&0&1&1&5&22&293&2,\!900&52,\!475&1,\!017,\!869\\\hline\hline\text{percent fixed by square rule (\%):}&100&67&86&86&90&89&93&94&95\\\hline\end{array}}$$\vspace{-18pt}\caption{Statistics of correlator coefficients fixed by the square rule through $\ell\!=\!11$ loops.\label{square_rule_strength_table}}\vspace{-35pt}\end{table}

\newpage
\vspace{-6pt}\subsection{The Triangle Rule: Collapsing Triangles and Edges}\label{subsec:triangle_shrink}\vspace{-6pt}
The triangle rule relates the coefficients of $f$-graphs at $\ell$ loops to those at $(\ell\mi1)$ loops. Simply stated, collapsing triangles (to points) at $\ell$ loops is equivalent to collapsing edges of graphs at $(\ell\mi1)$ loops. More specifically, we can define an operation $\mathcal{T}$ that projects all $f$-graphs onto their triangular faces (identifying the points of each face), and another operation $\mathcal{E}$ that collapses all edges of $f$-graphs (identifying points). Algebraically, the triangle rule corresponds to,
\vspace{-5pt}\eq{\mathcal{T}(\mathcal{F}^{(\ell)})=2\,\mathcal{E}(\mathcal{F}^{(\ell-1)}).\label{algebraic_but_figurative_triangle_rule}\vspace{-6pt}}
Under either operation, the result is some non-conformal (generally) multi-graph with fewer vertices, with each image coming from possibly many $f$-graphs; thus, (\ref{algebraic_but_figurative_triangle_rule}) gives a linear relation between the $\ell$ loop coefficients of $\mathcal{F}^{(\ell)}$---those that project under $\mathcal{T}$ to the same image---and the $(\ell\mi1)$ loop coefficients of $\mathcal{F}^{(\ell-1)}$. (It often happens that an image of $\mathcal{F}^{(\ell)}$ under $\mathcal{T}$ is not found among the images of $\mathcal{F}^{(\ell-1)}$ under $\mathcal{E}$; in this case, the right-hand side of (\ref{algebraic_but_figurative_triangle_rule}) will be zero.) 

One small subtlety that is worth mentioning is that we must be careful about symmetry factors---as the automorphism group of the pre-image may not align with the image. To be clear, $\mathcal{T}$ acts on {\it each} triangular face of a graph (not necessarily inequivalent), and $\mathcal{E}$ acts on {\it each} edge of a graph (again, not necessarily inequivalent); each term in the image is then summed with a factor equal to the ratio of symmetry factor of the image to that of the pre-image. In both cases, this amounts to including a symmetry factor that compensates for the difference between the symmetries of an ordinary $f$-graph and the symmetries of $f$-graphs with a {\it decorated} triangle or edge. 

Let us illustrate this with an example from the seven loop correlation function. The image of $\mathcal{F}^{(7)}$ under $\mathcal{T}$ includes $433$ graph-inequivalent images---each resulting in one identity among the coefficients $c_i^{7}$ and $c_i^6$. One of these inequivalent images results in the identity: 
\vspace{-7.5pt}\begin{align}\fwbox{0pt}{\mathcal{T}\hspace{-4pt}\left(\rule{0pt}{40pt}\right.\hspace{-5pt}c^{7}_1\!\!\fwbox{70pt}{\fig{-37.25pt}{1}{seven_loop_triangle_rule_1}}\!+\!c^7_2\!\!\fwbox{70pt}{\fig{-37.25pt}{1}{seven_loop_triangle_rule_2}}\!+\!c^7_3\fwbox{70pt}{\fig{-37.25pt}{1}{seven_loop_triangle_rule_3}}\,\,+\!\ldots\hspace{-5pt}\left.\rule{0pt}{40pt}\right)\hspace{-4pt}=2\,\mathcal{E}\hspace{-4pt}\left(\rule{0pt}{40pt}\right.\hspace{-6pt}c^6_1\!\!\fwbox{70pt}{\fig{-37.25pt}{1}{six_loop_triangle_rule_1}}\!+\!\ldots\hspace{-5pt}\left.\rule{0pt}{40pt}\right)}\nonumber\\[-32pt]\fwbox{0pt}{\hspace{-15pt}\bigger{\Rightarrow}\left(c_1^7+2\,c_2^7+c_3^7\right)\fwbox{70pt}{\fig{-37.25pt}{1}{seven_loop_triangle_image}}\hspace{-10pt}=2\,c_1^6\fwbox{70pt}{\fig{-37.25pt}{1}{seven_loop_triangle_image}}\hspace{-7.5pt}\bigger{\Rightarrow}\,c_1^7+2\,c_2^7+c_3^7=2\,c_1^6.}\label{triangle_rule_example}\\[-30pt]\nonumber
\end{align}
While not visually manifest, it is not hard to check that shrinking each highlighted triangle/edge in the first line of (\ref{triangle_rule_example}) results in graphs isomorphic to the one shown in the second line. And indeed, the coefficients of the six and seven loop correlators (obtained independently) satisfy this identity: $\{c_1^7,c^7_2,c^7_3,c^6_1\}\!=\!\{\!\pl1,\!\pl1,\!\mi1,\!\pl1\}$. (The coefficient of 2 appearing in front of $c_2^7$ results from the fact that the symmetry factor of the initial graph is 1, while its image under $\mathcal{T}$ has a symmetry factor of 2.)

\vspace{-6pt}\subsubsection*{Proof and Origins of the Triangle Rule}\label{subsubsec:proof_of_triangle_rule}\vspace{-4pt}
The triangle rule arises from a reformulation of the Euclidean short distance limit of correlation functions discussed in \mbox{ref.\ \cite{Eden:2012tu,Eden:2012fe}}. In the Euclidean short distance limit  $x_2\!\rightarrow\!x_1$, the operator product expansion dictates that the leading divergence of the logarithm of  the correlation function is proportional to the one loop divergence. More precisely,
\vspace{-5pt}\eq{\lim_{x_2\rightarrow x_1}\!\log\!\Big(1+\sum_{\ell\geq1}a^\ell\,F^{(\ell)}\Big)=\gamma(a)\!\lim_{x_2\rightarrow x_1}\!F^{(1)}+\ldots,\label{konishi_log_relation}\vspace{-7pt}}
where `$a$' refers to the coupling, $F$ is defined by,
\vspace{-5pt}\eq{F^{(\ell)}\equiv3\,\frac{\mathcal{G}^{(\ell)}_4(x_1,x_2,x_3,x_4)}{\mathcal{G}^{(0)}_4(x_1,x_2,x_3,x_4)}\,,\label{definition_of_F}\vspace{-5pt}}
and where the dots in (\ref{konishi_log_relation}) refer to subleading terms in this limit. The proportionality constant $\gamma(a)$ here is the anomalous dimension of the Konishi operator, and the factor 3 in (\ref{definition_of_F}) also has a physical origin---ultimately arising from the tree-level three-point function of two stress-energy multiplets and the Konishi multiplet.\footnote{See \mbox{refs.\ \cite{Eden:2012tu,Eden:2012fe}} for details. There, the double coincidence limit was taken $x_2\!\rightarrow\!x_1$, $x_4\!\rightarrow\!x_3$, but due to conformal invariance this is in fact equivalent to the single coincidence limit we consider.}

The important point for us from (\ref{konishi_log_relation}) is that the logarithm of the correlator has the same divergence as the one loop correlator, whereas the correlator itself at $\ell$ loops diverges as the $\ell^{\text{th}}$ power of the one loop correlator $\lim_{x_2\rightarrow x_1}\!\big(\mathcal{G}_4^{(\ell)}\big)\!\sim\!\log^\ell\!\left(\x{1}{2}\right)$. At the integrand level this divergence arises from loop integration variables approaching $x_2\!=\!x_1$. The only way for a loop integral of this form---with symmetrized integration variables---to be reduced to a single log divergence is if the integrand had reduced divergence in the simultaneous limit $x_5, x_2\!\rightarrow\!x_1$, where we recall that $x_5$ is one of the loop integration variables.\footnote{The weaker requirement that the integrand only had a reduced divergence in the limit where two integration variables both approach $x_1\!=\!x_2$ would result in a divergence of at most $\log^2$, etc.}

More precisely then, defining the relevant perturbative logarithm of the correlation function as ${g}^{(\ell)}$:
\vspace{-4pt}\eq{\sum_{\ell\geq1}a^\ell g^{(\ell)}\equiv\log\!\Big(1+\sum_{\ell\geq1}a^\ell\,F^{(\ell)}\Big),\label{definition_of_g}\vspace{-2pt}}
then at the integrand-level (\ref{konishi_log_relation}) implies:\footnote{In this section we are using the same notation for both integrated functions and integrands.}
\vspace{-4pt}\eq{\lim_{x_5,x_2\rightarrow x_1}\left(\frac{g^{(\ell)}(x_1, \dots, x_{4+\ell})}{g^{(1)}(x_1,\dots,x_{5})}\right)=0,\qquad\ell\!>\!1\,.\label{eq:3b}\vspace{-3pt}}
This equation gives a clean integrand-level consequence of the reduced divergence; however, it is phrased in terms of the logarithm of the integrand rather than the integrand itself, and this does not translate directly into a graphical rule. However, notice the relation between the $\log$-expansion $g$ and the correlator $F$,
\vspace{-6pt}\eq{g^{(\ell)} = F^{(\ell)} - \frac 1\ell g^{(1)}(x_{5}) F^{(\ell-1)} - \sum_{m=2}^{\ell-1} \frac{m}{\ell} g^{(m)}(x_{5}) F^{(\ell-m)}\, .\label{logarithm_expansion}\vspace{-3pt}}
This formula can be read at the level of the integrand, and we write the dependence of the loop variable $x_{5}$ explicitly, the dependence on all other loop variables is completely symmetrized.\footnote{Note that although not manifest, the loop  variable $x_{5}$ also appears completely symmetrically in the above formula. For example, consider terms of the form $F^{(1)}F^{(\ell-1)}$. One such term arises from the second term in (\ref{logarithm_expansion}), giving $1/\ell \times F^{(1)}(x_{5}) F^{(\ell-1)}$.   Other such terms arise from the sum with $m\!=\!\ell\mi1$, giving ${(\ell\mi1)}/\ell \times F^{(\ell-1)}(x_{5}) F^{(1)}$. We see that the integration variable appears with weight 1 in $F^{(1)}$ and weight $\ell\mi1$ in $F^{(\ell-1)}$---{\it i.e.}\ completely symmetrically.} From equation (\ref{logarithm_expansion}), it is straightforward to see (using an induction argument) that (\ref{eq:3b}) is equivalent to
\vspace{-3pt}\eq{\lim_{x_2,x_{5} \rightarrow x_1} \frac{F^{(\ell)}(x_1,\dots,x_{4+\ell})}{g^{(1)}(x_1,x_2,x_3,x_4,x_{5})}= \frac1\ell\,\lim_{x_2\rightarrow x_1} F^{(\ell-1)}(x_1, \dots,\hat x_5,\dots, x_{4+\ell})\,,\label{eq:6}\vspace{-3pt}}
where the variable $x_5$ is missing in the right-hand side. This is now a direct rewriting of the reduced divergence at the level of integrands and as a relation for the loop level correlator (rather than the more complicated logarithm).

Note that everything in the discussion of this section so far can be transferred straightforwardly onto the soft/collinear divergence constraint; and indeed, a rephrasing of the soft/collinear constraint similar to (\ref{eq:6}) was conjectured in \mbox{ref.\ \cite{Golden:2012hi}}, with the relevant limit being $x_5$ approaching the line joining $x_1$ and $x_2$, $\lim_{x_{5}\rightarrow [x_1,x_2]}$.

Now inputting the one loop correlator, $\lim_{x_2,x_{5}\rightarrow x_1}g^{(1)}(x_1,\dots,x_{5})= 6/(\x{1}{5}\x{2}{5})$, and rewriting this in terms of $\mathcal{F}^{(\ell)}$, (\ref{eq:6}) becomes simply
\vspace{-3pt}\eq{\lim_{x_2,x_{5}\rightarrow x_1}(\x{1}{2} \x{1}{5}\x{2}{5})\times {{\mathcal F}^{(\ell)}(x_1, \dots, x_{4+\ell})}= 6\lim_{x_2\rightarrow x_1} (\x{1}{2})\times {\mathcal F}^{(\ell-1)}(x_1, \dots, x_{3+\ell})\, .\label{eq:7}\vspace{-3pt}}

The final step in this rephrasing of the coincidence limit  is to view (\ref{eq:7}) graphically. Clearly the limit on the left-hand side will only be non-zero if the corresponding term in the labelled \mbox{$f$-graph} contains the triangle with vertices $x_1 ,x_2, x_5$. The limit then deletes this triangle and shrinks it to a point. On the right-hand side, we similarly choose terms in the labelled \mbox{$f$-graphs} containing the edge $x_1\!\!\leftrightarrow\!x_2$, delete this edge and then shrink to a point. The equation has to hold graphically and we no longer need to consider explicit labels. Simply shrink all inequivalent (up to automorphisms) triangles of the linear sum of graphs on the left-hand side and equate it to the result of shrinking all inequivalent (again, up to automorphisms) edges of the linear sum of graphs on the right-hand side. The different (non-isomorphic) shrunk graphs are independent, and thus for each shrunk graph we obtain an equation relating $\ell$ loop coefficients to $(\ell\mi1)$ loop coefficients. There are six different labelings of the triangle and two different labelings of the edge which all reduce to the same expression in this limit, thus the factor of 6 in the algebraic expression (\ref{eq:7}) becomes the factor of 2 in the equivalent graphical version (\ref{algebraic_but_figurative_triangle_rule}).

\newpage
\vspace{-6pt}\subsection{The Pentagon Rule: Equivalence of One Loop Pentagons}\label{subsec:pentagon_rule}\vspace{-6pt}
Let us now describe the pentagon rule. It is perhaps the hardest to describe (and derive), but it ultimately turns out to imply much simpler relations among coefficients than the triangle rule. In particular, the pentagon rule will always imply that the sum of some subset of coefficients $\{c_i^\ell\}$ vanishes---with no relative factors between terms in the sum. Let us first describe operationally how these identities are found graphically, and then describe how this rule can be deduced from considerations of 5-point light-like limits according to (\ref{f_to_5pt_amp_map2}). 

Graphically, each pentagon rule identity involves a relation between $f$-graphs involving the following topologies:
\vspace{-4pt}\eq{\fig{-34.75pt}{1}{pentagon_rule_seed}\;\bigger{\Rightarrow}\left\{\!\!\fig{-34.75pt}{1}{pentagon_rule_images}\right\}\label{topologies_of_the_pentagon_rule}}
Each pentagon rule identity involves an $f$-graph with a face with the topology on the left-hand side of the figure above, (\ref{topologies_of_the_pentagon_rule}). This sub-graph is easily identified as having the structure of $\mathcal{M}_{\text{even}}^{(1)}$---see equation (\ref{five_point_one_loop_terms}). (This is merely suggestive: we will soon see that it is the role these graphs play in $\mathcal{M}_{\text{odd}}^{(1)}$ that is critical.) Importantly, these $f$-graphs may involve any number of numerators of the form $\x{{\color{hred}a}}{{\color{hblue}b}}$---including some that are `implicit': any points ${\color{hblue}x_b}$ separated from ${\color{hred}x_a}$ by a face (not connected by an edge), because for such points ${\color{hblue}x_b}$, multiplication by $\x{{\color{hred}a}}{{\color{hblue}b}}/\x{{\color{hred}a}}{{\color{hblue}b}}$ would not affect planarity of the factors in the denominator. The graphs on the right-hand side of (\ref{topologies_of_the_pentagon_rule}), then, are the collection of those \mbox{$f$-graphs} obtained from that on the left-hand side by multiplication by a simple cross-ratio:
\eq{f_i^{(\ell)}({\color{hred}x_{a}},{\color{hblue}x}_{{\color{hblue}b}},x_c,x_d)\mapsto f_{i'}^{(\ell)}\equiv f_i^{(\ell)}\frac{\x{{\color{hred}a}}{d}\x{{\color{hblue}b}}{c}}{\x{{\color{hred}a}}{{\color{hblue}b}}\x{c}{d}}.\label{cross_ratio_relation_for_pentagon_rule}}
There is one final restriction that must be mentioned. The generators of pentagon rule identities---$f$-graphs including subgraphs with the topology shown on the left-hand side of (\ref{topologies_of_the_pentagon_rule})---must not involve any numerators connecting points on the pentagon {\it other than} between ${\color{hred}x_a}$ and $x_d$ (arbitrary powers of $\x{{\color{hred}a}}{d}$ are allowed). 

While the requirements for the graphs that participate in pentagon rule identities may seem stringent, each is important---as we will see when we describe the rule's proof. But the identities that result are very powerful: they always take the form that the sum of the coefficients of  the graphs involved (both the initial graph, and all its images in (\ref{topologies_of_the_pentagon_rule})) must vanish. 

Let us illustrate these relations with a concrete example from seven loops. Below, we have drawn an $f$-graph on the left, highlighting in blue the three points $\{{\color{hblue}x_b}\}$ that satisfy requirements described above; and on the right we have drawn the three $f$-graphs related to the initial graph according to (\ref{cross_ratio_relation_for_pentagon_rule}):
\vspace{-6pt}\eq{\hspace{-234pt}\fig{-54.75pt}{1}{seven_loop_pentagon_rule_example_seed}\bigger{\Rightarrow}\!\!\left\{\rule{0pt}{47.5pt}\right.\!\!\hspace{-7.5pt}\fig{-54.75pt}{1}{seven_loop_pentagon_rule_example_images_1},\hspace{-5pt}\fig{-54.75pt}{1}{seven_loop_pentagon_rule_example_images_2},\hspace{-5pt}\fig{-54.75pt}{1}{seven_loop_pentagon_rule_example_images_3}\hspace{-7.5pt}\left.\rule{0pt}{47.5pt}\right\}\hspace{-200pt}\label{seven_loop_pentagon_rule_example}\vspace{-6pt}}
Notice that two of the three points ${\color{hblue}x_b}$ are `implicit' in the manner described above. Labeling the coefficients of the $f$-graphs in (\ref{seven_loop_pentagon_rule_example}) from left to right as \mbox{$\{c_1^{7},c_2^7,c_3^7,c_4^7\}$}, the pentagon rule would imply that \mbox{$c_1^7\pl c_2^7\pl c_3^7\pl c_4^7\!=\!0$.} And indeed, these coefficients of terms in the seven loop correlator turn out to be: \mbox{$\{c_1^{7},c_2^7,c_3^7,c_4^7\}\!=\!\{0,0,\!\pl1,\!\mi1\},$} which do satisfy this identity. 

As usual, there are no symmetry factors to consider; but it is important that only {\it distinct} images are included in the set on the right-hand side of (\ref{topologies_of_the_pentagon_rule}). As will be discussed in \mbox{section \ref{sec:results}}, the pentagon rule is strong enough to fix all coefficients but one not already fixed by the square rule through seven loops. 

\vspace{-6pt}\subsubsection*{Proof of the Pentagon Rule}\label{subsubsec:proof_of_pentagon_rule}\vspace{-4pt}
The pentagon rule~(\ref{topologies_of_the_pentagon_rule}) arises from examining the 5-point light-like limit of the correlator and its relation to the five-particle amplitude (just as the square rule arises from the 4-point light-like limit and its relation to the four-particle amplitude explained in \mbox{section \ref{subsec:square_rule}}). As described in \mbox{section \ref{subsec:higher_point_amplitude_extraction}}, in the pentagonal light-like limit the correlator is directly related to the five-particle amplitude as in (\ref{f_to_5pt_amp_map2}).

In particular let us focus on the terms involving one loop amplitudes in (\ref{f_to_5pt_amp_map2}): ${\mathcal F}^{(\ell+1)}$ contains the terms,
\vspace{-5pt}\eq{\hspace{-75pt}\frac{1}{\xi^{(5)}}\left(\mathcal{M}_{\text{even}}^{(1)}\mathcal{M}_{\text{even}}^{(\ell-1)}+\epsilon_{123456}\epsilon_{12345(m+6)}\widehat{\mathcal{M}}_{\text{odd}}^{(1)}\widehat{\mathcal{M}}_{\text{odd}}^{(\ell-1)}\right)\,.\label{eq:24}\hspace{-40pt}\vspace{-5pt}}
Indeed any term in the correlator which graphically has a plane embedding with the topology of a 5-cycle whose `inside'  contains a single vertex and whose `outside' contains $\ell\mi1$ vertices has to arise from the above terms \cite{Ambrosio:2013pba}.

Inserting the one loop expressions~\eqref{five_point_one_loop_terms} and the algebraic identity (valid only in the pentagonal light-like limit),
\vspace{-0pt}\eq{\begin{split}&\hspace{-20pt}\phantom{=\,}\frac{\epsilon_{123456}\,\epsilon_{123457}}{\x{1}{2}\x{2}{3}\x{3}{4}\x{4}{5}\x{1}{5}}\\&\hspace{-20pt}=2\,\x{6}{7}+\left[\frac{\x{1}{6}\x{2}{7}\x{3}{5}+\x{1}{7}\x{2}{6}\x{3}{5}}{\x{1}{3}\x{2}{5}}-\frac{\x{1}{7}\x{3}{6}+\x{1}{6}\x{3}{7}}{\x{1}{3}}-\frac{\x{1}{6}\x{1}{7}\x{2}{4}\x{3}{5}}{\x{1}{3}\x{1}{4}\x{2}{5}}+\text{cyclic}\right]\,,\hspace{-24pt}\end{split}\label{eq:19}\vspace{-0pt}}
then \eqref{eq:24} becomes the following contribution to ${\mathcal F}^{(\ell+1)}$
\vspace{-5pt}\eq{\begin{split}&\hspace{-30pt}\frac{1}{\x{1}{2}\x{2}{3}\x{3}{4}\x{4}{5}\x{1}{5}}\Bigg(2\,\frac{\x{6}{7}}{\x{1}{6}\x{2}{6}\x{3}{6}\x{4}{6}\x{5}{6}}\hat{\mathcal{M}}^{(\ell-1)}_{\text{odd}}+\Bigg\{\frac{1}{\x{1}{6}\x{2}{6}\x{3}{6}\x{4}{6}}\Bigg[\frac{1}{\x{1}{4}\x{2}{5}\x{3}{5}}\mathcal{M}_{\text{even}}^{(\ell-1)}\hspace{-40pt}\\&\hspace{-30pt}+\Bigg(\frac{\x{1}{7}\x{2}{4}}{\x{1}{4}\x{2}{5}}+\frac{\x{4}{7}\x{1}{3}}{\x{1}{4}\x{3}{5}}-\frac{\x{3}{7}}{\x{3}{5}}-\frac{\x{2}{7}}{\x{2}{5}}-\frac{\x{5}{7}\x{1}{3}\x{2}{4}}{\x{2}{5}\x{3}{5}\x{1}{4}}\Bigg)\hat{\mathcal{M}}_{\text{odd}}^{(\ell-1)}\Bigg]+\text{cyclic}\Bigg\}\Bigg)\,.\hspace{-40pt}\end{split}\label{eq:12}\vspace{-5pt}}

We wish to now consider all terms in ${\mathcal F}^{(\ell+1)}$ containing the structure occurring in the pentagon rule, namely a `pentawheel' with a spoke missing,
\vspace{-5pt}\eq{\fig{-34.75pt}{1}{pentagon_proof_fig_1}\label{eq:23}\vspace{-5pt}}
with numerators (if present at all within this subgraph) allowed {\it only} between the vertex with the missing spoke and the marked point (as shown). A term in ${\mathcal F}^{(\ell+1)}$ containing this subgraph inevitably contributes to the pentagonal light-like limit and by its topology it has to arise from the ${\mathcal M}^{(1)} \times {\mathcal M}^{(\ell-1)}$ terms, {\it i.e.}\ somewhere in~\eqref{eq:12}. We now proceed to investigate all seven terms in~\eqref{eq:12} to show that this structure of interest can only arise from the fifth and sixth terms.  

We start with the second term of \eqref{eq:12}
\vspace{-5pt}\eq{\frac{1}{\x{1}{2}\x{2}{3}\x{3}{4}\x{4}{5}\x{5}{1}}\,\frac{1}{\x{1}{6}\x{2}{6}\x{3}{6}\x{4}{6}}\,\frac{1}{\x{1}{4}\x{2}{5}\x{3}{5}}{\mathcal M}_{\text{even}}^{(\ell-1)}\,,\label{eq:3}\vspace{-5pt}}
arising from the even part of the amplitude, which is the most subtle one. Graphically, this term can be displayed as:
\vspace{-5pt}\eq{\fig{-34.75pt}{1}{pentagon_proof_fig_2}\label{pentagon_proof_figure_2}\vspace{-5pt}}
In order for this to yield the structure \eqref{eq:23} in a planar \mbox{$f$-graph}, the amplitude ${\mathcal M}_{\text{even}}^{(\ell-1)}$ must either contain a numerator $\x{1}{4}$ (to cancel the corresponding propagator above) or alternatively it must contain the numerator terms $\x{2}{5}$ and $\x{3}{5}$ in order to allow the edge $\x{1}{4}$ to be drawn  outside the pentagon without any edge crossing. Analyzing these different possibilities one concludes that this requires all three numerators $\x{1}{4}\x{2}{5}\x{3}{5}$ to be present in a term of ${\mathcal M}_{\text{even}}^{(\ell-1)}$. Now using the amplitude/correlator duality again in a different way note that such a contribution to ${\mathcal M}_{\text{even}}^{(\ell-1)}$ must also contribute to the lower-loop correlator ${\mathcal F}^{(\ell)}$ through~(\ref{f_to_5pt_amp_map2})
\vspace{-5pt}\eq{\lim_{\substack{\text{5-point}\\\text{light-like}}}\!\!\left(\xi^{(5)}\mathcal{F}^{(\ell-1)}\right)={\mathcal M}_\text{even}^{(\ell-1)} + \ldots\,.\label{eq:28}\vspace{-5pt}}
So a term in  ${\mathcal M}_{\text{even}}^{(\ell-1)}$ with numerators $\x{1}{4}\x{2}{5}\x{3}{5}$ contributes a term with topology,
\vspace{-5pt}\eq{\fig{-34.75pt}{1}{pentagon_proof_fig_3}\label{pentagon_proof_figure_3}\vspace{-5pt}}
(Here the numerators $\x{1}{4}\x{2}{5}\x{3}{5}$ cancel three of the denominators of $1/\xi^{(5)}$, but they leave the pentagon and two further edges attached to the pentagon as shown.)

We see that this term can never be planar (this term in ${\mathcal M}_{\text{even}}^{(\ell-1)}$ has to be attached to all five external legs by conformal invariance so one cannot pull one of the offending edges outside the pentagon) {\it unless} there is a further numerator term, either $\x{2}{4}$ or $\x{1}{3}$ to cancel one of these edges. But in this case inserting this back into~\eqref{eq:3} we obtain the required structure~\eqref{eq:23} but with this further numerator which is of the type explicitly disallowed from our rule.

Having ruled out the second term, we consider the other terms of~\eqref{eq:12}.  The first term can clearly never give a pentawheel with a spoke missing. The contribution of the third term of~\eqref{eq:12} has the diagrammatic form:
\vspace{-5pt}\eq{\fig{-34.75pt}{1}{pentagon_proof_fig_4}\label{pentagon_proof_figure_4}\vspace{-2.5pt}}
and so could potentially give a contribution of the form of a pentawheel with a spoke missing if $\hat{\mathcal M}_\text{odd}^{(\ell-1)}$ has a numerator $\x{1}{4}$ to cancel the corresponding edge. However in any case such a term would also contain the numerator $\x{2}{4}$ which we disallow in~\eqref{eq:23}. The third and last terms are similarly ruled out as a source for the structure in question. So we conclude that the fifth and sixth terms are the only ones which can yield the structure we focus on in the pentagon rule.

Given this important fact, we are now in a position to understand the origin of the pentagon rule. Every occurrence of the structure~\eqref{eq:23} arises from the fifth or sixth terms in~\eqref{eq:12}, namely from $\x{3}{7}/\x{3}{5}\times \hat {\mathcal M}_{\text{odd}}^{(\ell-1)}$ (where $x_3$ is the marked point of the pentagon). But we also know \cite{Ambrosio:2013pba} that $\hat {\mathcal M}_{\text{odd}}^{(\ell-1)}$ is in direct one-to-one correspondence with pentawheel structures of $f^{(\ell+1)}$ (the first term in~\eqref{eq:12}). Thus there is a direct link between the pentawheel structures and the  structure~\eqref{eq:23} and this link appears with a sign due to the sign difference between the first and fifth/sixth terms in~\eqref{eq:12}. To get from the first term of~\eqref{eq:12} to the fifth term, one multiplies by $\x{3}{7}\x{5}{6}/(\x{3}{5}\x{6}{7})$---that is, deleting the two edges, $\x{3}{7}$ and $\x{5}{6}$, and deleting the two numerator lines $\x{6}{7},\x{3}{5}$. This is precisely the operation involved in the five-point rule described in more detail above (see~(\ref{cross_ratio_relation_for_pentagon_rule})).

\newpage
\vspace{-6pt}\section{Bootstrapping Amplitudes/Correlators to Many Loops}\label{sec:results}\vspace{-6pt}
In this section, we survey the relative strengths of the three rules described in the pervious section, and then some of the more noteworthy aspects of the forms found for the correlator through ten loops. Before we begin, however, it is worth emphasizing that the three rules we have used are only three among many which follow from the way in which lower loop (and higher point) amplitudes are encoded in the correlator $\mathcal{F}^{(\ell)}$ via equations (\ref{f_to_4pt_amp_map_with_series_expansion}) and (\ref{f_to_npt_amp_map}). The triangle, square, and pentagon rules merely represent those we implemented first, and which proved sufficient through ten loops. And finally, it is worth mentioning that we expect the soft-collinear bootstrap criterion to continue to prove sufficient to fix all coefficients at all loops, even if using this tool has proven computationally out of reach beyond eight loops. (If it were to be translated into a purely graphical rule, it may prove extraordinarily powerful.)

~\\[-34pt]\paragraph{The Square Rule:}~\\
\indent As described in the previous section, the square rule is undoubtedly the most powerful of the three, and results in the simplest possible relations between coefficients---namely, that certain $\ell$ loop coefficients are identical to particular $(\ell\mi1)$ loop coefficients. As illustrated in \mbox{Table \ref{square_rule_strength_table}}, the square rule is strong enough to fix $\sim\!95$\% of the $22,\!097,\!035$ $f$-graphs coefficients at eleven loops. The role of the triangle and pentagon rules, therefore, can be seen as tools to fix the coefficients not already fixed by the square rule. 

~\\[-34pt]\paragraph{The Triangle Rule:}~\\
\indent Similar to the square rule, the triangle rule is strong enough to fix all coefficients through three loops, but will leave one free coefficient at four loops. Conveniently, the relations required by the triangle rule are not the same as those of the square rule, and so the combination of the two fix everything. In fact, the square and triangle rule together immediately fix all correlation functions through seven loops, and all but 22 of the $2,\!709$ eight loop coefficients. (This fact was known when the eight loop correlator was found in \mbox{ref.\ \cite{Bourjaily:2015bpz}}, which is why we alluded to these new rules in the conclusions of that Letter.)

Interestingly, applying the triangle and square rules to nine loops fixes all but 3 of the $43,\!017$ {\it new coefficients}, including 20 of those not already fixed at eight loops. (To be clear, this means that, without any further input, there would be a total of $3\pl2$ unfixed coefficients at nine loops.) Motivated by this, we implemented the triangle and square rules at ten loops, and found that these rules sufficed to determine eight and nine loop correlators uniquely. At ten loops, we found the complete system of equations following from the two rules to fix all but $1,\!570$ of the coefficients of the $900,\!145$ $f$-graphs. 

These facts are summarized in \mbox{Table \ref{square_and_triangle_rules_strength_table}}. Notice that the number of unknowns quoted in that table for $\ell$ loops are the number of coefficients given the lower loop correlator. If the coefficients at lower loops were not assumed, then there would be $5$ unknowns at nine loops rather than 3; but the number quoted for ten loops would be the same---because all lower loop coefficients are fixed by the ten loop relations. 

\begin{table}[t]\caption{Statistics of coefficients fixed by the square \& triangle rules through $\ell\!=\!10$ loops.\label{square_and_triangle_rules_strength_table}}\vspace{-10pt}$$\fwbox{0pt}{\begin{array}{|r|r|r|r|r|r|r|r|r|r|}\cline{2-10}\multicolumn{1}{r}{\ell\!=}&\multicolumn{1}{|c|}{2}&\multicolumn{1}{c|}{3}&\multicolumn{1}{c|}{4}&\multicolumn{1}{c|}{5}&\multicolumn{1}{c|}{6}&\multicolumn{1}{c|}{7}&\multicolumn{1}{c|}{8}&\multicolumn{1}{c|}{9}&\multicolumn{1}{c|}{10}\\\hline\text{number of $f$-graph coefficients:}&\,1\,&\,1\,&\,3\,&\,7\,&\,36\,&\,220\,&\,2,\!709\,&\,43,\!017\,&\,900,\!145\,\\\hline\text{unknowns remaining after square rule:}&\,\,0\,&\,\,0\,&\,\,1\,&1\,&5\,&22\,&293\,&2,\!900\,&52,\!475\,\\\hline\text{unknowns after square \& triangle rules:}&\,\,0\,&\,\,0\,&\,\,0\,&\,\,0\,&\,\,0\,&\,\,0\,&22\,&3\,&1,\!570\,\\\hline\end{array}}$$\vspace{-18pt}\vspace{-0pt}\end{table}

~\\[-34pt]\paragraph{The Pentagon Rule:}~\\
\indent The pentagon rule is not quite as strong as the others, but the relations implied are much simpler to implement. In fact, there are no instances of $f$-graphs for which the pentagon rule applies until four loops, when it implies a single linear relation among the three coefficients. This relation, when combined with the square rule fixes the four loop correlator, and the same is true for five loops. However at six loops, the two rules combined leave 1 (of the $36$) $f$-graph coefficients undetermined. The reason for this is simple: there exists an $f$-graph at six loops which neither contributes to $\mathcal{A}^{(5)}_4\mathcal{A}^{(1)}_4$ nor to $\mathcal{M}_5^{(4)}\overline{\mathcal{M}}_5^{(1)}$. This is easily seen by inspection of the $f$-graph in question: 
\vspace{-5pt}\eq{\fig{-54.75pt}{1}{six_loop_prism_graph}\label{six_loop_prism_graph}\vspace{-5pt}}
We will have more to say about this graph and its coefficient below. There is one graph at seven loops related to (\ref{six_loop_prism_graph}) by the square rule that is also left undetermined, but all other coefficients (219 of the 220) are fixed by the combination of the square and pentagon rules. 

The number of coefficients fixed by the square and pentagon rules through nine loops is summarized in \mbox{Table \ref{square_and_pentagon_rules_strength_table}}. As before, only the number of {\it new} coefficients are quoted---assuming that the lower loop coefficients are known. 

\begin{table}[b]\vspace{-40pt}$$\fwbox{0pt}{\begin{array}{|r|r|r|r|r|r|r|r|r|}\cline{2-9}\multicolumn{1}{r}{\ell\!=}&\multicolumn{1}{|c|}{2}&\multicolumn{1}{c|}{3}&\multicolumn{1}{c|}{4}&\multicolumn{1}{c|}{5}&\multicolumn{1}{c|}{6}&\multicolumn{1}{c|}{7}&\multicolumn{1}{c|}{8}&\multicolumn{1}{c|}{9}\\\hline\text{number of $f$-graph coefficients:}&\,1\,&\,1\,&\,3\,&\,7\,&\,36\,&\,220\,&\,2,\!709\,&\,43,\!017\,\\\hline\text{unknowns remaining after square rule:}&\,\,0\,&\,\,0\,&\,\,1\,&\,\,1\,&\,\,5\,&22\,&293\,&2,\!900\,\\\hline\text{unknowns after square \& pentagon rules:}&\,\,0\,&\,\,0\,&\,\,0\,&\,\,0\,&\,\,\,1\,&\,\,\,0\,&\,\,\,17\,&\,\,\,64\,\\\hline\end{array}}$$\vspace{-18pt}\caption{Statistics of coefficients fixed by the square \& pentagon rules through $\ell\!=\!9$ loops.\label{square_and_pentagon_rules_strength_table}}\vspace{-30pt}\end{table}

\newpage
\vspace{-0pt}\subsection{Aspects of Correlators and Amplitudes at High Loop-Orders}\label{subsec:statistical_tour}\vspace{-0pt}
While no two of the three rules alone prove sufficient to determine the ten loop correlation function, the three in combination fix all coefficients uniquely---without any outside information about lower loops. As such, the reproduction of the eight (and lower) loop functions found in \mbox{ref.\ \cite{Bourjaily:2015bpz}} can be viewed as an independent check on the code being employed. Moreover, because the three rules each impose mutually overlapping (and individually over constrained) constraints on the coefficients, the existence of any solution is a source of considerable confidence in our results. 

One striking aspect of the correlation function exposed only at high loop-order is that the (increasingly vast) majority of coefficients are zero: while all possible $f$-graphs contribute through five loops, only 26 of the 36 graphs at six loops do; by ten loops, $85\%$ of the coefficients vanish. (At eleven loops, {\it at least} $19,\!388,\!448$ coefficients vanish ($88\%$) due to the square rule alone.) This pattern is illustrated in \mbox{Table \ref{correlator_contributions_table}}, where we count all contributions---both for $f$-graphs, and planar DCI integrands. 

The two principle novelties discovered for the eight loop correlator \cite{Bourjaily:2015bpz} also persist to higher loops. Specifically, we refer to the fact that there are contributions to the amplitude that are finite (upon integration) even on-shell, and contributions to the correlator that are (individually) divergent even off-shell. The meaning of the finite integrals remains unclear (although they would have prevented the use of the soft-collinear bootstrap without grouping terms according to $f$-graphs); but the existence of divergent contributions imposes an important constraint on the result: because the correlator is strictly finite off-shell, all such divergences must cancel in combination. (Moreover, these contributions impose an interesting technical obstruction to evaluation, as they cannot be easily regulated in four dimensions---such as by going to the Higgs branch of the theory \cite{Alday:2009zm}.)
 
\begin{table}[b]\vspace{-15pt}$$\hspace{1.5pt}\begin{array}{|@{$\,$}c@{$\,$}|@{$\,$}r@{$\,$}|@{$\,$}r@{$\,$}|@{$\,\,$}r@{$\,\,$}|@{$\;\;\;\;\;\;\;$}|@{$\,$}r@{$\,$}|@{$\,$}r@{$\,$}|@{$\,\,$}r@{$\,\,$}|}\multicolumn{1}{@{$\,$}c@{$\,$}}{\begin{array}{@{}l@{}}\\[-4pt]\text{$\ell\,$}\end{array}}&\multicolumn{1}{@{$\,$}c@{$\,$}}{\!\begin{array}{@{}c@{}}\text{number of}\\[-4pt]\text{$f$-graphs}\end{array}}\,&\multicolumn{1}{@{$\,$}c@{$\,$}}{\begin{array}{@{}c@{}}\text{no.\ of $f$-graph}\\[-4pt]\text{contributions}\end{array}}\,\,&\multicolumn{1}{@{$\,$}c@{$\,$}}{\begin{array}{@{}c@{$\,\,\,\,\;\;\;\;\;$}}\text{}\\[-4pt]\text{\!\!(\%)}\end{array}}&\multicolumn{1}{@{$\,$}c@{$\,$}}{\begin{array}{@{}c@{}}\text{number of}\\[-4pt]\text{DCI integrands}\end{array}}\,&\multicolumn{1}{@{$\,$}c@{$\,$}}{\begin{array}{@{}c@{}}\text{no.\ of integrand}\\[-4pt]\text{contributions}\end{array}}\,&\multicolumn{1}{@{$\,$}c@{$\,$}}{\begin{array}{@{}c@{}}\text{}\\[-4pt]\text{(\%)}\end{array}}\\[-0pt]\hline1&1&1&100&1&1&100\\\hline2&1&1&100&1&1&100\\\hline3&1&1&100&2&2&100\\\hline4&3&3&100&8&8&100\\\hline5&7&7&100&34&34&100\\\hline6&36&26&72&284&229&81\\\hline7&220&127&58&3,\!239&1,\!873&58\\\hline8&2,\!709&1,\!060&39&52,\!033&19,\!949&38\\\hline9&43,\!017&10,\!525&24&1,\!025,\!970&247,\!856&24\\\hline10&900,\!145&136,\!433&15&24,\!081,\!425&3,\!586,\!145&15\\\hline\end{array}\vspace{-16pt}$$\vspace{-6pt}\caption{Statistics of $f$-graph and DCI integrand {\it contributions} through $\ell\!=\!10$ loops.\label{correlator_contributions_table}}\vspace{-24pt}\end{table}
\newpage 

At eight loops there are exactly 4 $f$-graphs which lead to finite DCI integrands, and all 4 have non-vanishing coefficients. At nine loops there are 45, of which 33 contribute; at ten loops there are $1,\!287$, of which $570$ contribute. For the individually divergent contributions, their number and complexity grow considerably beyond eight loops. The first appearance of such divergences happened at eight loops---with terms that had a so-called `$k\!=\!5$' divergence (see \cite{Bourjaily:2015bpz} for details). Of the 662 $f$-graphs with a $k\!=\!5$ divergence at eight loops, only 60 contributed. At nine loops there are $15,\!781$, of which $961$ contribute; at ten loops, there are $424,\!348$, of which $21,\!322$ contribute. Notice that terms with these divergences grow proportionally in number---and even start to have the feel of being ubiquitous asymptotically. We have not enumerated all the divergent contributions for $k\!>\!5$, but essentially all categories of such divergences exist and contribute to the correlator. (For example, there are $971$ contributions at ten loops with (the simplest category of) a $k\!=\!7$ divergence.)

While the coefficients of $f$-graphs are encouragingly simple at low loop-orders, the variety of possible coefficients seems to grow considerably at higher orders. The distribution of these coefficients is given in \mbox{Table \ref{coefficient_statistics_table}}. While all coefficients through five loops were $\pm\!1$, those at higher loops include many novelties. (Of course, the increasing dominance of zeros among the coefficients is still rather encouraging.)

Interestingly, it is clear from \mbox{Table \ref{coefficient_statistics_table}} that new coefficients only appear at even loop-orders. The first term with coefficient $\!\mi1$ occurs at four loops, and the first appearance of $\!\pl2$ at six loops. At eight loops, we saw the first instances of $\pm\frac{1}{2}$, $\pm\frac{3}{2}$, and also $\!\mi5$. And there are many novel coefficients that first appear at ten loops. 
\begin{table}[t]$$\hspace{-1.2pt}\begin{array}{|c|r|r|r|r|r|r|r|r|r|r|r|r|r|r|}\multicolumn{1}{c}{}&\multicolumn{14}{c}{\text{number of $f$-graphs at $\ell$ loops having coefficient:}}\\\cline{2-15}\multicolumn{1}{c}{\ell}&\multicolumn{1}{|c|}{\pm1\phantom{}}&\multicolumn{1}{c|}{0}&\multicolumn{1}{c|}{\pm2}&\multicolumn{1}{c|}{\pm1/2}&\multicolumn{1}{c|}{\pm3/2}&\multicolumn{1}{c|}{\pm5}&\multicolumn{1}{c|}{\pm1/4}&\multicolumn{1}{c|}{\pm3/4}&\multicolumn{1}{c|}{\pm5/4}&\multicolumn{1}{c|}{+7/4}&\multicolumn{1}{c|}{\pm9/4}&\multicolumn{1}{c|}{\pm5/2}&\multicolumn{1}{c|}{+4}&\multicolumn{1}{c|}{+14}\\\hline1&1&{\color{dim}0}&{\color{dim}0}&{\color{dim}0}&{\color{dim}0}&{\color{dim}0}&{\color{dim}0}&{\color{dim}0}&{\color{dim}0}&{\color{dim}0}&{\color{dim}0}&{\color{dim}0}&{\color{dim}0}&{\color{dim}0}\\\hline2&1&{\color{dim}0}&{\color{dim}0}&{\color{dim}0}&{\color{dim}0}&{\color{dim}0}&{\color{dim}0}&{\color{dim}0}&{\color{dim}0}&{\color{dim}0}&{\color{dim}0}&{\color{dim}0}&{\color{dim}0}&{\color{dim}0}\\\hline3&1&{\color{dim}0}&{\color{dim}0}&{\color{dim}0}&{\color{dim}0}&{\color{dim}0}&{\color{dim}0}&{\color{dim}0}&{\color{dim}0}&{\color{dim}0}&{\color{dim}0}&{\color{dim}0}&{\color{dim}0}&{\color{dim}0}\\\hline4&3&{\color{dim}0}&{\color{dim}0}&{\color{dim}0}&{\color{dim}0}&{\color{dim}0}&{\color{dim}0}&{\color{dim}0}&{\color{dim}0}&{\color{dim}0}&{\color{dim}0}&{\color{dim}0}&{\color{dim}0}&{\color{dim}0}\\\hline5&7&{\color{dim}0}&{\color{dim}0}&{\color{dim}0}&{\color{dim}0}&{\color{dim}0}&{\color{dim}0}&{\color{dim}0}&{\color{dim}0}&{\color{dim}0}&{\color{dim}0}&{\color{dim}0}&{\color{dim}0}&{\color{dim}0}\\\hline6&25&10&1&{\color{dim}0}&{\color{dim}0}&{\color{dim}0}&{\color{dim}0}&{\color{dim}0}&{\color{dim}0}&{\color{dim}0}&{\color{dim}0}&{\color{dim}0}&{\color{dim}0}&{\color{dim}0}\\\hline7&126&93&1&{\color{dim}0}&{\color{dim}0}&{\color{dim}0}&{\color{dim}0}&{\color{dim}0}&{\color{dim}0}&{\color{dim}0}&{\color{dim}0}&{\color{dim}0}&{\color{dim}0}&{\color{dim}0}\\\hline8&906&1,\!649&9&141&3&1&{\color{dim}0}&{\color{dim}0}&{\color{dim}0}&{\color{dim}0}&{\color{dim}0}&{\color{dim}0}&{\color{dim}0}&{\color{dim}0}\\\hline9&7,\!919&32,\!492&54&2,\!529&22&1&{\color{dim}0}&{\color{dim}0}&{\color{dim}0}&{\color{dim}0}&{\color{dim}0}&{\color{dim}0}&{\color{dim}0}&{\color{dim}0}\\\hline10&78,\!949&763,\!712&490&50,\!633&329&9&5,\!431&559&18&5&4&4&1&1\\\hline\end{array}$$\vspace{-24pt}\caption{Statistics of $f$-graph coefficients in the expansion of $\mathcal{F}^{(\ell)}$ through $\ell\!=\!10$ loops.\label{coefficient_statistics_table}}\vspace{-10pt}\end{table}

While most of the `new' coefficients occur with sufficient multiplicity to require further consideration (more than warranted here), there is at least one class of contributions which seems predictably novel. Consider the following six, eight, and ten loop $f$-graphs: 
\vspace{-10pt}\eq{\fig{-54.75pt}{1}{six_loop_prism_graph}\quad\fig{-54.75pt}{1}{eight_loop_prism_graph}\quad\fig{-54.75pt}{1}{ten_loop_prism_graph}\label{prism_graph_figures}\vspace{-10pt}}
These graphs all have the topology of a $(\ell/2\pl2)$-gon anti-prism, and all represent contributions with unique (and always exceptional) coefficients. In particular, these graphs contribute to the correlator with coefficients $\!\pl2$, $\!\mi5$ and $\!\!\pl14$, respectively. (Notice also that the four loop $f$-graph $f_3^{(4)}$ shown in (\ref{one_through_four_loop_f_graphs}) is an anti-prism of this type---and is the first term having contribution $\!\mi1$---as is the only two loop $f$-graph (the octahedron), which also follows this pattern.) Each of the $f$-graphs in (\ref{prism_graph_figures}) contribute a unique DCI integrand to the $\ell$ loop amplitude,
\vspace{-12pt}\eq{\fig{-54.75pt}{1}{six_loop_coeff_2_dci_int}\qquad\fig{-54.75pt}{1}{eight_loop_coeff_5_dci_int}\qquad\fig{-54.75pt}{1}{ten_loop_coeff_14_dci_int}\label{prism_dci_integrands}\vspace{-12pt}}
with each drawn in momentum space as Feynman graphs for the sake intuition. From these, a clear pattern emerges---leading us to make a rather speculative guess for the coefficients of these terms. It seems plausible that the coefficients of anti-prism graphs are given by the Catalan numbers---leading us to predict that the coefficient of the octagonal anti-prism $f$-graph at twelve loops, for example, will be $\!\mi42$. Testing this conjecture---let alone proving it---however, must await further work. 

The only other term that contributes at ten loops with a unique coefficient is the following, which has coefficient $\!\pl4$: 
\vspace{-10pt}\eq{\fig{-54.75pt}{1}{ten_loop_coeff_4_graph}\;\;\bigger{\supset}\fig{-54.75pt}{1}{ten_loop_coeff_4_dci_int}\label{ten_loop_coefficient_4_graph},\ldots\vspace{-10pt}}

We hope that the explicit form of the correlation functions provided as part of this work's submission files to the {\tt arXiv} (see \mbox{Appendix \ref{appendix:mathematica_and_explicit_results}}) will provide sufficient data for other researchers to find new patterns within the structure of coefficients. 

\newpage
\vspace{-6pt}\section{Conclusions and Future Directions}\label{sec:conclusions}\vspace{-6pt}
In this work, we have described a small set of simple, graphical rules which prove to be extremely efficient in fixing the possible contributions to the $\ell$ loop four-point correlation function in planar maximally supersymmetric $(\mathcal{N}\!=\!4)$ Yang-Mills theory (SYM). And we have described the form that results when used to fix the correlation function through ten loop-order. While clearly this is merely the simplest non-trivial observable in (arguably) the simplest four-dimensional quantum field theory, it exemplifies many of the features (and possible tools) we expect will be applicable to more general quantum field theories. And even within the limited scope of planar SYM, this single function contains important information about higher-point amplitudes. 

It is important to reiterate that the rules we have described are merely necessary conditions---and not obviously sufficient to all orders. But these three rules are merely three among many that follow from the consistency of the amplitude/correlator duality. Even without extension beyond ten loops, it would be worthwhile (and very interesting) to explore the strengths of the various natural generalizations of the rules we have described. 

Another important open direction would be to explore the systematic extraction of higher-point (lower loop) amplitudes from the four-point correlator. This has proven exceptionally direct and straight-forward for five-point amplitudes, but further work should be done to better understand the systematics (and potential difficulties) of this procedure for higher multiplicity. (Even six-particle amplitude extraction remains largely unexplored.)

Finally, it is natural to wonder how far this programme can be extended beyond ten loops. Although the use of graphical rules essentially eliminates the challenges of setting up the linear algebra problem to be solved, solving the system of equations that result (with millions of unknowns) rapidly becomes rather non-trivial. However, such problems of linear algebra (involving (very) large systems of equations) arise in many areas of physics and computer science, and there is reason to expect that they may be surmounted through the use of programmes such as that described in \mbox{ref.\ \cite{vonManteuffel:2014ixa}} (an impressive implementation of Laporta's algorithm). At present, it is unclear where the next computational bottle-neck will be, but it is worth pushing these tools as far as they can go---certainly to eleven loops, and possibly even twelve.

\newpage
\vspace{-0pt}\section*{Acknowledgements}\vspace{-6pt}
The authors gratefully acknowledge helpful discussions with Zvi Bern, Simon Caron-Huot, JJ Carrasco, Dmitry Chicherin, Burkhard Eden, Gregory Korchemsky, Emery Sokatchev, and Marcus Spradlin. This work was supported in part by the Harvard Society of Fellows, a grant from the Harvard Milton Fund, by the Danish National Research Foundation (DNRF91), and by a MOBILEX research grant from the Danish Council for Independent Research (JLB); by an STFC studentship (VVT); and by an STFC Consolidated Grant ST/L000407/1 and the Marie Curie network GATIS (gatis.desy.eu) of the European Union's Seventh Framework Programme FP7/2007-2013 under REA Grant Agreement No.\ 317089 (PH). PH would also like to acknowledge the hospitality of Laboratoire dÕAnnecy-le-Vieux de Physique Th\'eorique, UMR 5108, where this work was completed.

\vspace{12pt}\appendix\section{Obtaining and Using the Explicit Results in {\sc Mathematica}}\label{appendix:mathematica_and_explicit_results}\vspace{-6pt}
Our full results, including all contributions to the amplitude and correlator $\mathcal{F}^{(\ell)}$ through ten loops, have been included as part of this work's submission files to the {\tt arXiv}. These can be obtained by downloading the `source files' for this paper on the {\tt arXiv} website. (On certain systems, it may be useful to manually append the extension `{\tt .tar.gz}' to the downloaded file name.) In this source file can be found several data files (encoded somewhat esoterically), a {\sc Mathematica} package, {\tt consolidated\rule[-0.75pt]{7.5pt}{0.5pt}multiloop\rule[-0.75pt]{7.5pt}{0.5pt}data.m}, and a notebook {\tt multiloop\rule[-0.75pt]{7.5pt}{0.5pt}demo.nb}. 

The demonstration notebook illustrates the principle data defined in the package, and examples of how these functions are represented. Also included in the package are several general-purpose functions that may be useful to the reader---for example, a functions that compute symmetry factors and check if two functions are isomorphic (as graphs). Principle among the data included in this package are the list of all $f$-graphs at $\ell$ loops with non-vanishing coefficients for $\ell\!=\!1,\ldots,10$, and the corresponding coefficients. Also included is a list of all $\ell$ loop DCI integrands obtained from each $f$-graph in the light-like limit. 

Importantly, we have only included terms with non-vanishing coefficients---in order to reduce the file size of the data. The complete list of $f$-graphs at each loop order can be obtained by contacting the authors.

\newpage
\providecommand{\href}[2]{#2}\begingroup\raggedright\endgroup

\end{document}